\documentclass{aa}
\usepackage{natbib}
\bibpunct{(}{)}{;}{a}{}{,}

\usepackage{lastpage}
\usepackage{color}
\usepackage{graphicx}
\usepackage{xcolor}
\usepackage{hyperref}
\usepackage[normalem]{ulem}
\usepackage{xurl}
\hypersetup{colorlinks=true,citecolor=blue,linkcolor=blue,anchorcolor=blue}

\usepackage{txfonts}

\begin{document}

    \title{When galaxies refuel: Evolution of the perturbed spiral galaxy NGC\,1385}
    \titlerunning{Evolution of NGC\,1385}
    \authorrunning{Kang et al.}

   \author{Xiaoyu~Kang
          \inst{1}\fnmsep\thanks{kxyysl@ynao.ac.cn}
          \and
          Eva~Sextl\inst{2,3}
          \and
          Rolf-Peter~Kudritzki\inst{2,4}
          \and
          Hassen~M.~Yesuf\inst{5}\fnmsep\thanks{yesufh@shao.ac.cn}
          \and
          Fenghui~Zhang\inst{1}
          \and
          Ruixiang~Chang\inst{5,6}
          \and
          Xiejin~Li\inst{1,7}
          \and
          Yunkun~Han\inst{1}
          }

   \institute{International Center of Supernovae (ICESUN), Yunnan Key Laboratory of Supernova Research, Yunnan Observatories, Chinese Academy of Sciences (CAS), Kunming 650216, People’s Republic of China
         \and University Observatory Munich,
         Ludwig-Maximilian-Universit$\rm \ddot{a}$t M$\rm \ddot{u}$nchen, Scheinerstr. 1, 81679 Munich, Germany
        \and Excellence Cluster ORIGINS, Boltzmannstr. 2, 85748 Garching, Germany
         \and Institute for Astronomy, University of Hawaii, 2680 Woodlawn Drive, Honolulu, HI96822, USA
         \and Shanghai Astronomical Observatory, Chinese Academy of Sciences, 80 Nandan Rd., Shanghai, 200030, People’s Republic of China
         \and Key Lab for Astrophysics, Shanghai, 200034, People’s Republic of China
         \and University of Chinese Academy of Sciences, Beijing 100049, People’s Republic of China
             }

   \date{Received 18 May 2026; accepted 21 July 2026}

 \abstract
{The spiral galaxy NGC\,1385 is characterized by a vigorous and protracted history of star formation, particularly in its central regions, that led to a current star formation rate that surpasses those of comparable systems. We analyze the evolution history of the galaxy using spatially resolved optical and submillimeter spectroscopy obtained from the PHANGS survey combined with WALLABY radio survey data. We constructed radial distributions of the star formation rate and the interstellar medium (ISM) neutral and molecular gas-mass surface densities and measured the metallicity distribution of the stellar populations using a refined full-spectrum fitting population synthesis method together with a determination of the ISM oxygen abundances using H\,II region emission lines. The metallicities of the young stars and the ISM are similar and show an almost flat distribution. This is crucially different from previous work, which found a positive gradient for the average stellar metallicities.
We fit a chemical evolution model (incorporating gas infall, outflow, and radial inflow) to the observed data of NGC\,1385. Based on this fit, the evolution of NGC\,1385 is characterized by the typical inside-out disk formation of spiral galaxies, even though our model does not assume an a priori shorter gas-infall timescale for the inner disk than for the outer disk.
However, the galaxy has experienced sustained star formation over gigayear timescales with a star formation efficiency that is higher by a factor of two than normal. This explains the high metallicity of the young stars, which is higher by 0.15 to 0.2\,dex than solar, and the flat metallicity distribution throughout the disk. The model predictions for the stellar metallicities agree well with the observed values, supporting the robustness of our refined spectral fitting analysis and the inferred evolutionary scenario.
}



   \maketitle

\section{Introduction}
\label{sec:intro}

Spatially resolved maps of key galactic properties such as chemical composition, star formation rates (SFRs), and gas conditions provide essential insights into the physical processes that drive galaxy evolution
\citep[e.g.,][]{Pessa2021, Pessa2023, Stuber2023, Groves2023, Lin2024, Sextl2024, Sextl2025, Sextl2026,Kreckel2025}.
Galaxies are complex ecosystems that evolve through the interplay of a few key physical quantities. One quantity is the chemical abundance: the gas-phase oxygen abundance, measured as $12 + \log(\mathrm{O/H})$, traces the present-day enrichment in the interstellar medium (ISM), while the average metallicity of all stars, old and young, reflects the chemical history of the galaxy over time. Another quantity is the star formation rate (SFR), which indicates how quickly new stars form, and which is regulated by gas cycling and feedback. The third quantity  is molecular gas, especially cold molecular hydrogen gas ($\rm{H}_2$), which serves as the direct fuel for star formation. The radial profiles of these parameters offer a powerful diagnostic for unraveling the formation history of a galaxy and the physical mechanisms at play.

In this context, the nearby barred spiral galaxy NGC\,1385 \citep{Buta2015}, located at a distance of
17.2\,Mpc \citep{Anand2021}, presents a particularly intriguing laboratory. Its basic observational properties are summarized in Table\,\ref{Tab:obs1}. NGC\,1385 is one of the targets of the project Physics at High Angular resolution in Nearby GalaxieS (PHANGS). This major multiwavelength survey is designed to resolve the key phases of the star formation cycle.
The project synergizes optical integral-field spectroscopy from the Multi Unit Spectroscopic Explorer (MUSE) with submillimeter interferometric imaging from the Atacama Large Millimeter/submillimeter Array (ALMA), providing a comprehensive view of stellar populations, ionized gas physics, and the cold molecular gas reservoir \citep{Leroy2021a, Emsellem2022}.
Studies of its resolved molecular gas and star formation at $\sim$100\,pc scales reveal a steeper-than-average resolved molecular gas main-sequence (rMGMS) slope, suggesting a higher star formation efficiency \citep[SFE,][]{Pessa2021} that might be linked to external gas accretion \citep{For2021}.
Most importantly, analyses of its stellar populations show strong positive radial gradients in both the  luminosity-weighted and mass-weighted average metallicities of the total stellar population. This is a significant departure from the typical inside-out formation scenario \citep{Pessa2023}. This picture contrasts sharply with the results from its ionized gas: H\,II region observations revealed a  negative gas-phase metallicity gradient \citep{Groves2023, Brazzini2024, Kreckel2025}, indicating a mismatch between the current chemical state of the star-forming gas and the long-term chemical history locked in the stars. The gas-phase metallicity provides a snapshot of the metal content at a given time, while the mean stellar metallicity reflects the time-averaged value of the ISM metal content over the star formation history (SFH) of the galaxy.

\begin{table}
\caption{Basic properties of NGC\,1385.}
\label{Tab:obs1}
\centering
\setlength{\tabcolsep}{2pt}
\begin{tabular}{lll}
\hline
Property            &    Value               &             \\
\hline
name                &    NGC\,1385           &              \\
RA                  &    $3^{\rm h}37^{\rm m}28^{\rm s}.6$      &  \citet{Leroy2021b} \\
Dec                 &    $-24^{\rm \circ}30^{\rm '}04^{\rm ''}$ &  \citet{Leroy2021b} \\
Morphology          &    SB(s)dm pec                 &     \citet{Buta2015} \\
Distance            &    $17.2\,\rm Mpc$     &   \citet{Anand2021}          \\
Inclination         &    $44^{\circ}$        &   \citet{Lang2020}            \\
Position Angle (PA)         &    $181.3^{\circ}$        &   \citet{Lang2020}           \\
${\rm R}_{25}$            &    $102^{\prime\prime}$  &   \citet{Makarov2014}   \\
$\log_{10}(\rm M_*/{\rm M}_{\odot})$  &   9.98   &  \citet{Leroy2021b}    \\
$\log_{10}(\rm M_{{\rm H}_2}/{\rm M}_{\odot})$  &   9.20   &  \citet{Leroy2021b}    \\
$\log_{10}(\rm M_{\rm HI}/{\rm M}_{\odot})$  &   9.19   &  \citet{Leroy2021b}    \\
$\log_{10}(\rm SFR/{\rm M}_{\odot}\,{\rm yr}^{-1})$  &   0.32   &  \citet{Leroy2021b}    \\
\hline
\end{tabular}\\
\end{table}

The apparent anticorrelation between its ionized gas metallicity gradient and the stellar metallicity gradient \citep{Pessa2023, Kreckel2025} suggests that NGC\,1385 might have experienced a complex and irregular chemical evolution history over gigayear timescales. This discrepancy is puzzling and
motivates this work. Furthermore, the neutral atomic hydrogen (HI) observations revealed a severely distorted gas morphology and a southern tidal debris field without an optical counterpart, providing clear evidence of a recent tidal interaction \citep{For2021}. \citet{Veronese2025} suggested a potential but weak interaction between NGC\,1385 and its group companion NGC\,1371 based on their estimated distances. This recent perturbation likely plays a key role in shaping the present-day properties of the galaxy, including its gas dynamics and star formation pattern.

To understand the origins of these observed features (particularly the potentially distinct metallicity gradients in stars and gas, and the impact of external accretion of metal-poor gas), we use a theoretical framework that accounts for the coupled effects of star formation, chemical enrichment, and internal gas transport. To this end, we first establish the present-day physical properties of NGC\,1385 using PHANGS (MUSE+ALMA) and the Widefield Australian Square Kilometre Array Pathfinder (ASKAP) L-band Legacy All-sky Blind Survey (WALLABY) data, namely the radial distributions of the metallicity, star formation rate surface density ($\Sigma_{\rm SFR}$), molecular hydrogen gas-mass surface density ($\Sigma_{\rm H2}$), molecular gas velocity dispersion ($\sigma_{\rm mol}$), and neutral atomic hydrogen gas-mass surface density ($\Sigma_{\rm HI}$).
We determine gas-phase metallicities using three independent strong-line methods and compare them with the auroral line measurements by \citet{Kreckel2025}. Most importantly, we apply our well-tested full-spectrum fitting population synthesis analysis \citep{Sextl2025,Sextl2026} to determine the metallicities of the young stars and the average of the total stellar population, and $\Sigma_{\rm SFR}$.
We then employ a chemical evolution model, which is constrained by the observed radial distributions of $\Sigma_{\rm H2}$, $\Sigma_{\rm HI}$, $\Sigma_{\rm SFR}$, and the metallicity of young stars and H\,II regions. We use this model and compare it with the observations to constrain the star formation and chemical enrichment history of NGC\,1385.

This paper is structured as follows. We start in Section\,\ref{sec:miscon} by reassessing the key properties of NGC\,1385. Sections\,\ref{sec:obv} and\,\ref{sec:analyzing} then describe the observational data, analysis, and results for NGC\,1385. Section\,\ref{sec:model}
presents the main ingredients of our chemical evolution model. Section\,\ref{sec:Model results} reports the model fit results and a discussion, and Section\,\ref{sec:sum} provides our conclusions.

\section{\texorpdfstring{NGC\,1385: basic properties}{NGC1385: Basic properties}}
\label{sec:miscon}

NGC\,1385 is often referred to as a relatively nearby face‑on spiral galaxy that is useful for detailed studies of star formation and galaxy structure. However, its basic properties such as the presence of a bar and its precise distance are not consistently agreed upon in the literature. It is mostly morphologically classified as a barred spiral galaxy \citep{Buta2015} with a total stellar mass of approximately $\log_{10}(M_*/{\rm M}_{\odot})=9.98$ \citep{Stuber2023}. However, it is not always straightforward to determine whether a galaxy possesses a bar. It can be quite ambiguous and can depend on the interpretation by the astronomer \citep{Iles2025}. NGC\,1385 is one such example. Well-known catalogs such as that by \citet{Vaucouleurs1991} (based on UBV photometric plates) and the S$^4$G infrared survey \citep{Buta2015} list this galaxy as Hubble type SB. In contrast, \citet{Stuber2023} and \citet{Querejeta2021} both concluded from H$\alpha$ and CO images that NGC\,1385 lacks a bar entirely. At least all studies describe its irregular patchy morphology as similar to the morphology of the Magellanic Clouds.

The galaxy distance is also still debated. Direct methods such as TRGB and Cepheid-based measurements are absent from the literature; only the Tully--Fisher relation, the planetary nebula luminosity function (PNLF), and the numerical action method (NAM) have been applied, with significantly different results. \citet{Scheuermann2022} reported a distance of $9.81^{+0.63}_{-1.46}$\,Mpc based on a PNLF, whereas \citet{Jacoby2024} obtained $25.0^{+1.4}_{-1.5}$\,Mpc using the PNLF applied to the same underlying PHANGS-MUSE data cube. Different authors commonly agree strongly on PNLF distances. The strong discrepancy with the earlier result is attributed to the improved extraction and classification of planetary nebulae within the data cube. \citet{For2021} and \citet{Wong2021} adopted in their study of neutral hydrogen gas related to NGC\,1385 a distance of 21\,Mpc based on a distance estimate for the most massive early-type galaxy NGC\,1395 in the Eridanus group and assumed that the other galaxies in the group and the observed HI clouds are at the same distance. \citet{Veronese2025} worked with a distance of 22.7\,Mpc, following \citet{Leroy2019}, who used the Extragalactic Distance Database (EDD) of \citet{Tully2009}. \citet{Leroy2019} reported an uncertainty of $32\%$ for their distance.

The computational technique NAM reconstructs galaxy orbits in the gravitational potential of the local  Universe by minimizing an action principle \citep{Shaya2017}. This yields an intermediate distance of $17.2 \pm 2.58$\,Mpc. This distance was also adopted in the main PHANGS compilation of galaxy distances \citep{Anand2021}, and we therefore also use it throughout the paper. We note that although the distance to NGC\,1385 is relatively uncertain, most of the results and related conclusions obtained in the following are not severely affected by this uncertainty.

\begin{figure}
  \centering
  \includegraphics[angle=0,scale=0.45]{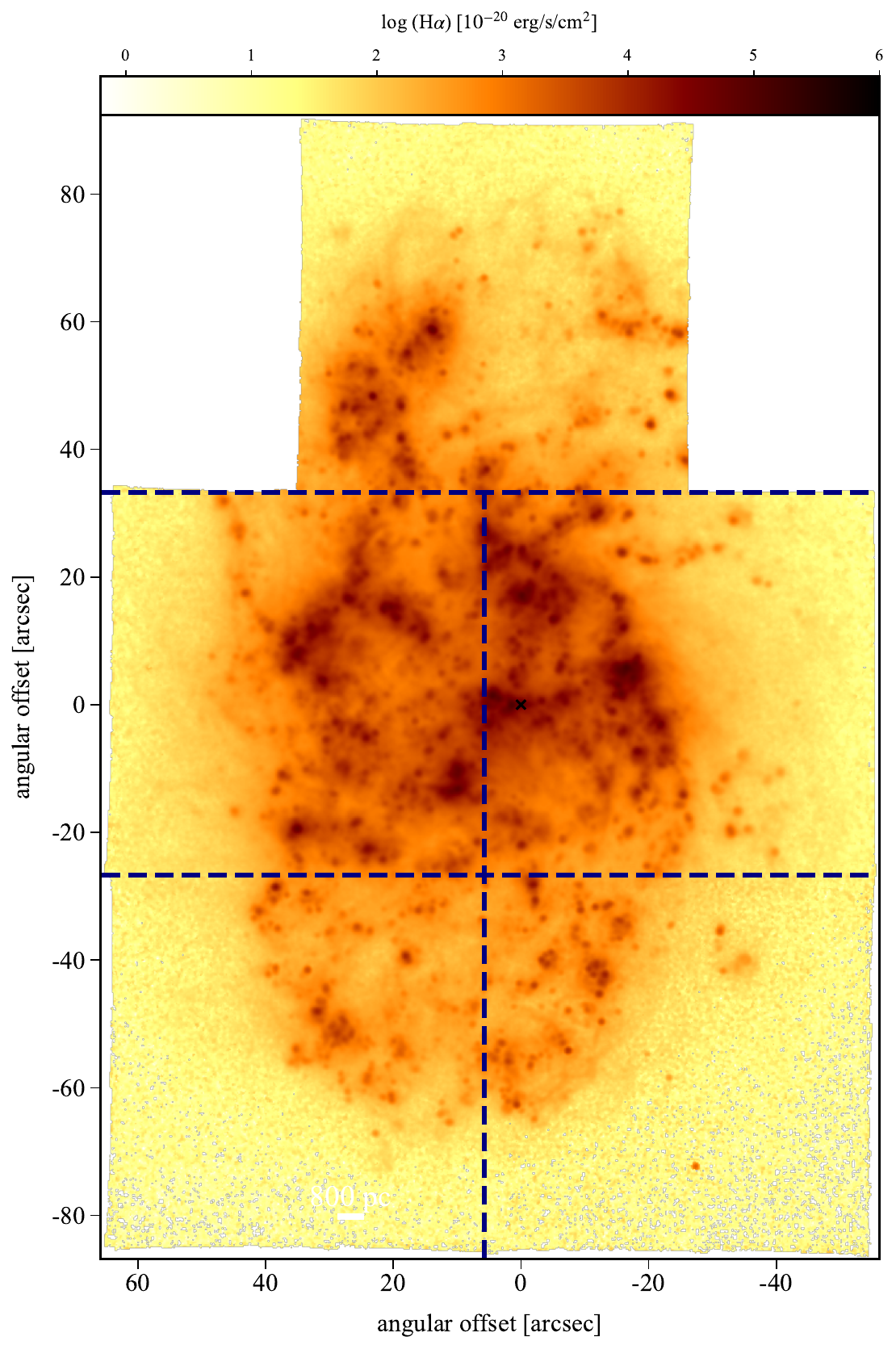}
    \caption{NGC\,1385 H$\alpha$ image constructed from the PHANGS-MUSE IFU survey. The borders of the five observed MUSE fields are indicated.}
\label{fig:Halpha_overview}
\end{figure}

\section{Observations and data analysis}
\label{sec:obv}

Our analysis of NGC\,1385 is primarily based on complementary spatially resolved datasets from the PHANGS survey. Specifically, we used optical integral-field unit (IFU) spectroscopy from the PHANGS-MUSE project \footnote{\url{https://www.phangs.org/home}} and (sub)millimeter interferometric imaging from the PHANGS-ALMA project. The MUSE data provide a spectral cube covering the wavelength range of $4800$--$9300$\,\AA\ at a spatial sampling of $0.2''$. The NGC\,1385 mosaic is composed of five separate spatial observations collected between October 2019 and December 2020. These data are publicly available and were accessed via the ESO Science Portal\footnote{\url{https://www.canfar.net/storage/vault/list/phangs/RELEASES/PHANGS-MUSE/DR1.0}}. We exclusively worked with the \texttt{COPT} data cubes, where the PSF between different observations of the mosaic was homogenized. These data cubes enable a better mapping of the spectra of the integrated stellar populations, ionized gas emission lines, and key diagnostics such as the $\Sigma_{\rm SFR}$ and the stellar and ionized gas-phase metallicity \citep{Emsellem2022, Pessa2023, Groves2023}.

For the cold molecular gas component, the ALMA data \footnote{\url{http://www.canfar.net/storage/list/phangs/RELEASES/PHANGS-ALMA/}} deliver maps of the $^{12}$CO($J=2$--$1$) line emission, which traces the distribution and kinematics of cold molecular gas.
The observations include data from the total power antennas and achieve a spatial resolution in the range of $1''.0$--$1''.5$ \citep{Leroy2021a}.

In addition, the HI data for NGC\,1385 were obtained from the WALLABY \citep{Koribalski2020, Westmeier2022, Deg2022, Murugeshan2024}.
The data cube for NGC\,1385 we used was retrieved in FITS format \footnote{\url{https://wallaby-survey.org/data/data-pilot-survey-dr2/}}.

\subsection[Gas-phase metallicity, 12+log(O/H)]{ISM gas-phase metallicity}
\label{sec:Zg}

To measure the gas-phase metallicities, we used the PHANGS-MUSE nebular catalog provided by \citet{Groves2023,Groves2025}. In this work, a H$\alpha$ emission map of NGC\,1385 was constructed and a total 1,029 candidate H\,II regions were identified. For each nebula, an integrated spectrum was constructed by combining all MUSE spectral pixels (spaxels) associated with the region. These spectra were fitted using the penalized pixel-fitting method \citep[pPXF, ][]{Cappellari2017} employing single stellar population (SSP) models from the E-MILES library \citep{Vazdekis2016} and Gaussian components for strong emission lines such as H$\beta$, H$\alpha$, [O\,III], [N\,II], and [S\,II]. The fitting also yielded line fluxes and kinematics (velocity and velocity dispersion) for each H\,II region.
All the measured line fluxes were corrected for dust reddening by determining $E(B-V)$ from the $\rm H\alpha/H\beta$ ratio (the intrinsic ratio was assumed to be 2.86) using the \citet{ODonnell1994} extinction curve with $R_{V}=3.1$. The dereddened H$\alpha$ fluxes were then converted into luminosities based on the distance to NGC\,1385.

To obtain a pure sample of H\,II regions without nebular sources associated with nonstellar ionization (e.g., supernova remnants, planetary nebulae, and AGN),
we applied the BPT diagram \citep{BPT1981} classification criteria. Specifically, the empirical demarcation from \citet{Kauffmann2003} was used for the BPT--N\,II diagram to isolate pure star-forming regions, while the theoretical extreme starburst line from \citet{Kewley2001} was used for the BPT--S\,II and BPT--O\,I diagrams to exclude AGN and shock-ionized regions. For our analysis, we defined star-forming H\,II regions as objects satisfying $\mathrm{BPT\_NII}=0$, $\mathrm{BPT\_SII}=0$, and $\mathrm{BPT\_OI}\leq 0$. Objects where [O\,I] is undetected ($\mathrm{S/N}<5$) but that were otherwise consistent with H\,II region classifications in the other two diagrams were also included.
Using these criteria, we obtained a final sample of 914 pure star-forming H\,II regions.

The strong emission lines from H\,II regions are easily detectable and provide empirical estimates of gas-phase metallicities, most importantly, the abundances of oxygen. However, as is well known (see, e.g., \citealt{Kewley2008, Bresolin2016, Teim2021}), these strong-line method abundances are subject to significant systematic uncertainties and depend heavily on the calibration used. Therefore, we used three different calibrations as an indication of the range of potential uncertainties. As an alternative, we also compared our results with oxygen abundances that were obtained with the direct method, which uses electron-temperature-dependent weak auroral lines.

As the first strong-line method, following \citet{Kreckel2019}, we adopted the S-calibration (Scal) prescription from \citet[][hereafter Scal-PG16]{Pilyugin2016} to determine the gas-phase metallicity. This choice was motivated by the small intrinsic scatter of this method when compared to metallicities derived from auroral lines.

The Scal-PG16 calibration relies on the following three standard diagnostic line ratios:
\begin{align*}
    N_2 &= I_{\rm [N\,II] \lambda 6548+ \lambda 6584} / I_{\rm H\beta}, \\
    S_2 &= I_{\rm [S\,II] \lambda 6717+ \lambda 6731} / I_{\rm H\beta}, \\
    R_3 &= I_{\rm [O\,III] \lambda 4959+ \lambda 5007} / I_{\rm H\beta}.
\end{align*}
We only measured the stronger line of each doublet for [O\,III] and [N\,II], adopting a fixed doublet ratio of approximately 3:1 to estimate the contribution of the weaker component
\citep{Emsellem2022}.

The Scal-PG16 calibration is defined separately for the upper and lower branches of $\log N_2$. For star-forming regions on the upper branch, where $\log N_2 \ge -0.6$, the metallicity was calculated as
\begin{align}
    \label{equation:ohsu}
    \begin{split}
        12+\log(\mathrm{O/H}) = & \; 8.424 + 0.030 \, \log (R_3/S_2) + 0.751 \, \log N_2 \\
                                & \; + (-0.349 + 0.182 \, \log (R_3/S_2) + 0.508 \log N_2) \\
                                & \; \times \log S_2,
    \end{split}
\end{align}
and for regions on the lower branch, where $\log N_2 \le -0.6$, the metallicity was calculated as
\begin{align}
    \label{equation:ohsl}
    \begin{split}
        12+\log(\mathrm{O/H}) = & \; 8.072 + 0.789 \, \log (R_3/S_2) + 0.726 \, \log N_2 \\
                                & \; + (1.069 - 0.170 \, \log (R_3/S_2) + 0.022 \log N_2) \\             & \; \times \log S_2.
    \end{split}
\end{align}

As the second strong-line method, we employed the N2S2H$\alpha$ calibrator from \citet{2016Ap&SS.361...61D},
\begin{equation}
    \mathrm{N2S2H\alpha} = \log\left( \frac{I_{[\rm N\,II]\lambda 6584}}{I_{[\rm S\,II](\lambda 6717+6731)}} \right)
                         + 0.264\,\log\left( \frac{I_{[\rm N\,II]\lambda 6584}}{I_{\rm H\alpha}} \right),
\end{equation}
involving the NII/H${\alpha}$ term, which is considered to be less sensitive to the ionization parameter \citep{ZhangK2017}, thereby complementing the Scal-PG16 calibration.

The empirical relation between the oxygen abundance and N2S2H$\alpha$ was then given by
\begin{equation}\label{eq:N2S2}
    12+\log(\mathrm{O/H}) = 8.77 + \mathrm{N2S2H\alpha} + 0.45\,(\mathrm{N2S2H\alpha}+0.3)^5.
\end{equation}

As the third strong-line approach, we adopted the method by \citet[][PP04]{PP04} based on the O3N2 ratio,
\begin{equation}\label{eq:o3n2}
    \mathrm{O3N2} = \log_{10}\left( \frac{I_{[\rm O\,III]\lambda 5007} / I_{\rm H\beta}}{I_{[\rm N\,II]\lambda 6584} / I_{\rm H\alpha}} \right).
\end{equation}
This ratio is largely insensitive to differential dust attenuation and shows a monotonic single-valued trend over its valid range. The gas-phase oxygen abundance was then derived through
\begin{equation}
    12+\log(\mathrm{O/H}) = 8.73 - 0.42 \times \mathrm{O3N2}.
\end{equation}
We chose the PP04 method from the many additionally available strong-line methods because it agrees reasonably well with accurate and well-tested spectroscopic stellar metallicity determinations of massive blue supergiant stars in star-forming galaxies \citep{Bresolin2009,Bresolin2016,Bresolin2025,Kudritzki2024}.

An alternative to the strong-line methods is the inclusion of weak auroral lines, which allow us to constrain the H\,II electron temperature. In their PHANGS-MUSE study of metallicity gradients in nearby galaxies, \cite{Kreckel2025} used the forbidden nitrogen lines [\rm N\,II]$\lambda$5755 and [\rm N\,II]$\lambda$6583 to measure gas-phase oxygen abundances. We included their results for NGC\,1385 as an additional comparison.

\subsection{Full-spectrum fitting}
\label{sec:fullspectral_fit}

In addition to ISM emission lines, the MUSE IFU spectra provide crucial spatially resolved information about the stellar population through the spectral energy distributions (SEDs) and absorption lines provided by the stars contributing to the light in each spectral pixel (spaxel). In principle, a full-spectrum fitting technique simultaneously fits all wavelength points of the spectra with a library of high-resolution SSP models across a range of ages and metallicities. This nonparametric SFH approach lets the data define complex mixtures of stellar populations of different ages and metallicities without imposing a predetermined functional form.
Through the fit, we can retrieve information about the metallicity of the young and old stars, the average of the total stellar population, and the distribution of interstellar dust in the disk of NGC\,1385. In practice, our workflow consisted of three consecutive steps: removal of strong flux outliers, followed by the determination of kinematic parameters, and finally, the determination of stellar population properties such as age, dust, and metallicity. The code we applied for the full-spectrum fitting is \texttt{pPXF} \citep{Cappellari2017, Cappellari2023}, and the overall method is described in detail in \citet{Sextl2024, Sextl2025, Sextl2026}.

For the data in PHANGS-MUSE, this workflow needs to be adapted further. The mosaic nature of this survey makes the sky subtraction even more challenging. The PHANGS-MUSE survey paper \citet{Emsellem2022} found that an incorrect sky-background is present for about $10\%$ to $20\%$ of the taken spectra even after extensive efforts to remove it. This does not affect the extraction of emission lines (i.e., for the H\,II region analysis in \citet{Barnes2026}, \citet{Groves2023}, \citet{Groves2025}), but a full-spectrum fitting, which uses the overall continuum of the spectra, ultimately runs into problems. Nonphysical offsets between adjacent mosaic fields plague the 2D maps of age, metallicity, and dust. \citet{Pessa2023} therefore developed a better approach for the official PHANGS-MUSE that splits the problem of stellar fitting into two steps. The spectra are dereddened, and ages and metallicities are only inferred in a second step. We built upon their work for our analysis and developed it further (see below).

An additional important aspect of our approach must be mentioned: in the set of SSP template spectra, which are used to calculate the contribution of stars of a specific age and metallicity to the total stellar spectrum of each bin, the young stars need to be well represented at all ages. The use of the popular MILES stellar library \citep{Sanchez2006, Barroso2011} alone is not sufficient for NGC\,1385 and needs to be improved with further spectra from hot and post-MS stars. This limitation does not reflect a shortcoming of the library itself, but rather arises because vigorously star-forming systems such as NGC\,1385 were not among its primary use cases. Consequently, hot-star spectra from \citet{Eldridge2017}, Wolf–Rayet spectra from \citet{Smith2002}, AGB stars from \citet{Lancon2000}, post-AGB from \citet{Rauch2003}, and carbon-rich stars from \citet{Aringer2009} were incorporated into the SSP templates. The underlying stellar evolution is encoded in the MIST isochrones \citep{Dotter2016}, and a Chabrier IMF was assumed \citep{Chabrier2003}. In our SSP set, which was then calculated with \texttt{fsps} \citep{Conroy2010}, roughly one-third of the template spectra correspond to ages of 20 Myr or younger. Using these templates, \citet{Sextl2025} found that the [Z]$_y$ derived from a full-spectrum fitting and independent stellar probes agreed well, including blue supergiant stars and star clusters in M83. This crucial test confirmed the reliability of stellar metallicities obtained with a full-spectrum fitting.

\subsubsection{New workflow for PHANGS-MUSE}

Because the accuracy and stability of the full-spectrum fitting approach depend critically on a high level of the signal-to-noise ratio (S/N) of the spectra, especially over the limited wavelength range considered here, we first optimized the data quality across the observed MUSE fields. To this end, we applied a Voronoi binning using the \texttt{PowerBin} package \citep{PowerBin2025} to obtain an S/N of $100$ in the stellar continuum at 5500~\AA. Individual observations of the southern fields $4$ and $5$ (see figure\,6 in \citet{Emsellem2022}) of the data cube show a substantially lower data quality than in the northern fields. We included these regions in our analysis, but the Voronoi bins in these regions were roughly three times larger to achieve the target S/N of $100$. In summary, all 19403 Voronoi bins contained fewer than 250 spaxels, corresponding to an maximum area of approximately 3.2\,arcsec $\times$ 3.2\,arcsec.

A spectral fitting mostly focuses on the stellar continuum and stellar absorption lines alone. We therefore masked H\,II emission lines and absorption features contaminated with emission. For PHANGS-MUSE, an additional masking was necessary for the regions [5900 6100]\,\AA~and [6900, 6940]\,\AA,~which still show atmospheric airglow emission. These are not visible in broad- or narrow-band colors, but the slight additional emission was visible in preliminary tests.
After this, we adopted a two-step approach: First, we ran the spectra through our pipeline to only obtain the value of the color excess E(B-V) caused by the interstellar dust. Tests indicated that a wavelength range of $4900 - 5900$\,\AA~was optimal for this purpose because the bias introduced by the sky subtraction is negligible in this region. Continuum-correction polynomials, bootstrapping, or outlier-clipping parameters played only a minor role at this stage. In the second part, the spectra were dereddened using the Calzetti law (\citet{Calzetti2000} with R$_V = 4.05$) as described in \citet{Pessa2023}. The wavelength region was expanded to $4850 - 7000$\,\AA. After this, we started again with a new outlier clipping, kinematic redetermination, and stellar population fitting. The stellar ages and metallicities were fitted this time with a 12th degree continuum-correction polynomial consistent with the PHANGS-MUSE pipeline in \citet{Pessa2023}. A bootstrapping procedure (30 iterations, implemented following \citet{Davidson2008}) finally yielded robust ages, metallicities, and their uncertainties. This procedure removed systematic offsets while remaining computationally efficient through parallelization with \texttt{joblib}\footnote{\url{https://github.com/joblib/joblib}}. In the second part of the pipeline, the bootstrapping procedure is a crucial element.

\subsubsection{A chemically consistent definition of the stellar metallicity}
\label{subsec:zdef}

To determine stellar metallicities through a population synthesis analysis, we used the metal content of the young stellar population (Z$_y$) instead of the commonly applied luminosity-weighted or mass-weighted metallicity averages over all stars, young and old. Specifically, Z$_y$ was determined as the ratio of the mass M$^y_Z$ of all metals confined in the young stars divided by their total mass M$^y$, that is, Z$_y$ = M$^y_Z$/M$^y$, and we used [Z]$_y$ = log Z$_y/$Z$_{\odot}$ (with Z$_{\odot}$ the metallicity of the \text{Sun}) to characterize the metal content of the young stars. For the metallicitiy of all stars, young and old, with a total mass M and metal content M$_Z$, we used Z$_{\rm star}$ = M$_Z$/M and [Z]$_{\rm star}$ = log Z$_{\rm star}$/Z$_{\odot}$. For the relative distribution of the element abundances in our model calculation of the spectra, we used the abundance pattern of the solar photosphere as described in \citet{Asplund2009}. The metallicity introduced in this way is consistent with the analysis of individual stars and with the chemical evolution, and it can be used for a direct comparison, whereas the frequently used luminosity- or mass-weighted averages of metallicities over masses and ages of all stars introduce very problematic biases for such comparisons \citep[see][for a detailed discussion]{Sextl2025, Sextl2026}. We therefore studied [Z]$_y$, the contribution of all stars younger than 100\,Myr, explicitly as another way to access the current chemical composition of a galaxy. We also investigated the average metallicity of all stars [Z]$_{\rm star}$. \\

\subsection[SFR surface density, Sigma SFR]{SFR surface density, $\Sigma_{\rm SFR}$}
\label{sec:SFR}

Through the full-spectrum fitting population synthesis described above, we separated the contribution of stars of different ages to the total observed spectrum of each Voronoi bin. This allowed us to measure the stellar mass growth as a function of time, and thus, the SFRs. The method is described in \cite{Sextl2025}. For a good average over sequences of recent star formation events, we considered the last 200\,Myr to obtain an estimate of the current SFR.

As an alternative, we also estimated the SFR from the dust-corrected H$\alpha$ luminosity using the calibration from \citet[][Equation 6]{Calzetti2007}, that is,
\begin{equation}
    \label{eq:sfr}
    \left( \frac{\rm SFR}{{\rm M}_{\odot}\,{\rm yr}^{-1}} \right)
    = 5.3\times10^{-42}\left( \frac{{L(\rm H}\alpha_{\rm corr})}{{\rm erg\,s}^{-1}} \right),
\end{equation}
where $L({\rm H}\alpha_{\rm corr})$ is the H$\alpha$ luminosity corrected for internal extinction \citep{Groves2023}. A MUSE H$\alpha$ image of NGC\,1385 is shown in Figure \ref{fig:Halpha_overview}. As we show in Section \ref{sec:analyzing}, the SFRs of the two methods roughly agree.

The $\Sigma_{\rm SFR}$ is then given by
\begin{equation}
    \label{eq:sfrd}
    \left( \frac{\Sigma_{\rm SFR}}{{\rm M}_{\odot}\,{\rm yr}^{-1}\,{\rm kpc}^{-2}} \right)
    = 4.814\left( \frac{\rm SFR}{\rm area_{phy}} \right),
\end{equation}
where $\rm area_{phy}$ is the surface area considered, converted from an angular area (in $\rm arcsec^{2}$) into a physical area (in $\rm kpc^{2}$),
using the distance of NGC\,1385.

We note that the SFR densities derived from H$\alpha$ depend weakly on the size of $\rm area_{phy}$. After some experiments, we settled on $6~\rm arcsec^{2}$, corresponding to $0.5~\rm kpc^{2}$. At this scale, the results become independent of the surface area. Similar experiments for the population synthesis SFR densities showed that the choice of $\rm area_{phy}$ has no significant effect. We therefore used the SFR densities obtained from the Voronoi bins.

\subsection[Molecular hydrogen gas mass surface density, Sigma H2]{Molecular hydrogen gas-mass surface density, $\Sigma_{\rm H2}$}
\label{sec:H2}

The direct measurement of the H$_{2}$ mass is challenging because its lacks a dipole moment. Therefore, we used carbon monoxide (CO), the second most abundant molecule, and its rotational transitions as a tracer for H$_{2}$ because it probes a similar region of the ISM. We focused on the CO(2-1) emission line, which offers observational efficiency by mapping a given mass surface density, sensitivity, and angular resolution approximately 2-4 times faster than CO(1-0) \citep{Leroy2021b} and used the integrated-intensity maps generated through the broad masking scheme, which is designed to maximize completeness and encompass the full extent of the galaxy emission.
The observing strategy, data reduction procedures, and product generation are detailed in \citet{Leroy2021a}.

The measured CO(2-1) intensity was obtained directly from the publicly released PHANGS-ALMA moment-0 map, and we subsequently derived the H$_{2}$ gas-mass surface density ($\Sigma_{\rm H_{2}}$) on a spaxel-by-spaxel basis across the full field of view, using the native resolution of the CO map.
The CO(2-1) intensity was first transformed to the CO(1-0) intensity using the line ratio $I_{\rm CO(1-0)} = R^{-1}_{21} \times I_{\rm CO(2-1)}$ (for the value of $R_{21}$, see below). The resulting CO(1-0) intensity was then converted into the total $\Sigma_{\rm H_{2}}$ by using the CO-to-H$_{2}$ conversion factor ($\alpha_{\rm CO}$). The conversion is given by
$\Sigma_{\rm H_{2}} = \alpha_{\rm CO} \times I_{\rm CO(1-0)} \times {\rm cos}(i)$, where $i$ is the inclination of the galaxy.
\citet{Accurso2017} and \citet{Sun2020} adopted a metallicity-dependent $\alpha_{\rm CO}$ conversion factor of $4.35\,Z^{\prime\,-1.6}\,\rm M_{\odot}\,\rm pc^{-2}\,(K\,km\,s^{-1})^{-1}$ (with $Z^{\prime}\equiv Z/Z_{\odot}$). For NGC\,1385, which has an inclination $i=44^{\circ}$ and a near-solar average metallicity ($Z^{\prime}\sim1$), this prescription converges to the canonical Galactic value $\alpha_{\rm CO}=4.35\,\rm M_{\odot}\,\rm pc^{-2}\,(K\,km\,s^{-1})^{-1}$. We therefore adopted this constant $\alpha_{\rm CO}$ together with the standard line ratio $R_{21}=0.65$ \citep{denBrok2021, Leroy2021b, Pessa2021}.
The $\Sigma_{\rm H_{2}}$ can then be calculated using the equation
\begin{equation}
    \label{eq:H2}
    \left( \frac{\Sigma_{\rm H_{2}}}{{\rm M}_{\odot}\,{\rm pc}^{-2}} \right)
    = 4.81\left( \frac{I_{\rm CO(2-1)}}{\rm K\,km\,s^{-1}} \right),
\end{equation}
where $\Sigma_{\rm H_{2}}$ is projected to face-on values and includes the contribution from helium.

In addition, we also extracted the $\sigma_{\rm mol}$ directly from the publicly released PHANGS-ALMA moment-2 map (the strictly masked product). This map provides the intensity-weighted root-mean-square scatter about the mean velocity in units of $\rm km\,s^{-1}$. The strict masking procedure effectively excludes noise-dominated pixels, yielding a reliable dispersion measurements for regions with a high signal-to-noise ratio of the galaxy disk.

\subsection[Neutral atomic hydrogen gas mass surface density, Sigma HI]{Neutral atomic hydrogen gas-mass surface density, $\Sigma_{\rm HI}$}
\label{sec:HI}

Under the standard assumption that the HI 21\,cm emission line is optically thin, the HI column density ($N_{\rm HI}$) can be directly derived from the observed flux density using a standard conversion formula that accounts for cosmological corrections \citep{Meyer2017}. The moment-0 flux maps obtained from WALLABY are calibrated in units of Jy\,Hz. In these maps, the flux $F$ at each spaxel represents the integrated line emission over that specific pixel. The conversion into $N_{\rm HI}$ is given by
\begin{equation}
    \left( \frac{N_{\rm HI}}{\rm cm^{-2}} \right) \approx 2.64 \times 10^{20} \, (1+z)^4 \left( \frac{F}{\rm Jy\,Hz} \right) \left( \frac{\Omega}{\rm arcsec^2} \right)^{-1},
    \label{eq:nhi}
\end{equation}
where $z$ is the cosmological redshift of the source, and $\Omega$ is the corresponding solid angle in arcsec$^2$.

The above $N_{\rm HI}$ expressions can be transformed into HI gas-mass surface densities ($\Sigma_{\rm HI}$) by simply multiplying by $m_{\rm H}$ and applying the appropriate unit conversions, for example,
\begin{align}
    \left( \frac{\Sigma_{\rm HI}}{\rm M_\odot\,pc^{-2}} \right) &= 8.01 \times 10^{-21} \left( \frac{N_{\rm HI}}{\rm cm^{-2}} \right), \\
    &= 2.12 \, (1+z)^4 \left( \frac{F}{\rm Jy\,Hz} \right) \left( \frac{\Omega}{\rm arcsec^2} \right)^{-1}.
\end{align}

\section{Observational results}
\label{sec:analyzing}

\subsection{Distribution of the ISM gas-phase oxygen abundance}
\label{sec:Zg1}

\begin{figure*}
  \centering
  \includegraphics[angle=0,scale=0.65]{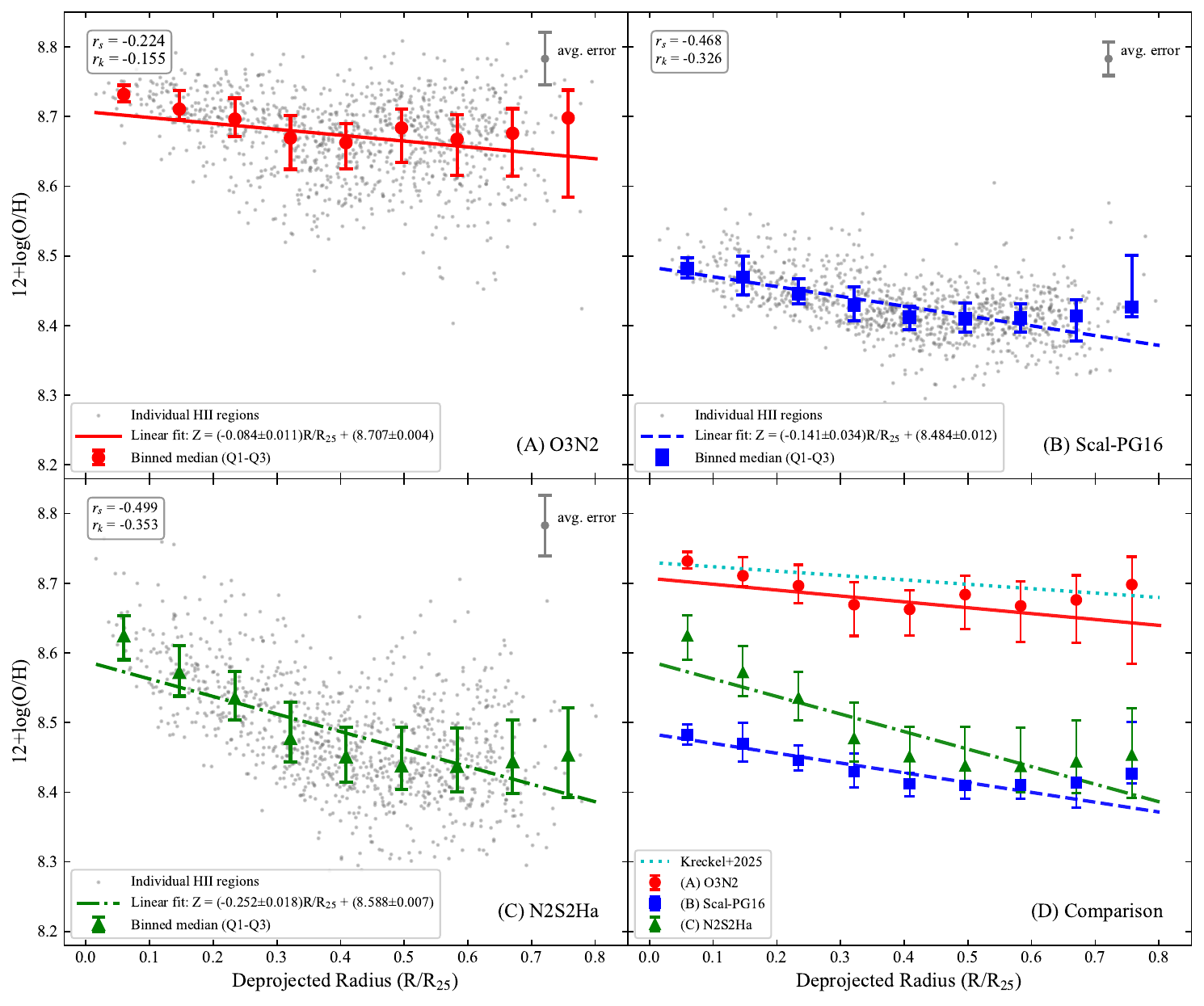}
    \caption{Deprojected radial distributions of $12+\log(\mathrm{O/H})$ for H\,II regions in NGC\,1385, derived from gas-phase metallicity calibrations using different methods: O3N2 (A), Scal-PG16 (B), and N2S2H$\alpha$ (C). In panels (A)--(C), the gray points represent individual H\,II regions, and the red, blue, and green points show the median and interquartile range (Q1-Q3) in nine radial bins.
    In each of these three panels, an average error bar is placed in the upper right corner to indicate the representative 1$\sigma$ measurement uncertainty.
    Panel (D) compares the binned radial trends from panels (A)--(C). The solid, dashed, and dash-dotted lines in all panels indicate linear fits to the binned data for each calibration method. In Panel D, we also add the oxygen abundance obtained by \cite{Kreckel2025} using the direct method with stacked spectra of auroral [N\,II] lines. This relation is shown as the dotted cyan line.
    }
\label{fig:metallicity_comparison}
\end{figure*}

Figure\,\ref{fig:metallicity_comparison} presents the radial distribution of ISM $12+\log(\mathrm{O/H})$ in NGC\,1385, where different calibration methods are distinguished by color and symbol. As shown in the figure, the central regions exhibit higher $12+\log(\mathrm{O/H})$ than the outer regions, clearly indicating the absence of a positive radial gradient. Although systematic offsets exist among the different calibrations, their radial trends remain mostly consistent. To quantitatively characterize this distribution to first order by a linear gradient, we performed a linear fit to the relation between $12+\log(\mathrm{O/H})$ and the deprojected galactocentric distance over the range $0.02 < R/R_{25} < 0.8$. For the three different strong-line calibrations. we obtained
\begin{equation}
\begin{aligned}[b]
&12+\log(\mathrm{O/H}) \\[0.15cm]
&= 8.484(\pm0.011)-0.141(\pm0.033) \, R/R_{25} \quad \text{(Scal-PG16)} \\[4pt]
&= 8.588(\pm0.007)-0.252(\pm0.018) \, R/R_{25} \quad \text{(N2S2H}\alpha\text{)} \\[4pt]
&= 8.707(\pm0.004)-0.084(\pm0.011) \, R/R_{25} \quad \text{(O3N2)}
\end{aligned}.
\label{equ:gradients}
\end{equation}

The Spearman rank correlation coefficient ($r_s$) and Kendall rank correlation coefficient ($r_k$) displayed in the figure indicate a mild negative radial $12+\log(\mathrm{O/H})$ gradient in NGC\,1385. These two coefficients are nonparametric statistics that assess the strength and direction of associations based on concordant and discordant pairs in the data, with values ranging from $-1$ to $+1$. Positive values denote a positive relation, negative values denote a negative relationship, and a value of zero indicates no association. $r_s$ is particularly suitable for smaller samples or weaker correlations, whereas $r_k$ is more robust to outliers and offers a higher precision for strong correlations in small datasets.
All three calibration methods yielded negative correlation coefficients ($r_s$ and $r_k$), confirming the weak outward radial decline in the metallicity in NGC\,1385, with higher $12+\log(\mathrm{O/H})$ values in the central regions and lower values in the outskirts. This is classic behavior as commonly observed in disk galaxies. We note, however, that the strengths of gradients resulting from the three different calibrations differ significantly (see below).

Using the direct method with weak nitrogen auroral lines, \cite{Kreckel2025} obtained a very weak negative gradient,
\begin{equation}
12+\log(\mathrm{O/H}) = 8.73-0.063 \, R/R_{25} \\
\label{equ:gradients2},
\end{equation}
from stacked MUSE spectra with [N\,II] $\lambda \lambda$5755, 6583.

The wide range of oxygen abundances obtained with the different calibrations is striking, but given the results shown, for instance, in \citet{Bresolin2016, Kreckel2025, Teimoorinia2021}, this is expected. It is encouraging, however, that the more reliable strong-line calibration PP04 (see section \ref{sec:Zg}) and the auroral line measurements agree reasonably well. They both show a very weak negative gradient. Interestingly, the N2S2H$\alpha$ calibration favored by \citet{Easeman2024} for MUSE data shows a stronger gradient and an offset of $\sim 0.2$ dex to the auroral line method. We took this into account in the comparison with our galaxy evolution model. We also compared our results with stellar metallicities derived from population synthesis methods (see below).

\subsection{Stellar metallicity and dust}
\label{sec:popsyn}

\begin{figure}
  \centering
  \vspace{1mm}
  \includegraphics[angle=0,width=0.98\linewidth]{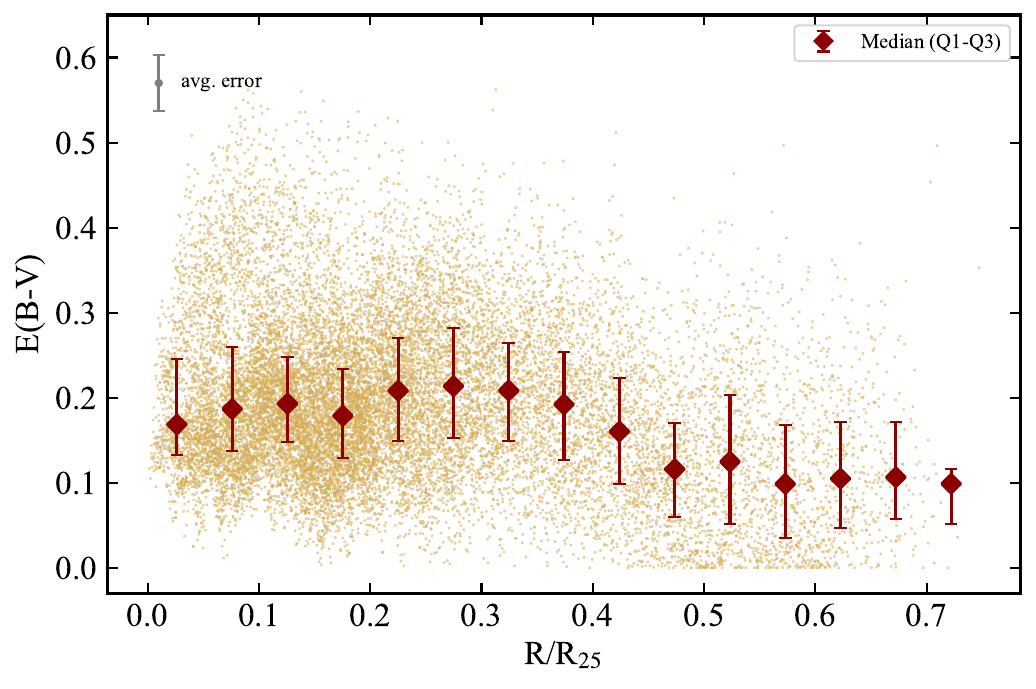}
    \caption{Deprojected radial distributions of interstellar reddening E(B-V). The median reddening
    and its interquartile range (Q1-Q3) of the radial bins are provided as diamonds with vertical error bars.
}
\label{fig:EBV}
\end{figure}

The MUSE full-spectrum fitting described in Section \ref{sec:fullspectral_fit} yields the spatial distribution of interstellar reddening E(B-V) from the comparison of observed and population synthesis model SEDs. Figure \ref{fig:EBV} shows the radial distribution, which increases slightly out to 0.3 R/R$_{25}$ and then drops to lower values. We note that the distribution of the molecular gas also drops between 0.25 to 0.4 R/R$_{25}$. E(B-V) is proportional to the surface density of interstellar dust. The similarity of the distributions in Figs.\,\ref{fig:H2} and \ref{fig:EBV} thus indicates a loose correlation between dust and molecular gas and reflects the fact that the formation of ISM molecules requires dust.

\begin{figure*}
  \centering
  \includegraphics[angle=0,scale=0.82]{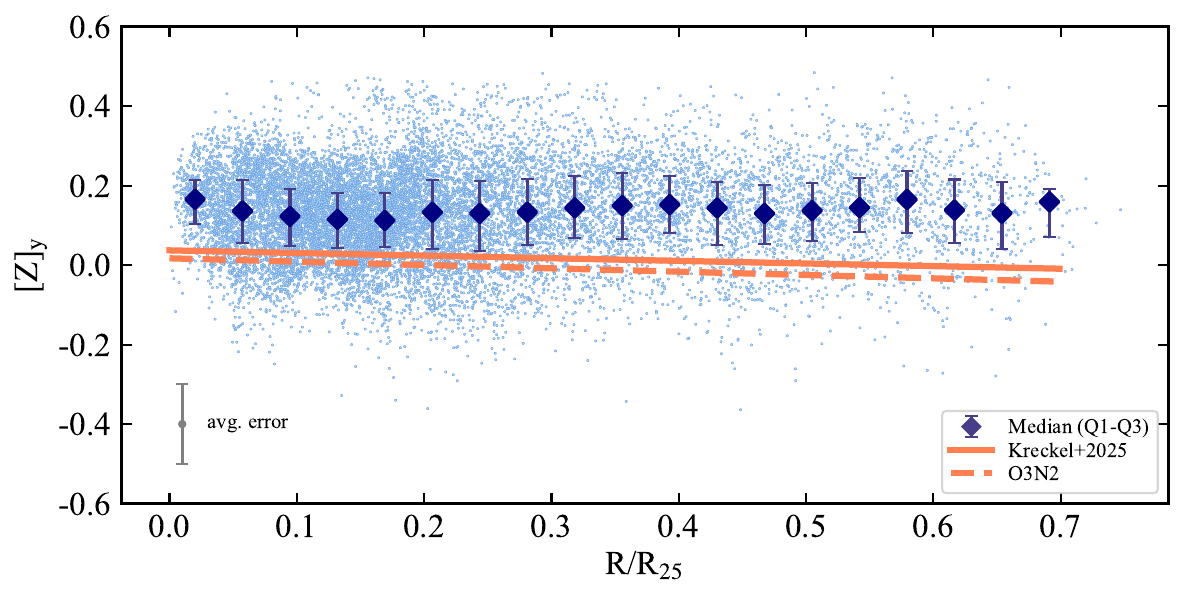}
    \caption{Deprojected radial distribution of the metallicity [Z]$_y$ of the young stellar population in NGC\,1385. The median values and their interquartile ranges are also plotted. The ISM oxygen abundance gradient based on the direct method with the auroral line [NII]$\lambda$5755 \citep{Kreckel2025} is also shown as the solid orange line. The dashed orange line represents the oxygen gradient based on the PP04 strong-line method. The ISM oxygen abundances are normalized to the solar value 12+log(O/H) = 8.69 \citep{Asplund2009} to allow for a direct comparison with the stellar metallicity.}
\label{fig:Zyoung}
\end{figure*}

The radial metallicity distribution of the young stellar population is provided in Figure\, \ref{fig:Zyoung}. The result is striking in many ways. The distribution is almost flat.
This is very different from the result shown in \cite{Pessa2023}, which indicated a very strong positive stellar metallicity gradient. (However, we note that their result described an average of all stars, young and old, based on the conventional luminosity- or mass-weighted averaging procedure, which can be problematic, as described in Section \ref{subsec:zdef}).
The comparison of our [Z]$_y$ with the ISM oxygen abundance reveals a very similar flat spatial distribution. The small offset of $\sim$0.15 dex between the stellar median and the ISM values can be explained by oxygen-dust depletion in H\,II regions \citep{Bresolin2025} or general systematic effects in the determination of the ISM oxygen abundances \citep{Bresolin2016}.

\subsection{\texorpdfstring{$\Sigma_{\rm SFR}$ distributions}{SFR surface density distributions}}
\label{subsec:SFR1}

\begin{figure*}
  \centering
  \includegraphics[angle=0,scale=0.6]{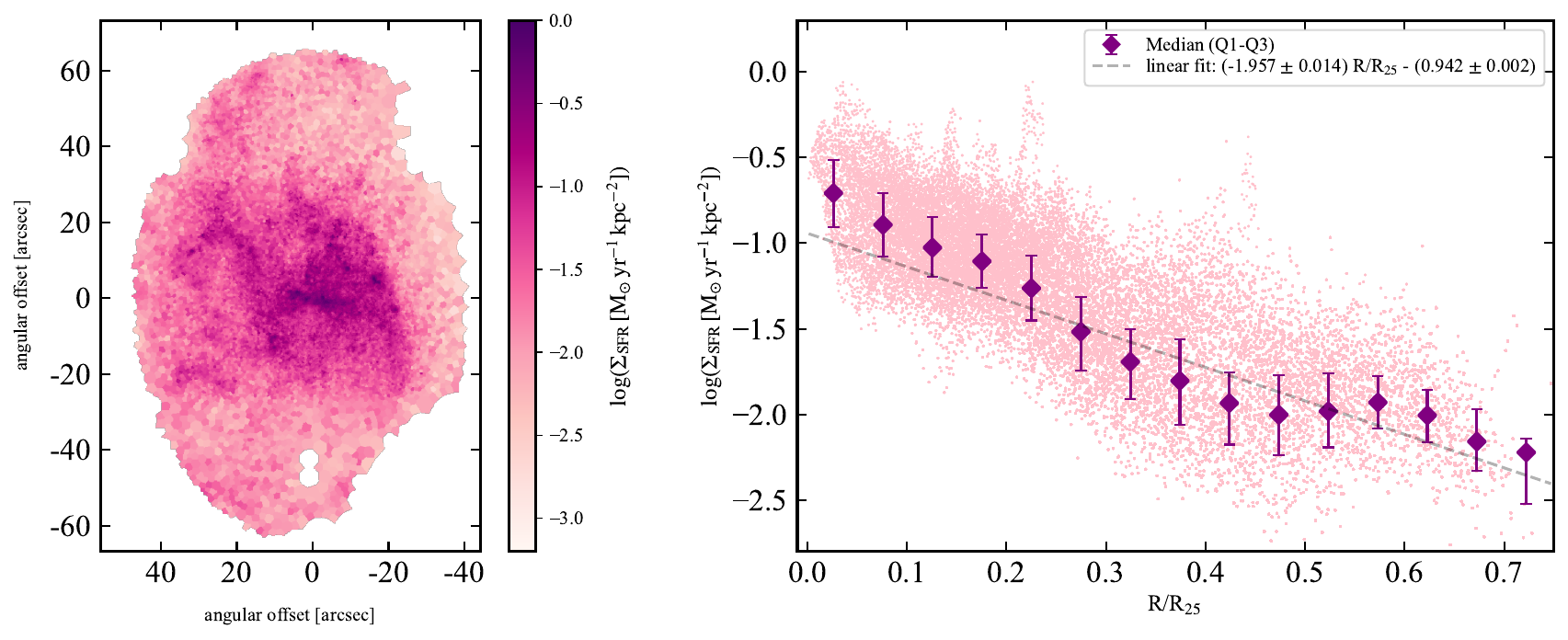}
    \caption{Two-dimensional and deprojected radial distributions of $\Sigma_{\rm SFR}$ from a full-spectrum fitting on a logarithmic scale. The median $\Sigma_{\rm SFR}$ and its interquartile range (Q1-Q3) within each radial bin are represented by violet diamonds with vertical error bars. The dashed gray line indicates a linear fit to the radial trend.}
\label{fig:SFR}
\end{figure*}

\begin{figure}
  \centering
  \includegraphics[angle=0,scale=0.45]{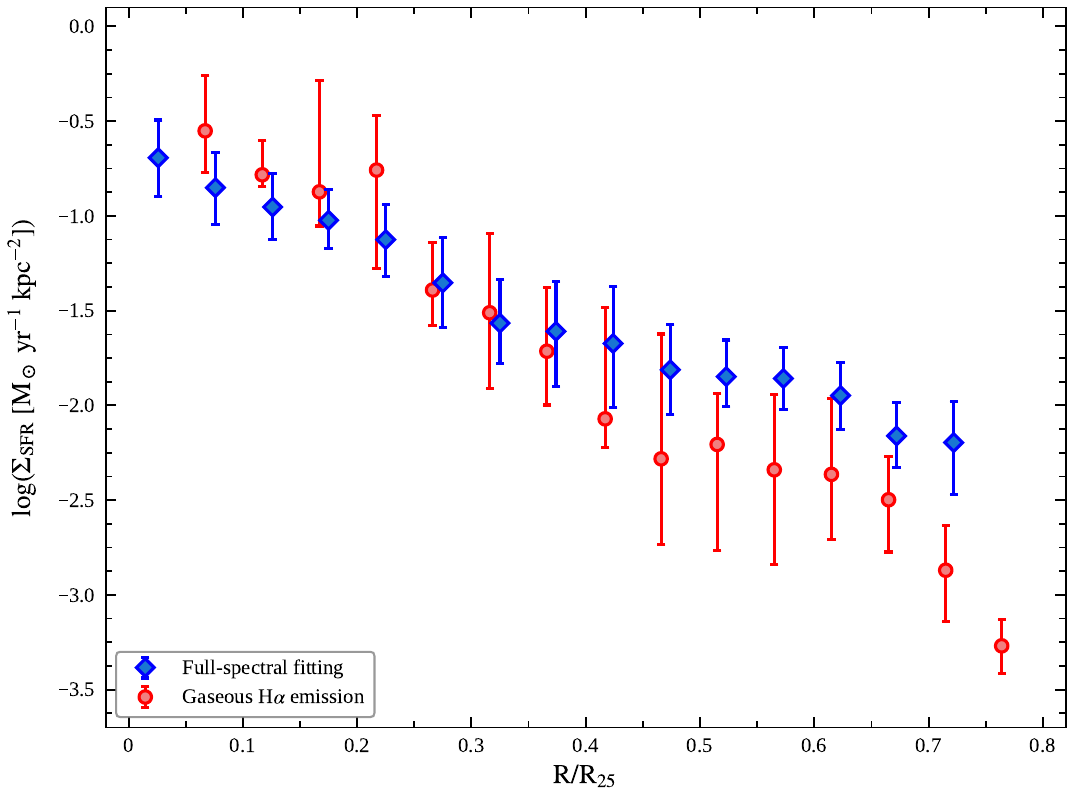}
    \caption{Comparison of $\Sigma_{\rm SFR}$ distributions obtained from full-spectrum fitting (blue diamonds) with those from gaseous H$\alpha$ emission (red circles).
}
\label{fig:SFR_comp}
\end{figure}

Figure\,\ref{fig:SFR} shows the two-dimensional and deprojected radial distribution of $\Sigma_{\rm SFR}$ across the NGC\,1385 as obtained from the full-spectrum fitting of the MUSE data. $\Sigma_{\rm SFR}$ clearly declines radially, with the inner disk exhibiting a substantially higher $\Sigma_{\rm SFR}$ than the outer disk.
The steady decline with radius is consistent with a model of an inside-out growth of the galactic disk.
This finding is consistent with extensive IFU studies of galaxies in the local Universe \citep[e.g., ][]{Gonzalez-Delgado2016, Sanchez2020, Barrera-Ballesteros2023} that investigated how $\Sigma_{\rm SFR}$ gradients relate to stellar mass and morphological type. These studies consistently reported that late-type galaxies, regardless of their stellar mass, exhibit similar negative gradients in $\Sigma_{\rm SFR}$ \citep{Barrera-Ballesteros2023}, which underscores the generality of our result for NGC\,1385.

A simple linear fit to the radial distribution of the median values yields a gradient of $-1.96 \pm 0.01\,\rm dex\,R_{25}^{-1}$.
This gradient is steeper than those reported by \citet{Schruba2011} for galaxies of comparable morphological type and stellar mass ($-1.52 \pm 0.16\, \rm dex\,R_{25}^{-1}$) and by \citet{Barrera-Ballesteros2023} for late-type galaxies ($-1.75\, \rm dex\,R_{25}^{-1}$).
In addition, the central intercept (i.e., $\log \Sigma_{\rm SFR}$ in the central region) from the fit is $-0.94$, which is significantly higher (less negative) than the corresponding value of $-1.85$ found by \citet{Schruba2011} for their similar-mass sample. This combination, a higher central $\Sigma_{\rm SFR}$ together with a steeper negative gradient, indicates a highly concentrated star-forming core in NGC\,1385.

In Figure \ref{fig:SFR_comp} we compare the full-spectrum fitting population synthesis SFR densities with those obtained from gaseous H$\alpha$ emission. The result is roughly similar within the error margins.
The radial gradient of $\Sigma_{\mathrm{SFR}}$ derived from the gaseous H$\alpha$ emission is slightly steeper.

\subsection{\texorpdfstring{$\Sigma_{\rm H_{2}}$ and $\sigma_{\rm mol}$ distributions}
{Molecular hydrogen gas-mass surface density and distributions of the molecular gas velocity dispersion}}
\label{subsec:H2_1}

\begin{figure*}
  \centering
  \includegraphics[angle=0,scale=0.55]{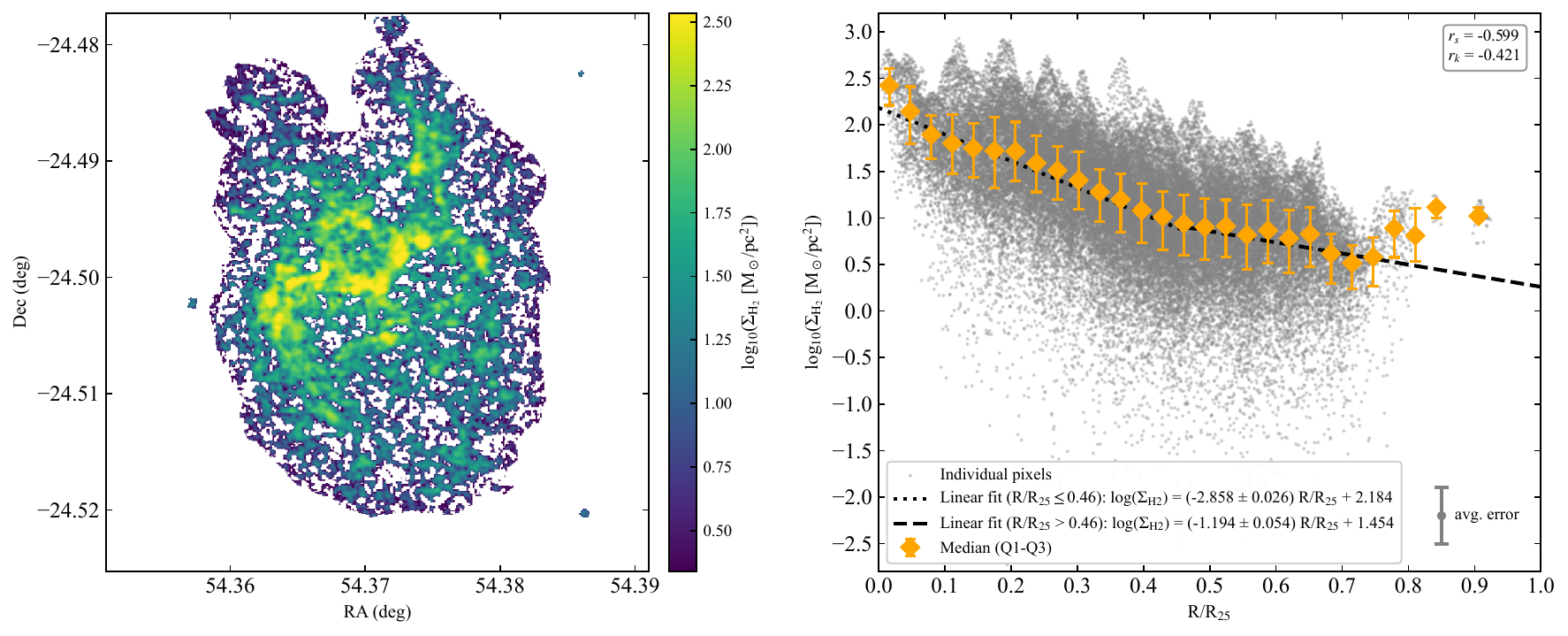}
    \caption{Two-dimensional and deprojected radial distributions of $\Sigma_{\rm H_{2}}$ on a logarithmic scale. The dotted and dashed lines represent the linear fits to the $\log \Sigma_{\rm H_2}$--$R/R_{25}$ relation for $R/R_{25} \le 0.46$ and $R/R_{25} > 0.46$, respectively. The median $\Sigma_{\rm H_{2}}$ and its interquartile range (Q1-Q3) within each radial bin are shown as orange diamonds with vertical error bars.
}
\label{fig:H2}
\end{figure*}

Figure\,\ref{fig:H2} shows the two-dimensional distribution and deprojected radial profile of $\Sigma_{\rm H_{2}}$ on a logarithmic scale across the disk of NGC\,1385.
 $\Sigma_{\rm H_2}$ clearly declines radially, with the inner disk showing higher values than the outer disk.
$\Sigma_{\rm H_2}$ of NGC\,1385 is overall significantly higher than that of mass-matching spiral galaxies. Its $\Sigma_{\rm H_2}$ profile is best described by piecewise linear fits,
\begin{align}
    \log \Sigma_{\rm H_2} &= (-2.858 \pm 0.026)\,R/R_{25} + 2.184, R/R_{25} \le 0.46, \\
    \log \Sigma_{\rm H_2} &= (-1.194 \pm 0.054)\,R/R_{25} + 1.454, R/R_{25} > 0.46.
\end{align}
At the break radius $R/R_{25} \approx 0.46$, the $\Sigma_{\rm H_2}$ of NGC\,1385 is about $7.7\,{\rm M_\odot\,\mathrm{pc}^{-2}}$ (from the inner fit: $7.4\,{\rm M_\odot\,\mathrm{pc}^{-2}}$, and from the outer fit: $8.0\,{\rm M_\odot\,\mathrm{pc}^{-2}}$), whereas the comparison sample (mass-matching spirals, which follow a single linear gradient of $-1.708 \pm 0.071\,\rm dex\,R_{25}^{-1}$) only has $3.61\,{\rm M_\odot\,\mathrm{pc}^{-2}}$ at the same radius, which is roughly half the value.
More strikingly,  NGC\,1385 reaches $\sim150\,{\rm M_\odot\,\mathrm{pc}^{-2}}$ in the central region, while the central-region values for other similar-mass galaxies are $23\,{\rm M_\odot\,\mathrm{pc}^{-2}}$ at most, differing by a factor of several.
Overall, the surface densities of NGC\,1385 are not only higher throughout the entire radial extent, but its central enrichment is particularly pronounced, indicating a much richer gas reservoir.
\citep[e.g.,][]{Leroy2008, Schruba2011, Casasola2017, Park2024}.

The enhanced central star formation in NGC\,1385 (Figure\,\ref{fig:SFR}) is accompanied by elevated molecular gas surface densities (Figure\,\ref{fig:H2}), suggesting that the abundant gas reservoir provides the fuel for vigorous star formation in the central galactic region. This picture is consistent with the global galaxy properties. NGC\,1385 indeed exhibits a high level of star formation activity: its total SFR is $2.089\ M_\odot\,{\rm yr}^{-1}$ and its stellar mass is $\log(M_*/{\rm M}_\odot) = 9.98$ (Table\,4 of \citet{Leroy2021b}). For a galaxy of the same stellar mass, the typical main-sequence SFR estimated from equation\,(7) of \citet{Leroy2021b} is only $0.662\ {\rm M}_\odot\,{\rm yr}^{-1}$. Thus, the observed SFR of NGC\,1385 is approximately three times higher than the main-sequence expectation, placing it well above the star formation main sequence \citep[see also Fig.\,18 in][]{Leroy2021b}.

\begin{figure*}
  \centering
  \includegraphics[angle=0,scale=0.55]{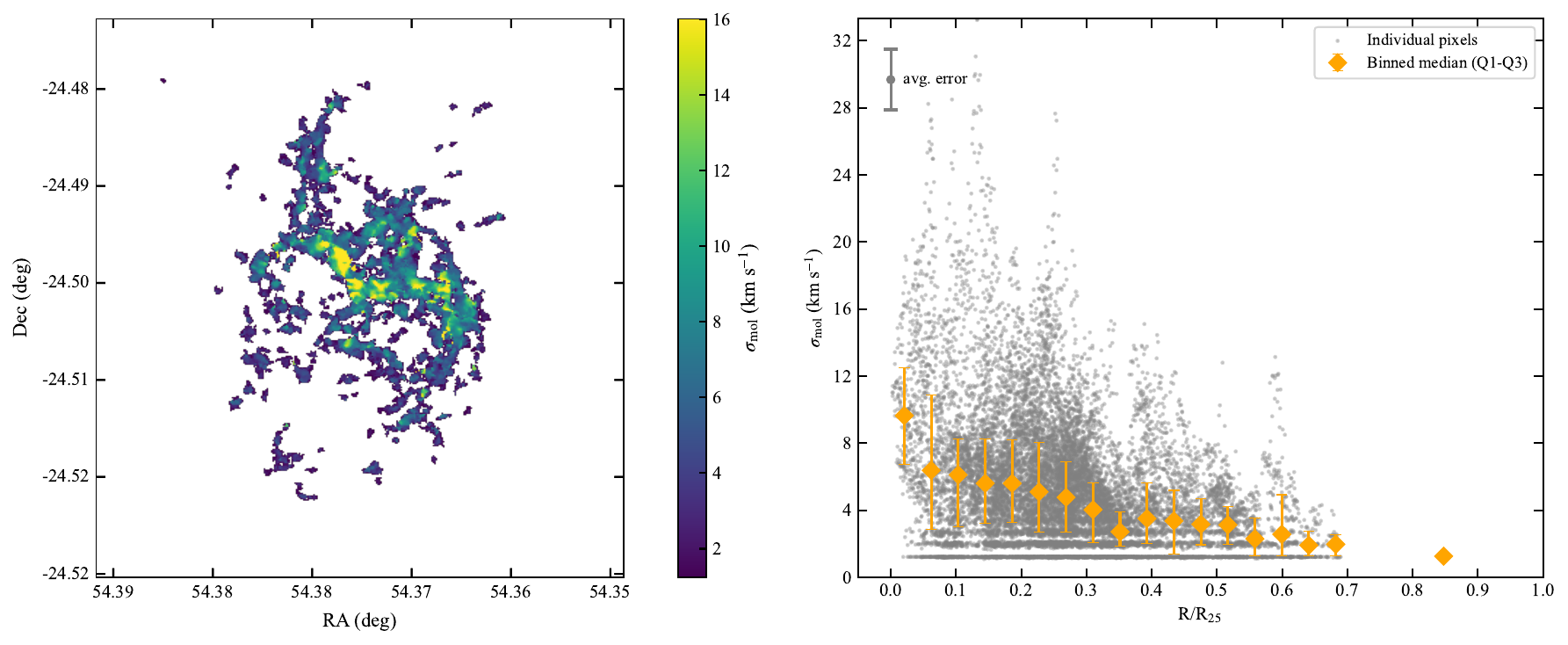}
    \caption{Two-dimensional and deprojected radial distributions of $\sigma_{\rm mol}$ derived from the PHANGS-ALMA (2-1) data. The median $\sigma_{\rm mol}$ and its interquartile range (Q1-Q3) within each radial bin are displayed as orange diamonds with vertical error bars.
}
\label{fig:Vilocity_dispersion}
\end{figure*}

To further elucidate the star-forming environment of NGC\,1385, we also analyzed the distribution of its molecular gas kinematic properties, such as the velocity dispersion profile. Figure\,\ref{fig:Vilocity_dispersion} presents $\sigma_{\rm mol}$ derived from the PHANGS-ALMA (2-1) data for reference. $\sigma_{\rm mol}$ in NGC\,1385 decreases with increasing radius, that is, its profile is characterized by a very low
velocity dispersion at the outer radii, with a rapid increase only toward the galactic center ($\sigma_{\rm mol}\,\leq\,10\, {\rm km\,s^{-1}}$).
This distribution is strictly consistent with the observed high $\Sigma_{\rm {H2}}$ and high $\Sigma_{\rm{SFR}}$ in the central region, collectively validating the causal chain of gas enrichment  $\rightarrow$ star formation triggering $\rightarrow$ dynamical chaos. It provides critical dynamical constraints for understanding the active star formation mechanism in the central region of NGC\,1385.

\subsection{\texorpdfstring{$\Sigma_{\mathrm{HI}}$}{Sigma HI} distributions}
\label{sec:HI_1}

\begin{figure}
  \centering
  \includegraphics[angle=0,scale=0.43]{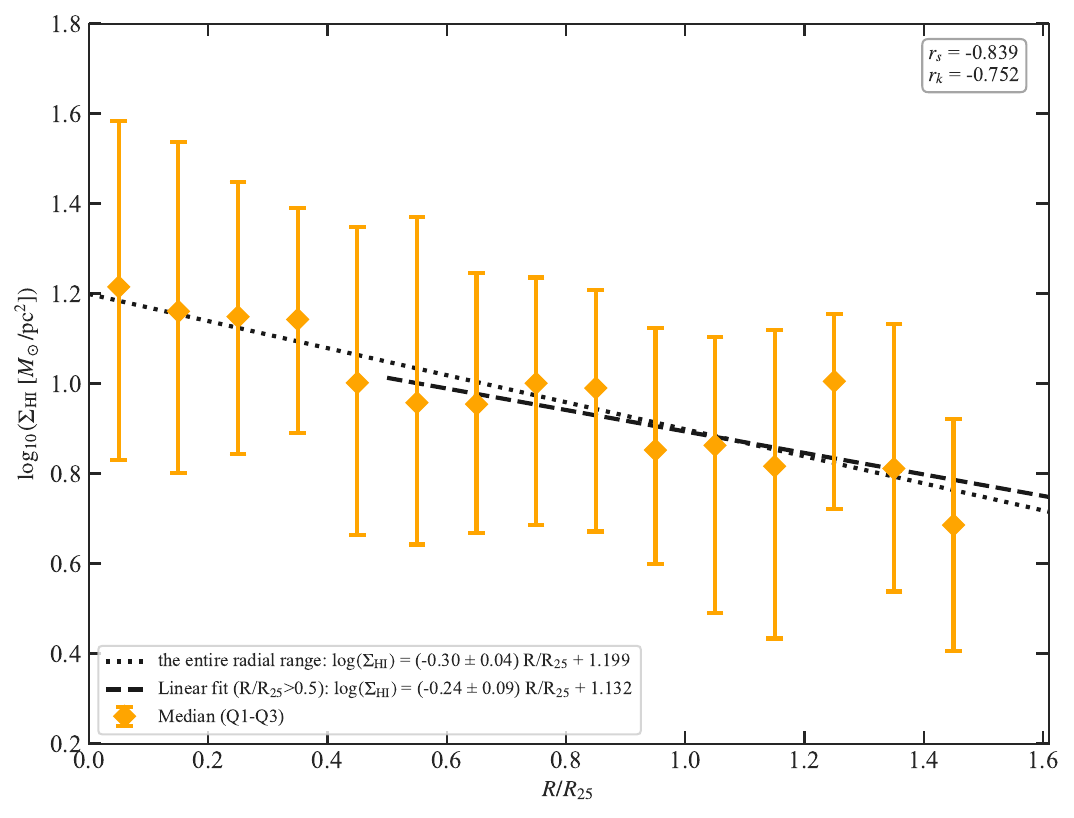}
    \caption{Deprojected radial distributions of $\Sigma_{\rm HI}$ on a logarithmic scale. The median $\Sigma_{\rm HI}$ and its interquartile range (Q1-Q3) within each radial bin are denoted by orange diamonds with vertical error bars. The dotted line represents the linear fit to the full radial range, and the dashed line represents the fit for the outer region ($R/R_{25}>0.5$).
}
\label{fig:HI-mass surface density}
\end{figure}

The deprojected radial profile of $\Sigma_{\rm HI}$ on a logarithmic scale across the disk of NGC\,1385 is shown in Figure\,\ref{fig:HI-mass surface density}.
Linear fits to the logarithmic $\Sigma_{\rm HI}$ profile yield slopes of $-0.30 \pm 0.04$ over the entire radial range and $-0.24 \pm 0.09$ for $R/R_{25}>0.5$. The two slopes agree within the uncertainties, indicating that the HI surface density declines at a similar rate throughout the disk.
Notably, significant HI emission is detected out to $\sim 1.5\,R_{25}$, demonstrating a gaseous disk extending beyond the optical radius.
This distribution is consistent with typical spiral galaxy profiles, although external factors such as tidal interactions, as discussed by \citet{For2021} and \citet{Veronese2025}, might affect the gas morphology and distribution at larger radii. Spiral galaxies with extended HI disks were also discussed in \citet{Bresolin2012} and \citet{Kudritzki2014}.

\section{A chemical evolution model for NGC\texorpdfstring{\,}{}1385}
\label{sec:model}

The chemical evolution model we adopted is based on \citet{Kang2025}.
In this framework, the NGC\,1385 disk is progressively built up by infall of primordial gas
($X\,=\,0.7571, Y_{\rm p}\,=\,0.2429, Z\,=\,0$) from its halo, combined with outflows of metal-enriched gas. The disk was modeled as a set of concentric rings, with radial gas inflows implemented following the prescriptions of \citet{Portinari2000} to account for the transport of gas and metals between adjacent rings. Within each annulus, we adopted the instantaneous recycling approximation (IRA), which assumes that stars more massive than $1\,M_{\odot}$ die instantaneously while lower-mass stars live forever, and that metals from the previous stellar generation are immediately and uniformly mixed with the existing ISM. IRA is good for oxygen, which is produced by short-lived massive stars, but potentially less acceptable
for elements from long-lived stars \citep[e.g., nitrogen, carbon, and iron; see][]{Vincenzo2016, Matteucci2021}.
Oxygen is the most abundant heavy element by mass and the best proxy for the global ISM metallicity; thus, we used oxygen to represent the metallicity in our model. The chemical evolution equations and main ingredients of the model are described in the subsections below.

\subsection{Equations of the chemical evolution}
\label{subsec:equations}

The evolution in each ring can be described by the following differential equations:
\begin{equation}
\frac{{\rm d}[\Sigma_{\rm tot}(r,t)]}{{\rm d}t}\,=\,f_{\rm{in}}(r,t)-f_{\rm{out}}(r,t),\\
\label{eq:tot}
\end{equation}
\begin{equation}
\frac{{\rm d}[\Sigma_{\rm gas}(r,t)]}{{\rm d}t}\,=\,-(1-R)\Psi(r,t)+f_{\rm{in}}(r,t)-f_{\rm{out}}(r,t)+\left[\frac{{\rm d}\Sigma_{\rm gas}(r,t)}{{\rm d}t}\right]_{rf},\\
\label{eq:gas}
\end{equation}
\begin{eqnarray}
\frac{{\rm d}[Z(r,t)\Sigma_{\rm gas}(r,t)]}{{\rm d}t}\,=\,y(1-R)\Psi(r,t)-Z(r,t)(1-R)\Psi(r,t) \nonumber\\
+Z_{\rm{in}}f_{\rm{in}}(r,t)-Z_{\rm{out}}(r,t)f_{\rm{out}}(r,t) \nonumber\\
+\left[\frac{{\rm d}[Z(r,t)\Sigma_{\rm gas}(r,t)]}{{\rm d}t}\right]_{rf},
\label{eq:metallicity}
\end{eqnarray}
where $\Sigma_{\rm tot}(r,t)$ and $\Sigma_{\rm gas}(r,t)$ are the total (star + gas) and gas-mass surface density at the galactocentric radius $r$ at the evolution time $t$, respectively; $\Psi(r,t)$ and $Z(r,t)$ describe the SFR and metallicity, and $f_{\rm{in}}(r,t)$ and $f_{\rm{out}}(r,t)$ are the gas infall and outflow rate, respectively.
$R$ is the return fraction, and $y$ is the nuclear synthesis yield. Following \citet{Vincenzo2016} with the \citet{Chabrier2003} initial mass function (IMF), we used the values $(R, y) = (0.451, 0.035)$ for oxygen. $Z_{\rm{in}}$ is the metallicity of the infalling gas and assumed to be primordial, that is, $Z_{\rm{in}}=0$, whereas $Z_{\rm{out}}(r,t)$ is the metallicity of the outflowing gas and assumed to have the same metallicity as the ISM, that is, $Z_{\rm{out} }(r,t)=Z(r,t)$ \citep{Ho2015}. $[\frac{{\rm d}\Sigma_{\rm gas}(r,t)}{{\rm d}t}]_{rf}$ and $[\frac{{\rm d}[Z(r,t)\Sigma_{\rm gas}(r,t)]}{{\rm d}t}]_{rf}$ are the radial flow terms of gas-mass surface density and metallicity, respectively.

\subsection{Main ingredients of the model}
\label{subsec:Model ingredients}

The gas-infall rate of primordial gas from the galactic halo at a given radius $r$ and time $t$, denoted above as $f_{\rm in}(r, t)$ (in units of $\rm M_\odot\,pc^{-2}\,Gyr^{-1}$), is expressed as
\begin{equation}
f_{\rm{in}}(r,t)=A(r)\cdot t\cdot e^{-t/\tau},
\label{eq:infall rate}
\end{equation}
where $\tau$ is the gas-infall timescale, a free parameter in the model.
The function $A(r)$ modifies the radial gas-infall rate (see \citet{Kang2025}). It was calculated iteratively by requiring that the model-predicted present-day stellar mass surface density $\Sigma_*(r,t_{\rm g})$ matches its observed value ($t_{\rm g}=13.5\rm\,Gyr$ is the cosmic age). NGC\,1385 is not an ideal exponential disk \citep{Leroy2021b}, and imposing an analytical exponential profile for $\Sigma_*(r,t_{\rm g})$ as in \citet{Kang2025} would introduce systematic uncertainties. To avoid this, we derived $\Sigma_*(r,t_{\rm g})$ directly from the observed 3.6$\,\mu$m surface brightness profile \citep{Bouquin2018} assuming a constant mass-to-light ratio $\Upsilon_{[3.6]}=0.6$ (based on the IMF of \citet{Chabrier2003}). This allowed the model to more accurately reflect the actual stellar mass distribution of NGC\,1385.

The star formation (SF) law is one of the key ingredients of our disk formation model. We adopted the $\Sigma_{\rm H_2}(r,t)$ based SF law \citep{Leroy2008}, where the SFR surface density $\Psi(r,t)$ (in units of $\rm{M_{\odot}}\,{pc}^{-2}\,{Gyr}^{-1}$) is linearly proportional to $\Sigma_{\rm H_2}(r,t)$,
\begin{equation}
\Psi(r,t)=\nu\Sigma_{\rm{H_2}}(r,t),
\label{eq:h2sfr}
\end{equation}
with $\nu$ being the SFE, treated as the second free parameter in this work. $\Sigma_{\rm H_2}(r,t)$ was derived from the total ISM gas-mass surface density in the same way as described in detail in \cite{Kang2023}.

The outflow of metal-enriched gas ($Z=Z_{\rm gas}$) occurs at a rate $f_{\rm out}(r,t)$ (in units of $\rm{M_{\odot}}\,{pc}^{-2}\,{Gyr}^{-1}$) that is proportional to $\Psi(r,t)$ \citep[see ][]{Recchi2008},
\begin{equation}
f_{\rm out}(r,t)=b_{\rm out}\Psi(r,t).
\label{eq:outflow}
\end{equation}
Here, $b_{\rm out}$ is the gas-outflow efficiency (dimensionless quantity), another free parameter in the model.

Infall of primordial gas from the halo onto the disk at a given radius does not necessarily share the same specific angular momentum as the disk at that radius. Consequently, radial gas inflows in disks naturally arise from angular momentum conservation \citep{Mayor1981, Lacey1985, Spitoni2011, Pezzulli2016}.
Radial gas inflows were implemented following the prescriptions of \citet{Portinari2000}. We therefore divided the galactic disk into discrete shells in our numerical solution of the chemical evolution equations. The $k$th shell is located at the galactocentric radius $r_k$, with its inner and outer boundaries denoted as $r_{k-\frac{1}{2}}$ and $r_{k+\frac{1}{2}}$, respectively. Gas flows across these boundaries at velocities $\upsilon_{k-\frac{1}{2}}$ and $\upsilon_{k+\frac{1}{2}}$, where positive velocities indicate outward flow and negative velocities indicate inward flow.
The radial flow term is expressed as
$\left[\frac{{\rm d}[Z(r_k,t)\Sigma_{\rm gas}(r_{k},t)]}{{\rm d}t} \right]_{rf}$ as follows:
\begin{eqnarray}
\left[\frac{{\rm d}[Z(r_k,t)\Sigma_{\rm gas}(r_{k},t)]}{{\rm d}t}\right]_{rf}\,=\,
-\beta_{k}Z(r_{k},t)\Sigma_{\rm gas}(r_{k},t) \nonumber\\
+ \gamma_{k}Z(r_{k+1},t)\Sigma_{\rm gas}(r_{k+1},t),
\label{eq:flow_term}
\end{eqnarray}
where
\begin{equation}
\beta_{k}\,=\,-\frac{2}{r_{k}+\frac{r_{k-1}+r_{k+1}}{2}}\times\left[\upsilon_{k-\frac{1}{2}}\frac{r_{k-1}+r_{k}}{r_{k+1}-r_{k-1}}\right]\\
\label{eq:beta}
\end{equation}
and
\begin{equation}
\gamma_{k}\,=\,-\frac{2}{r_{k}+\frac{r_{k-1}+r_{k+1}}{2}}\times\left[\upsilon_{k+\frac{1}{2}}\frac{r_{k}+r_{k+1}}{r_{k+1}-r_{k-1}}\right].\\
\label{eq:gama}
\end{equation}
Here, the radial inﬂow velocity $v_{R}$ is the last free parameter in the model.

\begin{figure*}
  \centering
  \includegraphics[angle=0,scale=0.9]{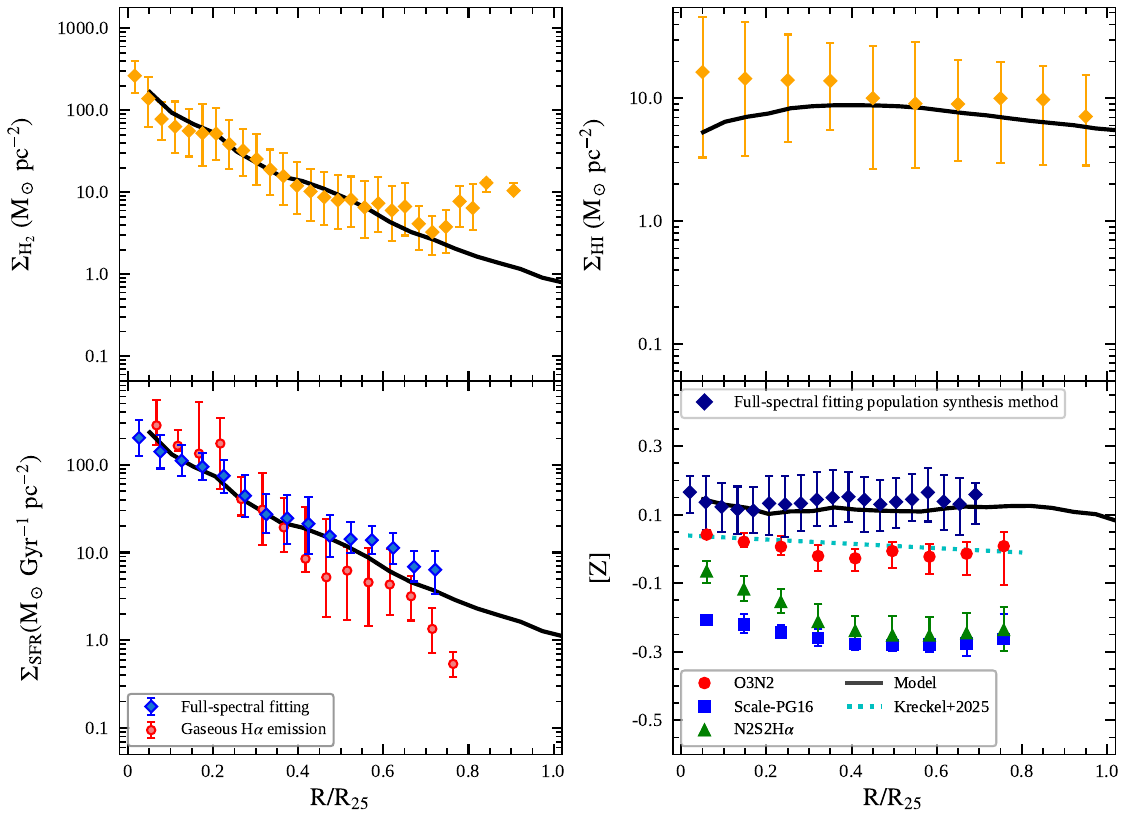}
    \caption{Comparison between observations of NGC\,1385 and our chemical evolution model. The model results are plotted as solid black curves in each panel. The left column shows the radial profiles of the H$_2$ mass surface density (top) and the SFR surface density (bottom).  The right column presents the radial profiles of the HI mass surface density (top) and the metallicity of the young stars and H\,II regions (bottom). The symbols in the bottom left, top left, and top right panels correspond to those in Figures \ref{fig:SFR_comp}, \ref{fig:H2}, and \ref{fig:HI-mass surface density}, respectively. In the bottom right panel, different colors and symbols indicate distinct metallicity determinations. The solid dark blue diamond data points represent the metallicity of the young stellar population measured by using the full-spectrum fitting population synthesis method. The other symbols represent H\,II region oxygen abundances (in units of solar values) obtained with different methods:
    O3N2 (filled red cycles), scale-PG16 (filled blue squares), and the N2S2H$\alpha$ (filled green triangle). The dotted cyan line shows the oxygen abundance obtained by \citet{Kreckel2025} with the direct method.
    }
  \label{fig:result}
\end{figure*}

\section{Model fit results and discussion}
\label{sec:Model results}

A robust chemical evolution model for NGC\,1385, particularly one involving free parameters, must reproduce a comprehensive set of observational constraints. As detailed in Section\,\ref{sec:obv}, the radial distributions of the gas-mass surface densities ($\Sigma_{\rm H_2}$ and $\Sigma_{\rm HI}$), $\Sigma_{\rm SFR}$, the present-day radial gas-phase metallicity ($12 + \log(\mathrm{O/H})$) distribution, along with the metallicity of the young stars, provide fundamental constraints on the evolutionary state of the galaxy. While the gas and SFR surface densities constrain the gas reservoir and current star formation activity, the metallicity distribution offers equally crucial insights because it encapsulates the complex evolutionary history shaped by the interplay between primordial gas infall, star formation, stellar feedback, and metal-enriched outflows. These processes collectively redistribute metals within and around the galactic disk, making the metallicity gradient a sensitive probe of chemical evolution models \citep[][and references therein]{BP2000, Kudritzki2015, Kubryk2015, Bresolin2019, Belfiore2019, Kang2025}.

As mentioned in Section\,\ref{sec:model}, the model included four free parameters: the gas-infall timescale $\tau$, the SFE $\nu$, the outflow efficiency $b_{\rm out}$, and the radial gas-flow velocity $v_{R}$. Guided by this parameter space exploration, we performed a quantitative search over the four-dimensional parameter space to identify the combination that best matched all observational constraints simultaneously. This procedure yielded a best-fitting model with parameters $(\tau, \nu, b_{\rm{out}}, v_{R}) = (5.5\,\rm{Gyr}, 1.4\,\rm{Gyr^{-1}}, 0.8, -0.15\, \rm{km\,s^{-1}})$, whose predictions are shown as solid lines in Figure\,\ref{fig:result}.
The model simultaneously reproduces the present-day observed radial profiles of $\Sigma_{\rm H_2}$, $\Sigma_{\rm HI}$, $\Sigma_{\rm SFR}$, and the metallicity of young stars, which agree excellently well with the data for NGC\,1385. Most importantly, the model yields a flat radial metallicity distribution that reproduces the observed trend. We note again that the oxygen abundance (in solar units) in H\,II regions is usually 0.1 to 0.2 dex lower than the metallicity of the young stars, which is interpreted as the result of oxygen depletion into H\,II region dust (see \citealt{Bresolin2025} for a detailed discussion).

To understand why this particular combination of parameters yields the best fit, it is constructive to consider the individual effect of each free parameter on the model predictions.
\citet{Kang2023} showed that $b_{\rm out}$ strongly affects the gas-phase metallicity, but only has a minor impact on the gas-mass surface density and the SFR surface density, while $\tau$ plays a key role in shaping the radial distributions of all four quantities because gas infall continuously replenishes the star formation reservoir, whereas outflows preferentially remove metal-enriched material from the disk.
\citet{Kang2025} further demonstrated that $v_{R}$ steepens the present-day radial profiles of the gas-mass surface density, the SFR surface density, and the gas-phase metallicity, which agrees well with \citet{Calura2023}.
For the effect of $\nu$ on the model results, given that the observed stellar mass surface density is taken as the input physical quantity in this work, a larger $\nu$ necessarily implies a lower gas surface density profile, and consequently, a flatter metallicity gradient, because a high $\nu$ mainly boosts star formation in the outer regions, allowing them to become chemically enriched. Hence, a high $\nu$ today yields flatter metallicity gradients, which correspond to the natural state of evolved systems in this model. This result was reported in our earlier work \citep{Kang2012} and was also supported by \citet{Belfiore2019}.

\begin{figure}
  \centering
  \includegraphics[angle=0,scale=0.65]{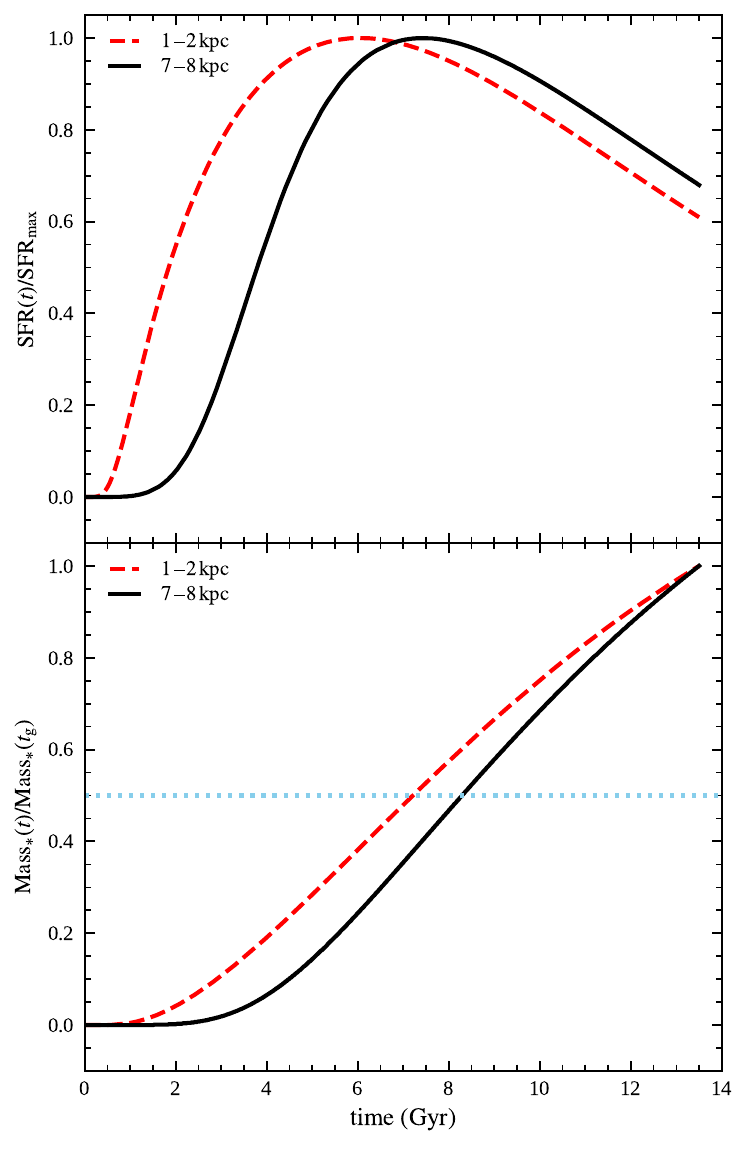}
    \caption{Star formation histories (top) and relative stellar mass
    growth (bottom) of two regions at different galactocentric distances
    (1-2\,kpc and 7-8\,kpc) in the disk of NGC\,1385. The SFH in each
    region is normalized to its maximum value, and the stellar mass is
    normalized to its present-day value. The dotted horizontal line in the bottom panel indicates the time at which each
    region reaches $50\%$ of its final stellar mass.
    }
  \label{fig:formation_scenario}
\end{figure}

To investigate the growth of stellar populations across the disk of NGC\,1385, Figure\,\ref{fig:formation_scenario} presents the best-fitting model predictions for two regions at different galactocentric distances. The upper panel reveals a systematic shift of the SFH peak to later times from the inner to the outer disk. The lower panel demonstrates that the stellar mass growth rate is higher in the inner than in the outer region. This means that the inner regions complete their mass assembly significantly earlier than the outskirts. Together, these features provide clear evidence of inside-out growth of the NGC\,1385 disk, consistent with a hierarchical inside-out structural formation scenario for galaxy evolution \citep{Pichon2011}.

\begin{figure}
  \centering
  \includegraphics[angle=0,scale=0.6]{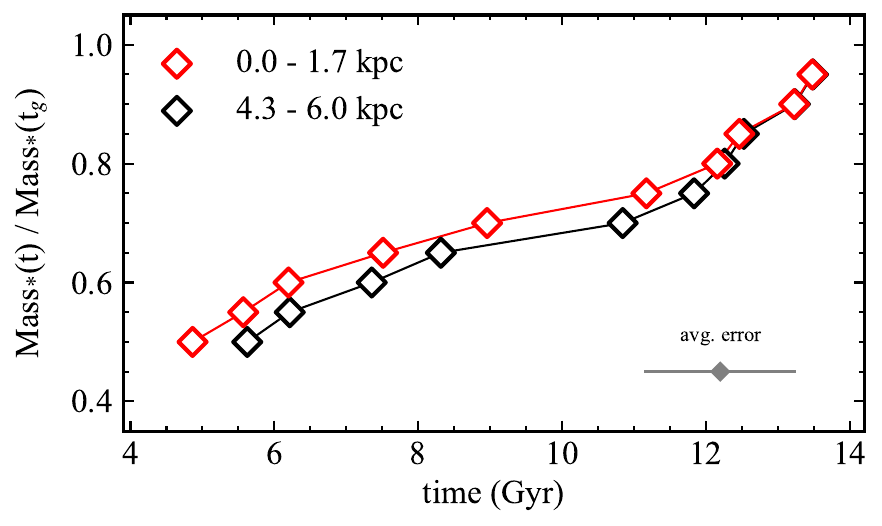}
    \caption{Relative stellar mass growth inferred from population synthesis and full spectral fitting. The inner region is shown in red, and the outermost accessible region in black (for a discussion, see the text).}
  \label{fig:massgrowth}
\end{figure}

The model prediction of inside-out growth can be tested with the results of our full-spectrum fitting analysis. Following equation (9) of \citet{Sextl2024}, we used the fitted luminosity contributions to the observed spectrum by the stellar populations of different ages. We converted them into mass contributions and constructed the cumulative mass assembly history in each population synthesis Voronoi bin (see also \citet{CidFernandes2013}). Only bins containing three or more SSPs were included to ensure a reliable result. Figure\,\ref{fig:massgrowth} shows the obtained mass-growth history and compares two different radial regions. The mass-growth of the innermost region ($<1.7$\,kpc) is shown in red and was obtained from the median growth in $9983$ individual Voronoi bins. Regions within $4.3 - 6.0$\,kpc represent the outermost galactocentric distances accessible with a full-spectrum fitting. Their mass growth determined from the median of $1565$ bins is shown in black. A mean error estimate is indicated in the bottom right corner. While the full-spectrum fitting result is quantitatively somewhat different compared to the model, the effect of inside-out growth is clearly visible. A mass growth of $50\%$ and $60\%$ is reached about $2$\,Gyr earlier than in the model. The time difference between the inner and outer zones is $\sim1.0$\,Gyr, similar to that in the model, but it varies substantially during the mass assembly. At a mass growth of $70\%$, the time difference is 1.8\,Gyr, twice as large as in the model, but it occurs at roughly the same time of $\sim9.5$\,Gyr. From $80\%$ on, the mass gain steeply increases to the final $100\%$.

\begin{figure}
  \centering
  \includegraphics[angle=0,scale=0.45]{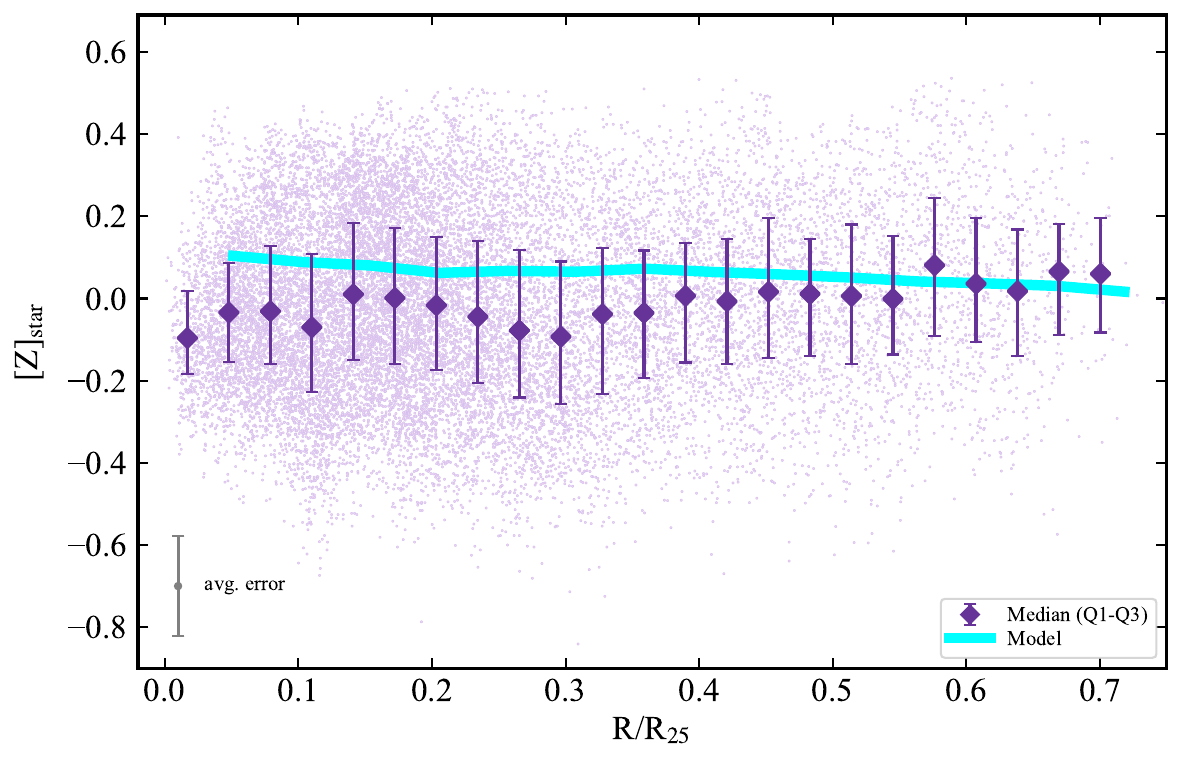}
    \caption{Radial distribution of the average metallicity of the entire stellar population as obtained from the full-spectrum fitting population synthesis analysis. The median values and their interquartile ranges are plotted in dark blue. The light blue curve shows the average metallicity predicted by our galaxy evolution model.}
  \label{fig:age_Z}
\end{figure}

An additional test of our model is provided by Figure\,\ref{fig:age_Z}. It shows the average [Z]$_{\rm star}$ of all stars, young and old, as obtained from our population synthesis analysis. It is consistently lower throughout the galaxy than the metallicity of the young stellar population.
This stems from the different ways in which the two metallicities record the chemical evolution history of galaxies. The total stellar metallicity indicates the cumulative state at the different times of the formation of all stars, young and old. In contrast, the young stars reflect the current instantaneous state.
The radial distribution of our measured average [Z]$_{\rm star}$ of all stars is nearly flat.
This result is strikingly different from that reported by \citet{Pessa2023}, who found a much steeper positive gradient of $\sim0.61\,\rm{dex}\,R_{25}^{-1}$ for the same galaxy. We attribute this difference to the use of the physical definition of metallicity as described in Section\,\ref{sec:fullspectral_fit} and to the extended set of our SSP template spectra.
The average stellar metallicity of our galaxy evolution model (see \citealt{Kang2025} eq.\,(17) for the calculation), which is shown as the light blue curve, agrees remarkably well with the observations.

\begin{figure}
  \centering
  \includegraphics[angle=0,scale=0.44]{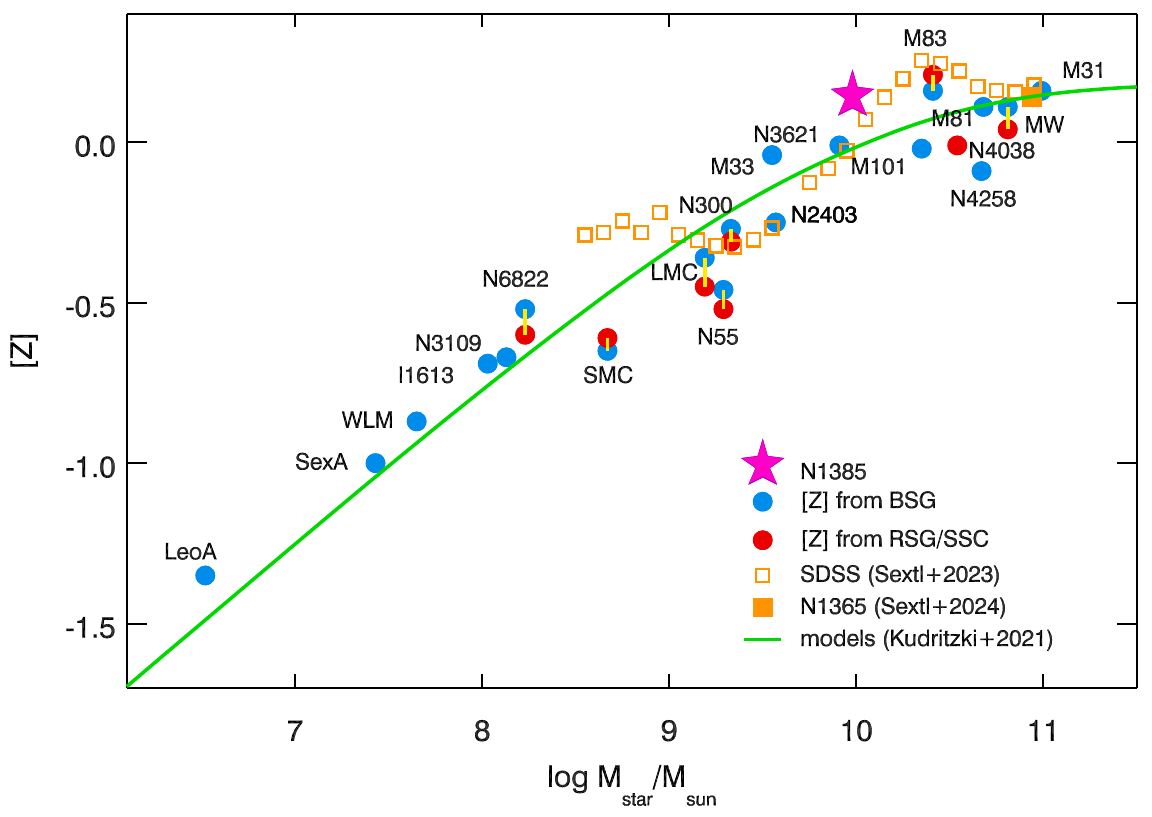}
    \caption{Position of NGC\,1385 on the MZR of star-forming galaxies. The data points for the individual galaxies result from a quantitative non-LTE spectroscopy of individual blue and red supergiant stars (BSG, RSG) and super star clusters (SSC) (see \citealt{Kudritzki2024, Bresolin2025} for references). In addition, results from the population synthesis analysis of 250000 SDSS galaxies \citep{Sextl2023} and the barred spiral NGC\,1365 \citep{Sextl2024} are shown together with the predictions by global galaxy evolution models \citep{Kudritzki2021a,Kudritzki2021b}. The metallicities for the SDSS galaxies, NGC\,1365, and NGC\,1385 correspond to the young stellar population.
   }
\label{fig:MZR}
\end{figure}

Intriguingly, our observations and model predictions suggest an inside-out formation scenario for NGC\,1385, consistent with a typical disk galaxy and resulting in a flat metallicity distribution. This finding contradicts the positive stellar metallicity gradient reported by \citet{Pessa2023}, who categorized it as a highly peculiar system. The apparent anticorrelation between its ionized gas and stellar metallicity gradients \citep{Pessa2023, Kreckel2025} presented a central puzzle and was originally the main motivation for this work. However, our new stellar population synthesis analysis reveals a very similar distribution of the young stellar population and ISM metallicities, which is well reproduced by our galactic evolution model. We also note that NGC\,1385 fits the observed mass-metallicity relation (MZR) of star-forming galaxies well, with our metallicity [Z]$_y$ for the young stellar population, as shown in Figure\,\ref{fig:MZR}.

Our model fit yields a very high SFE of $\nu$ = 1.4\,Gyr$^{-1}$. This is the main reason for the relatively flat spatial metallicity distribution, as discussed above. We note that \citet{Leroy2008} reported a typical SFE of $0.525\,\rm {Gyr}^{-1}$ for spiral galaxies. The value for NGC\,1385 is higher by about a factor of two.

NGC\,1385 is located approximately 160\,kpc south of the dark gas cloud WALLABY\,J033723-235753 \citep{Wong2021}. It also exhibits a significantly distorted HI morphology, along with a southern tidal debris field lacking any optical counterpart \citep{For2021}. Furthermore, \citet{Veronese2025} suggested the possibility of weak interactions between NGC\,1385 and its group companion NGC\,1371, located approximately 230\,kpc away. These observations suggest that cold gas accretion and/or tidal interactions might have supplied the inflowing gas that caused the enhanced recent star formation in NGC\,1385. However, more conclusive evidence is required to confirm this scenario.

\section{Conclusions}
\label{sec:sum}

We have combined spatially resolved multiphase observations from PHANGS (MUSE+ALMA) and WALLABY with chemical evolution modeling to investigate the evolution of the perturbed spiral galaxy NGC\,1385. Our main conclusions are listed below.

\begin{enumerate}

    \item \textbf{Inside-out disk formation scenario.}
    We developed a chemical evolution model incorporating the infall of metal-poor gas, outflow of metal-enriched gas, and radial gas inflows. Fitting our model to the observed radial profiles revealed that NGC\,1385 has undergone a typical inside-out disk formation, where the inner regions assembled earlier than the outer regions. Notably, this result does not rely on an assumed shorter gas-infall timescale for the inner disk. The model-predicted inside-out growth is supported by the results of our full-spectrum fitting analysis.

    \item \textbf{An exceptionally high and sustained SFE.}
    The best-fitting model requires an SFE that is higher by approximately a factor of two than that of mass-matched spiral galaxies. This high efficiency, sustained over gigayear timescales, has driven prolonged and vigorous star formation, particularly in the central regions, resulting in a current SFR that is approximately three times higher than that of comparable systems.

    \item \textbf{A flat metallicity gradient as a consequence of the high SFE.}
    Despite the inside-out formation mode, the radial metallicity distribution of young stars and the ISM is almost flat, with young stellar metallicities 0.15 to 0.2\,dex above solar. This flat gradient is not a contradiction but a direct consequence of the high SFE: the outer disk has been enriched more rapidly than usual, erasing the gradient that would otherwise arise from inside-out formation. The model predictions for stellar metallicities agree remarkably well with our refined spectral fitting measurements, supporting the observational analysis and the inferred evolutionary scenario. Full-spectrum fitting has reached a level of maturity that makes it comparable to H\,II region  analyses and spectroscopic studies of individual very luminous supergiant stars \citep{Kudritzki2024,Bresolin2025} for present-day metallicity measurements.

    \item \textbf{External gas supply as the enabler.}
    The critical evidence from WALLABY HI imagery (a severely distorted HI disk and a kinematically distinct tidal debris stream) indicates that this prolific history is not a product of isolated evolution. A recent gravitational interaction likely stripped gas from a companion or the intergalactic medium, providing the external fuel required to maintain the high SFE of the galaxy and sustained star formation over billions of years.

\end{enumerate}

In summary, our analysis of NGC\,1385 revealed a galaxy with sustained centralized star formation with a high SFE over billions of years, resulting in an elevated SFR and in an ISM enriched to the point of yielding young stars with high metallicities distributed throughout the galactic disk in a flat profile without a metallicity gradient.
NGC\,1385 thus presents a compelling case of how a perturbed spiral galaxy can maintain prolonged high-level star formation and chemical enrichment. Its evolution follows the classic inside-out disk formation sequence, but with an enhanced efficiency enabled by external gas accretion. The resulting flat metallicity gradient, distinct from the negative gradient predicted by the simplest inside-out models, reflects the combined effects of internal secular evolution and environmental interaction.
Future work should focus on detailed chemical tagging of stellar populations and dynamical modeling of the gas-inflow geometry to further unravel the history and impact of the accretion event.

\begin{acknowledgements}
We thank our referee for a very helpful and constructive report. This work was supported by the Basic Science Center project of the National Natural Science Foundation (NSF) of China (No. 12288102), the National Key R\&D Program of China with (Nos. 2021YFA1600403 and 2021YFA1600400), the international Center of Supernovae (ICESUN), Yunnan Key Laboratory of Supernova  Research (No. 202302AN360001),  and the New Cornerstone Science Foundation through the XPLORER PRIZE. Xiaoyu Kang thanks the support from the basic research program of Yunnan Province (No. 202401AT070142) and the Yunnan Revitalization Talent Support Program. Fenghui Zhang thanks the support from the NSF of China (No. 12573083). Eva Sextl, Rolf Kudritzki and Xiaoyu Kang acknowledge support by the Munich Excellence Cluster Origins and the Munich Instititute for Astro-, Particle and Biophysics (MIAPbP) both funded by the Deutsche Forschungsgemeinschaft (DFG, German Research Foundation) under the German Excellence Strategy EXC-2094 390783311.
We would like to thank Ismael Pessa and Kathryn Kreckel for sharing data that supported this research. Eva Sextl thanks Eric Emsellem, Enrico Congiu and Marco Troncoso Balbiano for insightful discussions on the sky subtraction problematic in PHANGS-MUSE.\\
Based on observations from the PHANGS-MUSE programme, collected at the European Southern Observatory under ESO programme $1100.B-0651$ (PHANGS-MUSE; PI: Schinnerer). This paper also makes use of the following ALMA data: ADS/JAO.ALMA\#$2015.1.00925$.S and ADS/JAO.ALMA\#$2018.1.01651$.S
ALMA is a partnership of ESO (representing its member states), NSF (USA) and NINS (Japan), together with NRC (Canada), MOST and ASIAA (Taiwan), and KASI (Republic of Korea), in cooperation with the Republic of Chile. The Joint ALMA Observatory is operated by ESO, AUI/NRAO and NAOJ. The National Radio Astronomy Observatory is a facility of the National Science Foundation operated under cooperative agreement by Associated Universities, Inc.
The ASKAP is part of the Australia Telescope National Facility that is funded by the Australian Government with support from the National Collaborative Research Infrastructure Strategy and Industry Endowment Fund. ASKAP uses the resources of the Pawsey Supercomputing Centre with funding provided by the Australian Government under the National Computational Merit Allocation Scheme (project JA3). We acknowledge the Wajarri Yamatji as the traditional owners of the Murchison Radio Observatory (MRO) site and thank the operational staff on-site.
\end{acknowledgements}

%
%

\bibliographystyle{aa} 
\bibliography{aa60983-26} 

@ARTICLE{Bresolin2009,
       author = {{Bresolin}, Fabio and {Gieren}, Wolfgang and {Kudritzki}, Rolf-Peter and {Pietrzy{\'n}ski}, Grzegorz and {Urbaneja}, Miguel A. and {Carraro}, Giovanni},
        title = "{Extragalactic Chemical Abundances: Do H II Regions and Young Stars Tell the Same Story? The Case of the Spiral Galaxy NGC 300}",
      journal = {\apj},
         year = 2009,
        month = jul,
       volume = {700},
       number = {1},
        pages = {309-330},
          doi = {10.1088/0004-637X/700/1/309},
archivePrefix = {arXiv},
       eprint = {0905.2791},
 primaryClass = {astro-ph.CO},
       adsurl = {https://ui.adsabs.harvard.edu/abs/2009ApJ...700..309B}
}

@ARTICLE{Leroy2019,
       author = {{Leroy}, Adam K. and {Sandstrom}, Karin M. and {Lang}, Dustin and {Lewis}, Alexia and {Salim}, Samir and {Behrens}, Erica A. and {Chastenet}, J{\'e}r{\'e}my and {Chiang}, I-Da and {Gallagher}, Molly J. and {Kessler}, Sarah and {Utomo}, Dyas},
        title = "{A z = 0 Multiwavelength Galaxy Synthesis. I. A WISE and GALEX Atlas of Local Galaxies}",
      journal = {\apjs},
         year = 2019,
        month = oct,
       volume = {244},
       number = {2},
          eid = {24},
        pages = {24},
          doi = {10.3847/1538-4365/ab3925},
archivePrefix = {arXiv},
       eprint = {1910.13470},
 primaryClass = {astro-ph.GA},
       adsurl = {https://ui.adsabs.harvard.edu/abs/2019ApJS..244...24L}
}

@ARTICLE{Tully2009,
       author = {{Tully}, R. Brent and {Rizzi}, Luca and {Shaya}, Edward J. and {Courtois}, H{\'e}l{\`e}ne M. and {Makarov}, Dmitry I. and {Jacobs}, Bradley A.},
        title = "{The Extragalactic Distance Database}",
      journal = {\aj},
         year = 2009,
        month = aug,
       volume = {138},
       number = {2},
        pages = {323-331},
          doi = {10.1088/0004-6256/138/2/323},
       adsurl = {https://ui.adsabs.harvard.edu/abs/2009AJ....138..323T}
}

@ARTICLE{Wong2021,
       author = {{Wong}, O.~I. and {Stevens}, A.~R.~H. and {For}, B.-Q. and {Westmeier}, T. and {Dixon}, M. and {Oh}, S.-H. and {J{\'o}zsa}, G.~I.~G. and {Reynolds}, T.~N. and {Lee-Waddell}, K. and {Rom{\'a}n}, J. and {Verdes-Montenegro}, L. and {Courtois}, H.~M. and {Pomar{\`e}de}, D. and {Murugeshan}, C. and {Whiting}, M.~T. and {Bekki}, K. and {Bigiel}, F. and {Bosma}, A. and {Catinella}, B. and {D{\'e}nes}, H. and {Elagali}, A. and {Holwerda}, B.~W. and {Kamphuis}, P. and {Kilborn}, V.~A. and {Kleiner}, D. and {Koribalski}, B.~S. and {Lelli}, F. and {Madrid}, J.~P. and {McQuinn}, K.~B.~W. and {Popping}, A. and {Rhee}, J. and {Roychowdhury}, S. and {Scott}, T.~C. and {Sengupta}, C. and {Spekkens}, K. and {Staveley-Smith}, L. and {Wakker}, B.~P.},
        title = "{WALLABY pre-pilot survey: two dark clouds in the vicinity of NGC 1395}",
      journal = {\mnras},
         year = 2021,
        month = oct,
       volume = {507},
       number = {2},
        pages = {2905-2921},
          doi = {10.1093/mnras/stab2262},
archivePrefix = {arXiv},
       eprint = {2108.04412},
 primaryClass = {astro-ph.GA},
       adsurl = {https://ui.adsabs.harvard.edu/abs/2021MNRAS.507.2905W}
}

@ARTICLE{Accurso2017,
       author = {{Accurso}, G. and {Saintonge}, A. and {Catinella}, B. and {Cortese}, L. and {Dav{\'e}}, R. and {Dunsheath}, S.~H. and {Genzel}, R. and {Gracia-Carpio}, J. and {Heckman}, T.~M. and {Jimmy} and {Kramer}, C. and {Li}, Cheng and {Lutz}, K. and {Schiminovich}, D. and {Schuster}, K. and {Sternberg}, A. and {Sturm}, E. and {Tacconi}, L.~J. and {Tran}, K.~V. and {Wang}, J.},
        title = "{Deriving a multivariate {\ensuremath{\alpha}}$_{CO}$ conversion function using the [C II]/CO (1-0) ratio and its application to molecular gas scaling relations}",
      journal = {\mnras},
         year = 2017,
        month = oct,
       volume = {470},
       number = {4},
        pages = {4750-4766},
          doi = {10.1093/mnras/stx1556},
archivePrefix = {arXiv},
       eprint = {1702.03888},
 primaryClass = {astro-ph.GA},
       adsurl = {https://ui.adsabs.harvard.edu/abs/2017MNRAS.470.4750A}
}

@ARTICLE{Sun2020,
       author = {{Sun}, Jiayi and {Leroy}, Adam K. and {Schinnerer}, Eva and {Hughes}, Annie and {Rosolowsky}, Erik and {Querejeta}, Miguel and {Schruba}, Andreas and {Liu}, Daizhong and {Saito}, Toshiki and {Herrera}, Cinthya N. and {Faesi}, Christopher and {Usero}, Antonio and {Pety}, J{\'e}r{\^o}me and {Kruijssen}, J.~M. Diederik and {Ostriker}, Eve C. and {Bigiel}, Frank and {Blanc}, Guillermo A. and {Bolatto}, Alberto D. and {Boquien}, M{\'e}d{\'e}ric and {Chevance}, M{\'e}lanie and {Dale}, Daniel A. and {Deger}, Sinan and {Emsellem}, Eric and {Glover}, Simon C.~O. and {Grasha}, Kathryn and {Groves}, Brent and {Henshaw}, Jonathan and {Jimenez-Donaire}, Maria J. and {Kim}, Jenny J. and {Klessen}, Ralf S. and {Kreckel}, Kathryn and {Lee}, Janice C. and {Meidt}, Sharon and {Sandstrom}, Karin and {Sardone}, Amy E. and {Utomo}, Dyas and {Williams}, Thomas G.},
        title = "{Molecular Gas Properties on Cloud Scales across the Local Star-forming Galaxy Population}",
      journal = {\apjl},
         year = 2020,
        month = sep,
       volume = {901},
       number = {1},
          eid = {L8},
        pages = {L8},
          doi = {10.3847/2041-8213/abb3be},
archivePrefix = {arXiv},
       eprint = {2009.01842},
 primaryClass = {astro-ph.GA},
       adsurl = {https://ui.adsabs.harvard.edu/abs/2020ApJ...901L...8S}
}

@ARTICLE{Sextl2023,
       author = {{Sextl}, Eva and {Kudritzki}, Rolf-Peter and {Zahid}, H. Jabran and {Ho}, I.-Ting},
        title = "{Mass-Metallicity Relationship of SDSS Star-forming Galaxies: Population Synthesis Analysis and Effects of Star Burst Length, Extinction Law, Initial Mass Function, and Star Formation Rate}",
      journal = {\apj},
         year = 2023,
        month = jun,
       volume = {949},
       number = {2},
          eid = {60},
        pages = {60},
          doi = {10.3847/1538-4357/acc579},
archivePrefix = {arXiv},
       eprint = {2303.11024},
 primaryClass = {astro-ph.GA},
       adsurl = {https://ui.adsabs.harvard.edu/abs/2023ApJ...949...60S}
}

@ARTICLE{Kudritzki2021a,
       author = {{Kudritzki}, Rolf-Peter and {Teklu}, Adelheid F. and {Schulze}, Felix and {Remus}, Rhea-Silvia and {Dolag}, Klaus and {Burkert}, Andreas and {Zahid}, H. Jabran},
        title = "{Erratum: ``Galaxy Lookback Evolution Models: A Comparison with Magneticum Cosmological Simulations and Observations'' (2021, ApJ, 910 , 87)}",
      journal = {\apj},
         year = 2021,
        month = dec,
       volume = {922},
       number = {2},
          eid = {274},
        pages = {274},
          doi = {10.3847/1538-4357/ac32cf},
       adsurl = {https://ui.adsabs.harvard.edu/abs/2021ApJ...922..274K}
}

@ARTICLE{Kudritzki2021b,
       author = {{Kudritzki}, Rolf-Peter and {Teklu}, Adelheid F. and {Schulze}, Felix and {Remus}, Rhea-Silvia and {Dolag}, Klaus and {Burkert}, Andreas and {Zahid}, H. Jabran},
        title = "{Galaxy Look-back Evolution Models: A Comparison with Magneticum Cosmological Simulations and Observations}",
      journal = {\apj},
         year = 2021,
        month = apr,
       volume = {910},
       number = {2},
          eid = {87},
        pages = {87},
          doi = {10.3847/1538-4357/abe40c},
archivePrefix = {arXiv},
       eprint = {2102.04135},
 primaryClass = {astro-ph.GA},
       adsurl = {https://ui.adsabs.harvard.edu/abs/2021ApJ...910...87K}
}

@ARTICLE{Kang2012,
       author = {{Kang}, Xiaoyu and {Chang}, Ruixiang and {Yin}, Jun and {Hou}, Jinliang and {Zhang}, Fenghui and {Zhang}, Yu and {Han}, Zhanwen},
        title = "{The evolution and star-formation history of M33}",
      journal = {\mnras},
         year = 2012,
        month = oct,
       volume = {426},
       number = {2},
        pages = {1455-1464},
          doi = {10.1111/j.1365-2966.2012.21778.x},
archivePrefix = {arXiv},
       eprint = {1207.5280},
 primaryClass = {astro-ph.CO},
       adsurl = {https://ui.adsabs.harvard.edu/abs/2012MNRAS.426.1455K}
}

@ARTICLE{Calzetti2007,
       author = {{Calzetti}, D. and {Kennicutt}, R.~C. and {Engelbracht}, C.~W. and {Leitherer}, C. and {Draine}, B.~T. and {Kewley}, L. and {Moustakas}, J. and {Sosey}, M. and {Dale}, D.~A. and {Gordon}, K.~D. and {Helou}, G.~X. and {Hollenbach}, D.~J. and {Armus}, L. and {Bendo}, G. and {Bot}, C. and {Buckalew}, B. and {Jarrett}, T. and {Li}, A. and {Meyer}, M. and {Murphy}, E.~J. and {Prescott}, M. and {Regan}, M.~W. and {Rieke}, G.~H. and {Roussel}, H. and {Sheth}, K. and {Smith}, J.~D.~T. and {Thornley}, M.~D. and {Walter}, F.},
        title = "{The Calibration of Mid-Infrared Star Formation Rate Indicators}",
      journal = {\apj},
         year = 2007,
        month = sep,
       volume = {666},
       number = {2},
        pages = {870-895},
          doi = {10.1086/520082},
archivePrefix = {arXiv},
       eprint = {0705.3377},
 primaryClass = {astro-ph},
       adsurl = {https://ui.adsabs.harvard.edu/abs/2007ApJ...666..870C}
}

@ARTICLE{Chabrier2003,
       author = {{Chabrier}, Gilles},
        title = "{Galactic Stellar and Substellar Initial Mass Function}",
      journal = {\pasp},
         year = 2003,
        month = jul,
       volume = {115},
       number = {809},
        pages = {763-795},
          doi = {10.1086/376392},
archivePrefix = {arXiv},
       eprint = {astro-ph/0304382},
 primaryClass = {astro-ph},
       adsurl = {https://ui.adsabs.harvard.edu/abs/2003PASP..115..763C}
}

@ARTICLE{Matteucci2021,
       author = {{Matteucci}, Francesca},
        title = "{Modelling the chemical evolution of the Milky Way}",
      journal = {\aapr},
         year = 2021,
        month = dec,
       volume = {29},
       number = {1},
          eid = {5},
        pages = {5},
          doi = {10.1007/s00159-021-00133-8},
archivePrefix = {arXiv},
       eprint = {2106.13145},
 primaryClass = {astro-ph.GA},
       adsurl = {https://ui.adsabs.harvard.edu/abs/2021A&ARv..29....5M}
}

@ARTICLE{Pezzulli2016,
       author = {{Pezzulli}, Gabriele and {Fraternali}, Filippo},
        title = "{Accretion, radial flows and abundance gradients in spiral galaxies}",
      journal = {\mnras},
         year = 2016,
        month = jan,
       volume = {455},
       number = {3},
        pages = {2308-2322},
          doi = {10.1093/mnras/stv2397},
archivePrefix = {arXiv},
       eprint = {1510.04289},
 primaryClass = {astro-ph.GA},
       adsurl = {https://ui.adsabs.harvard.edu/abs/2016MNRAS.455.2308P}
}

@ARTICLE{Lacey1985,
       author = {{Lacey}, C.~G. and {Fall}, S.~M.},
        title = "{Chemical evolution of the galactic disk with radial gas flows.}",
      journal = {\apj},
         year = 1985,
        month = mar,
       volume = {290},
        pages = {154-170},
          doi = {10.1086/162970},
       adsurl = {https://ui.adsabs.harvard.edu/abs/1985ApJ...290..154L}
}

@ARTICLE{Mayor1981,
       author = {{Mayor}, M. and {Vigroux}, L.},
        title = "{Effect of the infall of matter on the dynamical structure and chemical evolution of a spiral galaxy}",
      journal = {\aap},
         year = 1981,
        month = may,
       volume = {98},
       number = {1},
        pages = {1-8},
       adsurl = {https://ui.adsabs.harvard.edu/abs/1981A&A....98....1M}
}

@ARTICLE{Calura2023,
       author = {{Calura}, Francesco and {Palla}, Marco and {Morselli}, Laura and {Spitoni}, Emanuele and {Casasola}, Viviana and {Verma}, Kuldeep and {Enia}, Andrea and {Meneghetti}, Massimo and {Bianchi}, Simone and {Pozzi}, Francesca and {Gruppioni}, Carlotta},
        title = "{A Bayesian chemical evolution model of the DustPedia galaxy M74}",
      journal = {\mnras},
         year = 2023,
        month = aug,
       volume = {523},
       number = {2},
        pages = {2351-2368},
          doi = {10.1093/mnras/stad1316},
archivePrefix = {arXiv},
       eprint = {2305.05680},
 primaryClass = {astro-ph.GA},
       adsurl = {https://ui.adsabs.harvard.edu/abs/2023MNRAS.523.2351C}
}

@ARTICLE{Belfiore2019,
       author = {{Belfiore}, F. and {Vincenzo}, F. and {Maiolino}, R. and {Matteucci}, F.},
        title = "{From `bathtub' galaxy evolution models to metallicity gradients}",
      journal = {\mnras},
         year = 2019,
        month = jul,
       volume = {487},
       number = {1},
        pages = {456-474},
          doi = {10.1093/mnras/stz1165},
archivePrefix = {arXiv},
       eprint = {1903.05105},
 primaryClass = {astro-ph.GA},
       adsurl = {https://ui.adsabs.harvard.edu/abs/2019MNRAS.487..456B}
}

@ARTICLE{Recchi2008,
       author = {{Recchi}, S. and {Spitoni}, E. and {Matteucci}, F. and {Lanfranchi}, G.~A.},
        title = "{The effect of differential galactic winds on the chemical evolution of galaxies}",
      journal = {\aap},
         year = 2008,
        month = oct,
       volume = {489},
       number = {2},
        pages = {555-565},
          doi = {10.1051/0004-6361:200809879},
archivePrefix = {arXiv},
       eprint = {0808.0118},
 primaryClass = {astro-ph},
       adsurl = {https://ui.adsabs.harvard.edu/abs/2008A&A...489..555R}
}

@ARTICLE{Kudritzki2014,
       author = {{Kudritzki}, Rolf-Peter and {Urbaneja}, Miguel A. and {Bresolin}, Fabio and {Hosek}, Jr., Matthew W. and {Przybilla}, Norbert},
        title = "{Stellar Metallicity of the Extended Disk and Distance of the Spiral Galaxy NGC 3621}",
      journal = {\apj},
         year = 2014,
        month = jun,
       volume = {788},
       number = {1},
          eid = {56},
        pages = {56},
          doi = {10.1088/0004-637X/788/1/56},
archivePrefix = {arXiv},
       eprint = {1404.7244},
 primaryClass = {astro-ph.GA},
       adsurl = {https://ui.adsabs.harvard.edu/abs/2014ApJ...788...56K}
}

@ARTICLE{Bresolin2012,
       author = {{Bresolin}, Fabio and {Kennicutt}, Robert C. and {Ryan-Weber}, Emma},
        title = "{Gas Metallicities in the Extended Disks of NGC 1512 and NGC 3621. Chemical Signatures of Metal Mixing or Enriched Gas Accretion?}",
      journal = {\apj},
         year = 2012,
        month = may,
       volume = {750},
       number = {2},
          eid = {122},
        pages = {122},
          doi = {10.1088/0004-637X/750/2/122},
archivePrefix = {arXiv},
       eprint = {1203.0956},
 primaryClass = {astro-ph.CO},
       adsurl = {https://ui.adsabs.harvard.edu/abs/2012ApJ...750..122B}
}

@ARTICLE{Kudritzki2024,
       author = {{Kudritzki}, Rolf-Peter and {Urbaneja}, Miguel A. and {Bresolin}, Fabio and {Macri}, Lucas M. and {Yuan}, Wenlong and {Li}, Siyang and {Anand}, Gagandeep S. and {Riess}, Adam G.},
        title = "{The Hubble Constant Anchor Galaxy NGC 4258: Metallicity and Distance from Blue Supergiants}",
      journal = {\apj},
         year = 2024,
        month = dec,
       volume = {977},
       number = {2},
          eid = {217},
        pages = {217},
          doi = {10.3847/1538-4357/ad9279},
archivePrefix = {arXiv},
       eprint = {2411.07974},
 primaryClass = {astro-ph.GA},
       adsurl = {https://ui.adsabs.harvard.edu/abs/2024ApJ...977..217K}
}

@ARTICLE{Teim2021,
       author = {{Teimoorinia}, Hossen and {Jalilkhany}, Mansoureh and {Scudder}, Jillian M. and {Jensen}, Jaclyn and {Ellison}, Sara L.},
        title = "{A reassessment of strong line metallicity conversions in the machine learning era}",
      journal = {\mnras},
         year = 2021,
        month = may,
       volume = {503},
       number = {1},
        pages = {1082-1095},
          doi = {10.1093/mnras/stab466},
archivePrefix = {arXiv},
       eprint = {2102.07058},
 primaryClass = {astro-ph.GA},
       adsurl = {https://ui.adsabs.harvard.edu/abs/2021MNRAS.503.1082T}
}

@ARTICLE{Kewley2008,
       author = {{Kewley}, Lisa J. and {Ellison}, Sara L.},
        title = "{Metallicity Calibrations and the Mass-Metallicity Relation for Star-forming Galaxies}",
      journal = {\apj},
         year = 2008,
        month = jul,
       volume = {681},
       number = {2},
        pages = {1183-1204},
          doi = {10.1086/587500},
archivePrefix = {arXiv},
       eprint = {0801.1849},
 primaryClass = {astro-ph},
       adsurl = {https://ui.adsabs.harvard.edu/abs/2008ApJ...681.1183K}
}

@ARTICLE{Bresolin2016,
       author = {{Bresolin}, Fabio and {Kudritzki}, Rolf-Peter and {Urbaneja}, Miguel A. and {Gieren}, Wolfgang and {Ho}, I.-Ting and {Pietrzy{\'n}ski}, Grzegorz},
        title = "{Young Stars and Ionized Nebulae in M83: Comparing Chemical Abundances at High Metallicity.}",
      journal = {\apj},
         year = 2016,
        month = oct,
       volume = {830},
       number = {2},
          eid = {64},
        pages = {64},
          doi = {10.3847/0004-637X/830/2/64},
archivePrefix = {arXiv},
       eprint = {1607.06840},
 primaryClass = {astro-ph.GA},
       adsurl = {https://ui.adsabs.harvard.edu/abs/2016ApJ...830...64B}
}

@ARTICLE{Groves2025,
       author = {{Groves}, B. and {Kreckel}, K. and {Santoro}, F. and {Belfiore}, F. and {Zavodnik}, E. and {Congiu}, E. and {Egorov}, O.~V. and {Emsellem}, E. and {Grasha}, K. and {Leroy}, A. and {Scheuermann}, F. and {Schinnerer}, E. and {Watkins}, E.~J. and {Barnes}, A.~T. and {Bigiel}, F. and {Dale}, D.~A. and {Glover}, S.~C.~O. and {Pessa}, I. and {Sanchez-Blazquez}, P. and {Williams}, T.~G.},
        title = "{Correction to: The PHANGS─MUSE nebular catalogue}",
      journal = {\mnras},
         year = 2025,
        month = may,
       volume = {539},
       number = {2},
        pages = {1850-1855},
          doi = {10.1093/mnras/staf603},
       adsurl = {https://ui.adsabs.harvard.edu/abs/2025MNRAS.539.1850G}
}

@ARTICLE{Sextl2026,
       author = {{Sextl}, Eva and {Kudritzki}, Rolf-Peter},
        title = "{The Hidden Story of Chemical Evolution in Local Star-forming Nuclear Rings}",
      journal = {\apj},
         year = 2026,
        month = feb,
       volume = {997},
       number = {2},
          eid = {329},
        pages = {329},
          doi = {10.3847/1538-4357/ae29ee},
archivePrefix = {arXiv},
       eprint = {2510.14757},
 primaryClass = {astro-ph.GA},
       adsurl = {https://ui.adsabs.harvard.edu/abs/2026ApJ...997..329S}
}

@ARTICLE{Kang2023,
       author = {{Kang}, Xiaoyu and {Kudritzki}, Rolf-Peter and {Zhang}, Fenghui},
        title = "{The growth history of local M 33-mass bulgeless spiral galaxies}",
      journal = {\aap},
         year = 2023,
        month = nov,
       volume = {679},
          eid = {A83},
        pages = {A83},
          doi = {10.1051/0004-6361/202347677},
archivePrefix = {arXiv},
       eprint = {2309.07480},
 primaryClass = {astro-ph.GA},
       adsurl = {https://ui.adsabs.harvard.edu/abs/2023A&A...679A..83K}
}

@ARTICLE{Leroy2008,
   author = {{Leroy}, A.~K. and {Walter}, F. and {Brinks}, E. and {Bigiel}, F. and
	{de Blok}, W.~J.~G. and {Madore}, B. and {Thornley}, M.~D.},
    title = "{The Star Formation Efficiency in Nearby Galaxies: Measuring Where Gas Forms Stars Effectively}",
  journal = {\aj},
archivePrefix = "arXiv",
   eprint = {0810.2556},
     year = 2008,
    month = dec,
   volume = 136,
    pages = {2782-2845},
      doi = {10.1088/0004-6256/136/6/2782},
   adsurl = {http://adsabs.harvard.edu/abs/2008AJ....136.2782L}
}

@ARTICLE{Asplund2009,
   author = {{Asplund}, M. and {Grevesse}, N. and {Sauval}, A.~J. and {Scott}, P.
	},
    title = "{The Chemical Composition of the Sun}",
  journal = {\araa},
archivePrefix = "arXiv",
   eprint = {0909.0948},
 primaryClass = "astro-ph.SR",
     year = 2009,
    month = sep,
   volume = 47,
    pages = {481-522},
      doi = {10.1146/annurev.astro.46.060407.145222},
   adsurl = {http://adsabs.harvard.edu/abs/2009ARA%26A..47..481A}
}

@ARTICLE{Pessa2023,
       author = {{Pessa}, I. and {Schinnerer}, E. and {Sanchez-Blazquez}, P. and {Belfiore}, F. and {Groves}, B. and {Emsellem}, E. and {Neumann}, J. and {Leroy}, A.~K. and {Bigiel}, F. and {Chevance}, M. and {Dale}, D.~A. and {Glover}, S.~C.~O. and {Grasha}, K. and {Klessen}, R.~S. and {Kreckel}, K. and {Kruijssen}, J.~M.~D. and {Pinna}, F. and {Querejeta}, M. and {Rosolowsky}, E. and {Williams}, T.~G.},
        title = "{Resolved stellar population properties of PHANGS-MUSE galaxies}",
      journal = {\aap},
         year = 2023,
        month = may,
       volume = {673},
          eid = {A147},
        pages = {A147},
          doi = {10.1051/0004-6361/202245673},
archivePrefix = {arXiv},
       eprint = {2303.13676},
 primaryClass = {astro-ph.GA},
       adsurl = {https://ui.adsabs.harvard.edu/abs/2023A&A...673A.147P}
}

@ARTICLE{Querejeta2021,
       author = {{Querejeta}, M. and {Schinnerer}, E. and {Meidt}, S. and {Sun}, J. and {Leroy}, A.~K. and {Emsellem}, E. and {Klessen}, R.~S. and {Mu{\~n}oz-Mateos}, J.~C. and {Salo}, H. and {Laurikainen}, E. and {Be{\v{s}}li{\'c}}, I. and {Blanc}, G.~A. and {Chevance}, M. and {Dale}, D.~A. and {Eibensteiner}, C. and {Faesi}, C. and {Garc{\'\i}a-Rodr{\'\i}guez}, A. and {Glover}, S.~C.~O. and {Grasha}, K. and {Henshaw}, J. and {Herrera}, C. and {Hughes}, A. and {Kreckel}, K. and {Kruijssen}, J.~M.~D. and {Liu}, D. and {Murphy}, E.~J. and {Pan}, H. -A. and {Pety}, J. and {Razza}, A. and {Rosolowsky}, E. and {Saito}, T. and {Schruba}, A. and {Usero}, A. and {Watkins}, E.~J. and {Williams}, T.~G.},
        title = "{Stellar structures, molecular gas, and star formation across the PHANGS sample of nearby galaxies}",
      journal = {\aap},
         year = 2021,
        month = dec,
       volume = {656},
          eid = {A133},
        pages = {A133},
          doi = {10.1051/0004-6361/202140695},
archivePrefix = {arXiv},
       eprint = {2109.04491},
 primaryClass = {astro-ph.GA},
       adsurl = {https://ui.adsabs.harvard.edu/abs/2021A&A...656A.133Q}
}

@ARTICLE{Stuber2023,
       author = {{Stuber}, Sophia K. and {Schinnerer}, Eva and {Williams}, Thomas G. and {Querejeta}, Miguel and {Meidt}, Sharon and {Emsellem}, {\'E}ric and {Barnes}, Ashley and {Klessen}, Ralf S. and {Leroy}, Adam K. and {Neumann}, Justus and {Sormani}, Mattia C. and {Bigiel}, Frank and {Chevance}, M{\'e}lanie and {Dale}, Danny and {Faesi}, Christopher and {Glover}, Simon C.~O. and {Grasha}, Kathryn and {Kruijssen}, J.~M. Diederik and {Liu}, Daizhong and {Pan}, Hsi-an and {Pety}, J{\'e}r{\^o}me and {Pinna}, Francesca and {Saito}, Toshiki and {Usero}, Antonio and {Watkins}, Elizabeth J.},
        title = "{The gas morphology of nearby star-forming galaxies}",
      journal = {\aap},
         year = 2023,
        month = aug,
       volume = {676},
          eid = {A113},
        pages = {A113},
          doi = {10.1051/0004-6361/202346318},
archivePrefix = {arXiv},
       eprint = {2305.17172},
 primaryClass = {astro-ph.GA},
       adsurl = {https://ui.adsabs.harvard.edu/abs/2023A&A...676A.113S}
}

@ARTICLE{Sextl2025,
       author = {{Sextl}, Eva and {Kudritzki}, Rolf-Peter and {Bresolin}, Fabio and {Grasha}, Kathryn and {Park}, Hye-Jin and {Chen}, Qian-Hui and {Battisti}, Andrew J. and {Seibert}, Mark and {Madore}, Barry F. and {Rich}, Jeffrey A.},
        title = "{The TYPHOON Stellar Population Synthesis Survey. II. Pushing Full Spectral Fitting to the Limit in the Nearby Grand Design Barred Spiral M83}",
      journal = {\apj},
         year = 2025,
        month = jul,
       volume = {987},
       number = {2},
          eid = {138},
        pages = {138},
          doi = {10.3847/1538-4357/addec2},
archivePrefix = {arXiv},
       eprint = {2505.21127},
 primaryClass = {astro-ph.GA},
       adsurl = {https://ui.adsabs.harvard.edu/abs/2025ApJ...987..138S}
}

@ARTICLE{Sextl2024,
       author = {{Sextl}, Eva and {Kudritzki}, Rolf-Peter and {Burkert}, Andreas and {Ho}, I. -Ting and {Zahid}, H. Jabran and {Seibert}, Mark and {Battisti}, Andrew J. and {Madore}, Barry F. and {Rich}, Jeffrey A.},
        title = "{The TYPHOON Stellar Population Synthesis Survey. I. The Young Stellar Population of the Great Barred Spiral NGC 1365}",
      journal = {\apj},
         year = 2024,
        month = jan,
       volume = {960},
       number = {1},
          eid = {83},
        pages = {83},
          doi = {10.3847/1538-4357/ad08b3},
archivePrefix = {arXiv},
       eprint = {2311.01140},
 primaryClass = {astro-ph.GA},
       adsurl = {https://ui.adsabs.harvard.edu/abs/2024ApJ...960...83S}
}

@ARTICLE{Groves2023,
       author = {{Groves}, B. and {Kreckel}, K. and {Santoro}, F. and {Belfiore}, F. and {Zavodnik}, E. and {Congiu}, E. and {Egorov}, O.~V. and {Emsellem}, E. and {Grasha}, K. and {Leroy}, A. and {Scheuermann}, F. and {Schinnerer}, E. and {Watkins}, E.~J. and {Barnes}, A.~T. and {Bigiel}, F. and {Dale}, D.~A. and {Glover}, S.~C.~O. and {Pessa}, I. and {Sanchez-Blazquez}, P. and {Williams}, T.~G.},
        title = "{The PHANGS-MUSE nebular catalogue}",
      journal = {\mnras},
         year = 2023,
        month = apr,
       volume = {520},
       number = {4},
        pages = {4902-4952},
          doi = {10.1093/mnras/stad114},
archivePrefix = {arXiv},
       eprint = {2301.03811},
 primaryClass = {astro-ph.GA},
       adsurl = {https://ui.adsabs.harvard.edu/abs/2023MNRAS.520.4902G}
}

@ARTICLE{Pessa2021,
       author = {{Pessa}, I. and {Schinnerer}, E. and {Belfiore}, F. and {Emsellem}, E. and {Leroy}, A.~K. and {Schruba}, A. and {Kruijssen}, J.~M.~D. and {Pan}, H. -A. and {Blanc}, G.~A. and {Sanchez-Blazquez}, P. and {Bigiel}, F. and {Chevance}, M. and {Congiu}, E. and {Dale}, D. and {Faesi}, C.~M. and {Glover}, S.~C.~O. and {Grasha}, K. and {Groves}, B. and {Ho}, I. and {Jim{\'e}nez-Donaire}, M. and {Klessen}, R. and {Kreckel}, K. and {Koch}, E.~W. and {Liu}, D. and {Meidt}, S. and {Pety}, J. and {Querejeta}, M. and {Rosolowsky}, E. and {Saito}, T. and {Santoro}, F. and {Sun}, J. and {Usero}, A. and {Watkins}, E.~J. and {Williams}, T.~G.},
        title = "{Star formation scaling relations at {\ensuremath{\sim}}100 pc from PHANGS: Impact of completeness and spatial scale}",
      journal = {\aap},
         year = 2021,
        month = jun,
       volume = {650},
          eid = {A134},
        pages = {A134},
          doi = {10.1051/0004-6361/202140733},
archivePrefix = {arXiv},
       eprint = {2104.09536},
 primaryClass = {astro-ph.GA},
       adsurl = {https://ui.adsabs.harvard.edu/abs/2021A&A...650A.134P}
}

@ARTICLE{Lin2024,
       author = {{Lin}, Lin and {Shen}, Shiyin and {Yesuf}, Hassen M. and {Mao}, Ye-Wei and {Hao}, Lei},
        title = "{Radial Profiles of {\ensuremath{\Sigma}}$_{*}$, {\ensuremath{\Sigma}}$_{SFR}$, Gas Metallicity, and Their Correlations across the Galactic Mass{\textendash}Size Plane}",
      journal = {\apj},
         year = 2024,
        month = dec,
       volume = {977},
       number = {2},
          eid = {175},
        pages = {175},
          doi = {10.3847/1538-4357/ad8a61},
archivePrefix = {arXiv},
       eprint = {2410.16651},
 primaryClass = {astro-ph.GA},
       adsurl = {https://ui.adsabs.harvard.edu/abs/2024ApJ...977..175L}
}

@ARTICLE{Bouquin2018,
       author = {{Bouquin}, Alexandre Y.~K. and {Gil de Paz}, Armando and {Mu{\~n}oz-Mateos}, Juan Carlos and {Boissier}, Samuel and {Sheth}, Kartik and {Zaritsky}, Dennis and {Peletier}, Reynier F. and {Knapen}, Johan H. and {Gallego}, Jes{\'u}s},
        title = "{The GALEX/S$^{4}$G Surface Brightness and Color Profiles Catalog. I. Surface Photometry and Color Gradients of Galaxies}",
      journal = {\apjs},
         year = 2018,
        month = feb,
       volume = {234},
       number = {2},
          eid = {18},
        pages = {18},
          doi = {10.3847/1538-4365/aaa384},
archivePrefix = {arXiv},
       eprint = {1710.00955},
 primaryClass = {astro-ph.GA},
       adsurl = {https://ui.adsabs.harvard.edu/abs/2018ApJS..234...18B}
}

@ARTICLE{Leroy2021a,
       author = {{Leroy}, Adam K. and {Hughes}, Annie and {Liu}, Daizhong and {Pety}, J{\'e}r{\^o}me and {Rosolowsky}, Erik and {Saito}, Toshiki and {Schinnerer}, Eva and {Schruba}, Andreas and {Usero}, Antonio and {Faesi}, Christopher M. and {Herrera}, Cinthya N. and {Chevance}, M{\'e}lanie and {Hygate}, Alexander P.~S. and {Kepley}, Amanda A. and {Koch}, Eric W. and {Querejeta}, Miguel and {Sliwa}, Kazimierz and {Will}, David and {Wilson}, Christine D. and {Anand}, Gagandeep S. and {Barnes}, Ashley and {Belfiore}, Francesco and {Be{\v{s}}li{\'c}}, Ivana and {Bigiel}, Frank and {Blanc}, Guillermo A. and {Bolatto}, Alberto D. and {Boquien}, M{\`e}d{\`e}ric and {Cao}, Yixian and {Chandar}, Rupali and {Chastenet}, J{\'e}r{\'e}my and {Chiang}, I-Da and {Congiu}, Enrico and {Dale}, Daniel A. and {Deger}, Sinan and {den Brok}, Jakob S. and {Eibensteiner}, Cosima and {Emsellem}, Eric and {Garc{\'\i}a-Rodr{\'\i}guez}, Axel and {Glover}, Simon C.~O. and {Grasha}, Kathryn and {Groves}, Brent and {Henshaw}, Jonathan D. and {Jim{\'e}nez Donaire}, Mar{\'\i}a J. and {Kim}, Jaeyeon and {Klessen}, Ralf S. and {Kreckel}, Kathryn and {Kruijssen}, J.~M. Diederik and {Larson}, Kirsten L. and {Lee}, Janice C. and {Mayker}, Ness and {McElroy}, Rebecca and {Meidt}, Sharon E. and {Mok}, Angus and {Pan}, Hsi-An and {Puschnig}, Johannes and {Razza}, Alessandro and {S{\'a}nchez-Bl'azquez}, Patricia and {Sandstrom}, Karin M. and {Santoro}, Francesco and {Sardone}, Amy and {Scheuermann}, Fabian and {Sun}, Jiayi and {Thilker}, David A. and {Turner}, Jordan A. and {Ubeda}, Leonardo and {Utomo}, Dyas and {Watkins}, Elizabeth J. and {Williams}, Thomas G.},
        title = "{PHANGS-ALMA Data Processing and Pipeline}",
      journal = {\apjs},
         year = 2021,
        month = jul,
       volume = {255},
       number = {1},
          eid = {19},
        pages = {19},
          doi = {10.3847/1538-4365/abec80},
archivePrefix = {arXiv},
       eprint = {2104.07665},
 primaryClass = {astro-ph.IM},
       adsurl = {https://ui.adsabs.harvard.edu/abs/2021ApJS..255...19L}
}

@ARTICLE{Leroy2021b,
       author = {{Leroy}, Adam K. and {Schinnerer}, Eva and {Hughes}, Annie and {Rosolowsky}, Erik and {Pety}, J{\'e}r{\^o}me and {Schruba}, Andreas and {Usero}, Antonio and {Blanc}, Guillermo A. and {Chevance}, M{\'e}lanie and {Emsellem}, Eric and {Faesi}, Christopher M. and {Herrera}, Cinthya N. and {Liu}, Daizhong and {Meidt}, Sharon E. and {Querejeta}, Miguel and {Saito}, Toshiki and {Sandstrom}, Karin M. and {Sun}, Jiayi and {Williams}, Thomas G. and {Anand}, Gagandeep S. and {Barnes}, Ashley T. and {Behrens}, Erica A. and {Belfiore}, Francesco and {Benincasa}, Samantha M. and {Be{\v{s}}li{\'c}}, Ivana and {Bigiel}, Frank and {Bolatto}, Alberto D. and {den Brok}, Jakob S. and {Cao}, Yixian and {Chandar}, Rupali and {Chastenet}, J{\'e}r{\'e}my and {Chiang}, I-Da and {Congiu}, Enrico and {Dale}, Daniel A. and {Deger}, Sinan and {Eibensteiner}, Cosima and {Egorov}, Oleg V. and {Garc{\'\i}a-Rodr{\'\i}guez}, Axel and {Glover}, Simon C.~O. and {Grasha}, Kathryn and {Henshaw}, Jonathan D. and {Ho}, I. -Ting and {Kepley}, Amanda A. and {Kim}, Jaeyeon and {Klessen}, Ralf S. and {Kreckel}, Kathryn and {Koch}, Eric W. and {Kruijssen}, J.~M. Diederik and {Larson}, Kirsten L. and {Lee}, Janice C. and {Lopez}, Laura A. and {Machado}, Josh and {Mayker}, Ness and {McElroy}, Rebecca and {Murphy}, Eric J. and {Ostriker}, Eve C. and {Pan}, Hsi-An and {Pessa}, Ismael and {Puschnig}, Johannes and {Razza}, Alessandro and {S{\'a}nchez-Bl{\'a}zquez}, Patricia and {Santoro}, Francesco and {Sardone}, Amy and {Scheuermann}, Fabian and {Sliwa}, Kazimierz and {Sormani}, Mattia C. and {Stuber}, Sophia K. and {Thilker}, David A. and {Turner}, Jordan A. and {Utomo}, Dyas and {Watkins}, Elizabeth J. and {Whitmore}, Bradley},
        title = "{PHANGS-ALMA: Arcsecond CO(2-1) Imaging of Nearby Star-forming Galaxies}",
      journal = {\apjs},
         year = 2021,
        month = dec,
       volume = {257},
       number = {2},
          eid = {43},
        pages = {43},
          doi = {10.3847/1538-4365/ac17f3},
archivePrefix = {arXiv},
       eprint = {2104.07739},
 primaryClass = {astro-ph.GA},
       adsurl = {https://ui.adsabs.harvard.edu/abs/2021ApJS..257...43L}
}

@ARTICLE{Lang2020,
       author = {{Lang}, Philipp and {Meidt}, Sharon E. and {Rosolowsky}, Erik and {Nofech}, Joseph and {Schinnerer}, Eva and {Leroy}, Adam K. and {Emsellem}, Eric and {Pessa}, Ismael and {Glover}, Simon C.~O. and {Groves}, Brent and {Hughes}, Annie and {Kruijssen}, J.~M. Diederik and {Querejeta}, Miguel and {Schruba}, Andreas and {Bigiel}, Frank and {Blanc}, Guillermo A. and {Chevance}, M{\'e}lanie and {Colombo}, Dario and {Faesi}, Christopher and {Henshaw}, Jonathan D. and {Herrera}, Cinthya N. and {Liu}, Daizhong and {Pety}, J{\'e}r{\^o}me and {Puschnig}, Johannes and {Saito}, Toshiki and {Sun}, Jiayi and {Usero}, Antonio},
        title = "{PHANGS CO Kinematics: Disk Orientations and Rotation Curves at 150 pc Resolution}",
      journal = {\apj},
         year = 2020,
        month = jul,
       volume = {897},
       number = {2},
          eid = {122},
        pages = {122},
          doi = {10.3847/1538-4357/ab9953},
archivePrefix = {arXiv},
       eprint = {2005.11709},
 primaryClass = {astro-ph.GA},
       adsurl = {https://ui.adsabs.harvard.edu/abs/2020ApJ...897..122L}
}

@ARTICLE{Makarov2014,
       author = {{Makarov}, Dmitry and {Prugniel}, Philippe and {Terekhova}, Nataliya and {Courtois}, H{\'e}l{\`e}ne and {Vauglin}, Isabelle},
        title = "{HyperLEDA. III. The catalogue of extragalactic distances}",
      journal = {\aap},
         year = 2014,
        month = oct,
       volume = {570},
          eid = {A13},
        pages = {A13},
          doi = {10.1051/0004-6361/201423496},
archivePrefix = {arXiv},
       eprint = {1408.3476},
 primaryClass = {astro-ph.GA},
       adsurl = {https://ui.adsabs.harvard.edu/abs/2014A&A...570A..13M}
}

@ARTICLE{Anand2021,
       author = {{Anand}, Gagandeep S. and {Lee}, Janice C. and {Van Dyk}, Schuyler D. and {Leroy}, Adam K. and {Rosolowsky}, Erik and {Schinnerer}, Eva and {Larson}, Kirsten and {Kourkchi}, Ehsan and {Kreckel}, Kathryn and {Scheuermann}, Fabian and {Rizzi}, Luca and {Thilker}, David and {Tully}, R. Brent and {Bigiel}, Frank and {Blanc}, Guillermo A. and {Boquien}, M{\'e}d{\'e}ric and {Chandar}, Rupali and {Dale}, Daniel and {Emsellem}, Eric and {Deger}, Sinan and {Glover}, Simon C.~O. and {Grasha}, Kathryn and {Groves}, Brent and {S. Klessen}, Ralf and {Kruijssen}, J.~M. Diederik and {Querejeta}, Miguel and {S{\'a}nchez-Bl{\'a}zquez}, Patricia and {Schruba}, Andreas and {Turner}, Jordan and {Ubeda}, Leonardo and {Williams}, Thomas G. and {Whitmore}, Brad},
        title = "{Distances to PHANGS galaxies: New tip of the red giant branch measurements and adopted distances}",
      journal = {\mnras},
         year = 2021,
        month = mar,
       volume = {501},
       number = {3},
        pages = {3621-3639},
          doi = {10.1093/mnras/staa3668},
archivePrefix = {arXiv},
       eprint = {2012.00757},
 primaryClass = {astro-ph.GA},
       adsurl = {https://ui.adsabs.harvard.edu/abs/2021MNRAS.501.3621A}
}

@ARTICLE{BPT1981,
       author = {{Baldwin}, J.~A. and {Phillips}, M.~M. and {Terlevich}, R.},
        title = "{Classification parameters for the emission-line spectra of extragalactic objects.}",
      journal = {\pasp},
         year = 1981,
        month = feb,
       volume = {93},
        pages = {5-19},
          doi = {10.1086/130766},
       adsurl = {https://ui.adsabs.harvard.edu/abs/1981PASP...93....5B}
}

@ARTICLE{For2021,
       author = {{For}, B. -Q. and {Wang}, J. and {Westmeier}, T. and {Wong}, O.~I. and {Murugeshan}, C. and {Staveley-Smith}, L. and {Courtois}, H.~M. and {Pomar{\`e}de}, D. and {Spekkens}, K. and {Catinella}, B. and {McQuinn}, K.~B.~W. and {Elagali}, A. and {Koribalski}, B.~S. and {Lee-Waddell}, K. and {Madrid}, J.~P. and {Popping}, A. and {Reynolds}, T.~N. and {Rhee}, J. and {Bekki}, K. and {D{\`e}nes}, H. and {Kamphuis}, P. and {Verdes-Montenegro}, L.},
        title = "{WALLABY pre-pilot survey: H I content of the Eridanus supergroup}",
      journal = {\mnras},
         year = 2021,
        month = oct,
       volume = {507},
       number = {2},
        pages = {2300-2317},
          doi = {10.1093/mnras/stab2257},
archivePrefix = {arXiv},
       eprint = {2108.04410},
 primaryClass = {astro-ph.GA},
       adsurl = {https://ui.adsabs.harvard.edu/abs/2021MNRAS.507.2300F}
}

@ARTICLE{Kewley2001,
       author = {{Kewley}, L.~J. and {Heisler}, C.~A. and {Dopita}, M.~A. and {Lumsden}, S.},
        title = "{Optical Classification of Southern Warm Infrared Galaxies}",
      journal = {\apjs},
         year = 2001,
        month = jan,
       volume = {132},
       number = {1},
        pages = {37-71},
          doi = {10.1086/318944},
       adsurl = {https://ui.adsabs.harvard.edu/abs/2001ApJS..132...37K}
}

@ARTICLE{ODonnell1994,
       author = {{O'Donnell}, James E.},
        title = "{R v-dependent Optical and Near-Ultraviolet Extinction}",
      journal = {\apj},
         year = 1994,
        month = feb,
       volume = {422},
        pages = {158},
          doi = {10.1086/173713},
       adsurl = {https://ui.adsabs.harvard.edu/abs/1994ApJ...422..158O}
}

@ARTICLE{Kreckel2019,
       author = {{Kreckel}, K. and {Ho}, I. -T. and {Blanc}, G.~A. and {Groves}, B. and {Santoro}, F. and {Schinnerer}, E. and {Bigiel}, F. and {Chevance}, M. and {Congiu}, E. and {Emsellem}, E. and {Faesi}, C. and {Glover}, S.~C.~O. and {Grasha}, K. and {Kruijssen}, J.~M.~D. and {Lang}, P. and {Leroy}, A.~K. and {Meidt}, S.~E. and {McElroy}, R. and {Pety}, J. and {Rosolowsky}, E. and {Saito}, T. and {Sandstrom}, K. and {Sanchez-Blazquez}, P. and {Schruba}, A.},
        title = "{Mapping Metallicity Variations across Nearby Galaxy Disks}",
      journal = {\apj},
         year = 2019,
        month = dec,
       volume = {887},
       number = {1},
          eid = {80},
        pages = {80},
          doi = {10.3847/1538-4357/ab5115},
archivePrefix = {arXiv},
       eprint = {1910.07190},
 primaryClass = {astro-ph.GA},
       adsurl = {https://ui.adsabs.harvard.edu/abs/2019ApJ...887...80K}
}

@ARTICLE{Kauffmann2003,
       author = {{Kauffmann}, Guinevere and {Heckman}, Timothy M. and {Tremonti}, Christy and {Brinchmann}, Jarle and {Charlot}, St{\'e}phane and {White}, Simon D.~M. and {Ridgway}, Susan E. and {Brinkmann}, Jon and {Fukugita}, Masataka and {Hall}, Patrick B. and {Ivezi{\'c}}, {\v{Z}}eljko and {Richards}, Gordon T. and {Schneider}, Donald P.},
        title = "{The host galaxies of active galactic nuclei}",
      journal = {\mnras},
         year = 2003,
        month = dec,
       volume = {346},
       number = {4},
        pages = {1055-1077},
          doi = {10.1111/j.1365-2966.2003.07154.x},
archivePrefix = {arXiv},
       eprint = {astro-ph/0304239},
 primaryClass = {astro-ph},
       adsurl = {https://ui.adsabs.harvard.edu/abs/2003MNRAS.346.1055K}
}

@ARTICLE{Cappellari2017,
       author = {{Cappellari}, Michele},
        title = "{Improving the full spectrum fitting method: accurate convolution with Gauss-Hermite functions}",
      journal = {\mnras},
         year = 2017,
        month = apr,
       volume = {466},
       number = {1},
        pages = {798-811},
          doi = {10.1093/mnras/stw3020},
archivePrefix = {arXiv},
       eprint = {1607.08538},
 primaryClass = {astro-ph.GA},
       adsurl = {https://ui.adsabs.harvard.edu/abs/2017MNRAS.466..798C}
}

@ARTICLE{Pilyugin2016,
       author = {{Pilyugin}, L.~S. and {Grebel}, E.~K.},
        title = "{New calibrations for abundance determinations in H II regions}",
      journal = {\mnras},
         year = 2016,
        month = apr,
       volume = {457},
       number = {4},
        pages = {3678-3692},
          doi = {10.1093/mnras/stw238},
archivePrefix = {arXiv},
       eprint = {1601.08217},
 primaryClass = {astro-ph.GA},
       adsurl = {https://ui.adsabs.harvard.edu/abs/2016MNRAS.457.3678P}
}

@ARTICLE{Vazdekis2016,
       author = {{Vazdekis}, A. and {Koleva}, M. and {Ricciardelli}, E. and {R{\"o}ck}, B. and {Falc{\'o}n-Barroso}, J.},
        title = "{UV-extended E-MILES stellar population models: young components in massive early-type galaxies}",
      journal = {\mnras},
         year = 2016,
        month = dec,
       volume = {463},
       number = {4},
        pages = {3409-3436},
          doi = {10.1093/mnras/stw2231},
archivePrefix = {arXiv},
       eprint = {1612.01187},
 primaryClass = {astro-ph.GA},
       adsurl = {https://ui.adsabs.harvard.edu/abs/2016MNRAS.463.3409V}
}

@ARTICLE{Vincenzo2016,
       author = {{Vincenzo}, F. and {Matteucci}, F. and {Belfiore}, F. and {Maiolino}, R.},
        title = "{Modern yields per stellar generation: the effect of the IMF}",
      journal = {\mnras},
         year = 2016,
        month = feb,
       volume = {455},
       number = {4},
        pages = {4183-4190},
          doi = {10.1093/mnras/stv2598},
archivePrefix = {arXiv},
       eprint = {1503.08300},
 primaryClass = {astro-ph.GA},
       adsurl = {https://ui.adsabs.harvard.edu/abs/2016MNRAS.455.4183V}
}

@ARTICLE{denBrok2021,
       author = {{den Brok}, J.~S. and {Chatzigiannakis}, D. and {Bigiel}, F. and {Puschnig}, J. and {Barnes}, A.~T. and {Leroy}, A.~K. and {Jim{\'e}nez-Donaire}, M.~J. and {Usero}, A. and {Schinnerer}, E. and {Rosolowsky}, E. and {Faesi}, C.~M. and {Grasha}, K. and {Hughes}, A. and {Kruijssen}, J.~M.~D. and {Liu}, D. and {Neumann}, L. and {Pety}, J. and {Querejeta}, M. and {Saito}, T. and {Schruba}, A. and {Stuber}, S.},
        title = "{New constraints on the $^{12}$CO(2-1)/(1-0) line ratio across nearby disc galaxies}",
      journal = {\mnras},
         year = 2021,
        month = jul,
       volume = {504},
       number = {3},
        pages = {3221-3245},
          doi = {10.1093/mnras/stab859},
archivePrefix = {arXiv},
       eprint = {2103.10442},
 primaryClass = {astro-ph.GA},
       adsurl = {https://ui.adsabs.harvard.edu/abs/2021MNRAS.504.3221D}
}

@ARTICLE{Gonzalez-Delgado2016,
       author = {{Gonz{\'a}lez Delgado}, R.~M. and {Cid Fernandes}, R. and {P{\'e}rez}, E. and {Garc{\'\i}a-Benito}, R. and {L{\'o}pez Fern{\'a}ndez}, R. and {Lacerda}, E.~A.~D. and {Cortijo-Ferrero}, C. and {de Amorim}, A.~L. and {Vale Asari}, N. and {S{\'a}nchez}, S.~F. and {Walcher}, C.~J. and {Wisotzki}, L. and {Mast}, D. and {Alves}, J. and {Ascasibar}, Y. and {Bland-Hawthorn}, J. and {Galbany}, L. and {Kennicutt}, R.~C. and {M{\'a}rquez}, I. and {Masegosa}, J. and {Moll{\'a}}, M. and {S{\'a}nchez-Bl{\'a}zquez}, P. and {V{\'\i}lchez}, J.~M.},
        title = "{Star formation along the Hubble sequence. Radial structure of the star formation of CALIFA galaxies}",
      journal = {\aap},
         year = 2016,
        month = may,
       volume = {590},
          eid = {A44},
        pages = {A44},
          doi = {10.1051/0004-6361/201628174},
archivePrefix = {arXiv},
       eprint = {1603.00874},
 primaryClass = {astro-ph.GA},
       adsurl = {https://ui.adsabs.harvard.edu/abs/2016A&A...590A..44G}
}

@ARTICLE{Sanchez2020,
       author = {{S{\'a}nchez}, Sebasti{\'a}n F.},
        title = "{Spatially Resolved Spectroscopic Properties of Low-Redshift Star-Forming Galaxies}",
      journal = {\araa},
         year = 2020,
        month = aug,
       volume = {58},
        pages = {99-155},
          doi = {10.1146/annurev-astro-012120-013326},
archivePrefix = {arXiv},
       eprint = {1911.06925},
 primaryClass = {astro-ph.GA},
       adsurl = {https://ui.adsabs.harvard.edu/abs/2020ARA&A..58...99S}
}

@ARTICLE{Barrera-Ballesteros2023,
       author = {{Barrera-Ballesteros}, J.~K. and {S{\'a}nchez}, S.~F. and {Espinosa-Ponce}, C. and {L{\'o}pez-Cob{\'a}}, C. and {Carigi}, L. and {Lugo-Aranda}, A.~Z. and {Lacerda}, E. and {Bruzual}, G. and {Hernandez-Toledo}, H. and {Boardman}, N. and {Drory}, N. and {Lane}, Richard R. and {Brownstein}, J.~R.},
        title = "{SDSS-IV MaNGA: The Radial Distribution of Physical Properties within Galaxies in the Nearby Universe}",
      journal = {\rmxaa},
         year = 2023,
        month = oct,
       volume = {59},
        pages = {213-258},
          doi = {10.22201/ia.01851101p.2023.59.02.06},
archivePrefix = {arXiv},
       eprint = {2206.07058},
 primaryClass = {astro-ph.GA},
       adsurl = {https://ui.adsabs.harvard.edu/abs/2023RMxAA..59..213B}
}

@ARTICLE{Schruba2011,
       author = {{Schruba}, Andreas and {Leroy}, Adam K. and {Walter}, Fabian and {Bigiel}, Frank and {Brinks}, Elias and {de Blok}, W.~J.~G. and {Dumas}, Gaelle and {Kramer}, Carsten and {Rosolowsky}, Erik and {Sandstrom}, Karin and {Schuster}, Karl and {Usero}, Antonio and {Weiss}, Axel and {Wiesemeyer}, Helmut},
        title = "{A Molecular Star Formation Law in the Atomic-gas-dominated Regime in Nearby Galaxies}",
      journal = {\aj},
         year = 2011,
        month = aug,
       volume = {142},
       number = {2},
          eid = {37},
        pages = {37},
          doi = {10.1088/0004-6256/142/2/37},
archivePrefix = {arXiv},
       eprint = {1105.4605},
 primaryClass = {astro-ph.CO},
       adsurl = {https://ui.adsabs.harvard.edu/abs/2011AJ....142...37S}
}

@ARTICLE{Casasola2017,
       author = {{Casasola}, V. and {Cassar{\`a}}, L.~P. and {Bianchi}, S. and {Verstocken}, S. and {Xilouris}, E. and {Magrini}, L. and {Smith}, M.~W.~L. and {De Looze}, I. and {Galametz}, M. and {Madden}, S.~C. and {Baes}, M. and {Clark}, C. and {Davies}, J. and {De Vis}, P. and {Evans}, R. and {Fritz}, J. and {Galliano}, F. and {Jones}, A.~P. and {Mosenkov}, A.~V. and {Viaene}, S. and {Ysard}, N.},
        title = "{Radial distribution of dust, stars, gas, and star-formation rate in DustPedia face-on galaxies}",
      journal = {\aap},
         year = 2017,
        month = sep,
       volume = {605},
          eid = {A18},
        pages = {A18},
          doi = {10.1051/0004-6361/201731020},
archivePrefix = {arXiv},
       eprint = {1706.05351},
 primaryClass = {astro-ph.GA},
       adsurl = {https://ui.adsabs.harvard.edu/abs/2017A&A...605A..18C}
}

@ARTICLE{Park2024,
       author = {{Park}, Hye-Jin and {Battisti}, Andrew J. and {Wisnioski}, Emily and {Cortese}, Luca and {Seibert}, Mark and {Grasha}, Kathryn and {Madore}, Barry F. and {Groves}, Brent and {Rich}, Jeff A. and {Beaton}, Rachael L. and {Chen}, Qian-Hui and {Mun}, Marcie and {McClure-Griffiths}, Naomi M. and {de Blok}, W.~J.~G. and {Kewley}, Lisa J.},
        title = "{The spatially resolved relation between dust, gas, and metal abundance with the TYPHOON survey}",
      journal = {\mnras},
         year = 2024,
        month = nov,
       volume = {535},
       number = {1},
        pages = {729-752},
          doi = {10.1093/mnras/stae2298},
archivePrefix = {arXiv},
       eprint = {2410.02222},
 primaryClass = {astro-ph.GA},
       adsurl = {https://ui.adsabs.harvard.edu/abs/2024MNRAS.535..729P}
}

@ARTICLE{Meyer2017,
       author = {{Meyer}, Martin and {Robotham}, Aaron and {Obreschkow}, Danail and {Westmeier}, Tobias and {Duffy}, Alan R. and {Staveley-Smith}, Lister},
        title = "{Tracing HI Beyond the Local Universe}",
      journal = {\pasa},
         year = 2017,
        month = nov,
       volume = {34},
        pages = {52},
          doi = {10.1017/pasa.2017.31},
archivePrefix = {arXiv},
       eprint = {1705.04210},
 primaryClass = {astro-ph.CO},
       adsurl = {https://ui.adsabs.harvard.edu/abs/2017PASA...34...52M}
}

@ARTICLE{Koribalski2020,
       author = {{Koribalski}, B{\"a}rbel S. and {Staveley-Smith}, L. and {Westmeier}, T. and {Serra}, P. and {Spekkens}, K. and {Wong}, O.~I. and {Lee-Waddell}, K. and {Lagos}, C.~D.~P. and {Obreschkow}, D. and {Ryan-Weber}, E.~V. and {Zwaan}, M. and {Kilborn}, V. and {Bekiaris}, G. and {Bekki}, K. and {Bigiel}, F. and {Boselli}, A. and {Bosma}, A. and {Catinella}, B. and {Chauhan}, G. and {Cluver}, M.~E. and {Colless}, M. and {Courtois}, H.~M. and {Crain}, R.~A. and {de Blok}, W.~J.~G. and {D{\'e}nes}, H. and {Duffy}, A.~R. and {Elagali}, A. and {Fluke}, C.~J. and {For}, B.-Q. and {Heald}, G. and {Henning}, P.~A. and {Hess}, K.~M. and {Holwerda}, B.~W. and {Howlett}, C. and {Jarrett}, T. and {Jones}, D.~H. and {Jones}, M.~G. and {J{\'o}zsa}, G.~I.~G. and {Jurek}, R. and {J{\"u}tte}, E. and {Kamphuis}, P. and {Karachentsev}, I. and {Kerp}, J. and {Kleiner}, D. and {Kraan-Korteweg}, R.~C. and {L{\'o}pez-S{\'a}nchez}, {\'A}. R. and {Madrid}, J. and {Meyer}, M. and {Mould}, J. and {Murugeshan}, C. and {Norris}, R.~P. and {Oh}, S.-H. and {Oosterloo}, T.~A. and {Popping}, A. and {Putman}, M. and {Reynolds}, T.~N. and {Rhee}, J. and {Robotham}, A.~S.~G. and {Ryder}, S. and {Schr{\"o}der}, A.~C. and {Shao}, Li and {Stevens}, A.~R.~H. and {Taylor}, E.~N. and {van{\^A} der Hulst}, J.~M. and {Verdes-Montenegro}, L. and {Wakker}, B.~P. and {Wang}, J. and {Whiting}, M. and {Winkel}, B. and {Wolf}, C.},
        title = "{WALLABY {\textendash} an SKA Pathfinder H I survey}",
      journal = {\apss},
         year = 2020,
        month = jul,
       volume = {365},
       number = {7},
          eid = {118},
        pages = {118},
          doi = {10.1007/s10509-020-03831-4},
archivePrefix = {arXiv},
       eprint = {2002.07311},
 primaryClass = {astro-ph.GA},
       adsurl = {https://ui.adsabs.harvard.edu/abs/2020Ap&SS.365..118K}
}

@ARTICLE{Murugeshan2024,
       author = {{Murugeshan}, C. and {Deg}, N. and {Westmeier}, T. and {Shen}, A.~X. and {For}, B.-Q. and {Spekkens}, K. and {Wong}, O.~I. and {Staveley-Smith}, L. and {Catinella}, B. and {Lee-Waddell}, K. and {D{\'e}nes}, H. and {Rhee}, J. and {Cortese}, L. and {Goliath}, S. and {Halloran}, R. and {van der Hulst}, J.~M. and {Kamphuis}, P. and {Koribalski}, B.~S. and {Kraan-Korteweg}, R.~C. and {Lelli}, F. and {Venkataraman}, P. and {Verdes-Montenegro}, L. and {Yu}, N.},
        title = "{WALLABY Pilot Survey: Public data release of {\ensuremath{\sim}} 1800 H I sources and high-resolution cut-outs from Pilot Survey Phase 2}",
      journal = {\pasa},
         year = 2024,
        month = nov,
       volume = {41},
          eid = {e088},
        pages = {e088},
          doi = {10.1017/pasa.2024.91},
archivePrefix = {arXiv},
       eprint = {2409.13130},
 primaryClass = {astro-ph.GA},
       adsurl = {https://ui.adsabs.harvard.edu/abs/2024PASA...41...88M}
}

@ARTICLE{Westmeier2022,
       author = {{Westmeier}, T. and {Deg}, N. and {Spekkens}, K. and {Reynolds}, T.~N. and {Shen}, A.~X. and {Gaudet}, S. and {Goliath}, S. and {Huynh}, M.~T. and {Venkataraman}, P. and {Lin}, X. and {O'Beirne}, T. and {Catinella}, B. and {Cortese}, L. and {D{\'e}nes}, H. and {Elagali}, A. and {For}, B.-Q. and {J{\'o}zsa}, G.~I.~G. and {Howlett}, C. and {van der Hulst}, J.~M. and {Jurek}, R.~J. and {Kamphuis}, P. and {Kilborn}, V.~A. and {Kleiner}, D. and {Koribalski}, B.~S. and {Lee-Waddell}, K. and {Murugeshan}, C. and {Rhee}, J. and {Serra}, P. and {Shao}, L. and {Staveley-Smith}, L. and {Wang}, J. and {Wong}, O.~I. and {Zwaan}, M.~A. and {Allison}, J.~R. and {Anderson}, C.~S. and {Ball}, Lewis and {Bock}, D.~C.-J. and {Brodrick}, D. and {Bunton}, J.~D. and {Cooray}, F.~R. and {Gupta}, N. and {Hayman}, D.~B. and {Mahony}, E.~K. and {Moss}, V.~A. and {Ng}, A. and {Pearce}, S.~E. and {Raja}, W. and {Roxby}, D.~N. and {Voronkov}, M.~A. and {Warhurst}, K.~A. and {Courtois}, H.~M. and {Said}, K.},
        title = "{WALLABY pilot survey: Public release of H I data for almost 600 galaxies from phase 1 of ASKAP pilot observations}",
      journal = {\pasa},
         year = 2022,
        month = nov,
       volume = {39},
          eid = {e058},
        pages = {e058},
          doi = {10.1017/pasa.2022.50},
archivePrefix = {arXiv},
       eprint = {2211.07094},
 primaryClass = {astro-ph.GA},
       adsurl = {https://ui.adsabs.harvard.edu/abs/2022PASA...39...58W}
}

@ARTICLE{Deg2022,
       author = {{Deg}, N. and {Spekkens}, K. and {Westmeier}, T. and {Reynolds}, T.~N. and {Venkataraman}, P. and {Goliath}, S. and {Shen}, A.~X. and {Halloran}, R. and {Bosma}, A. and {Catinella}, B. and {de Blok}, W.~J.~G. and {D{\'e}nes}, H. and {Di Teodoro}, E.~M. and {Elagali}, A. and {For}, B.-Q. and {Howlett}, C. and {J{\'o}zsa}, G.~I.~G. and {Kamphuis}, P. and {Kleiner}, D. and {Koribalski}, B. and {Lee-Waddell}, K. and {Lelli}, F. and {Lin}, X. and {Murugeshan}, C. and {Oh}, S. and {Rhee}, J. and {Scott}, T.~C. and {Staveley-Smith}, L. and {van der Hulst}, J.~M. and {Verdes-Montenegro}, L. and {Wang}, J. and {Wong}, O.~I.},
        title = "{WALLABY Pilot Survey: Public release of HI kinematic models for more than 100 galaxies from phase 1 of ASKAP pilot observations}",
      journal = {\pasa},
         year = 2022,
        month = nov,
       volume = {39},
          eid = {e059},
        pages = {e059},
          doi = {10.1017/pasa.2022.43},
archivePrefix = {arXiv},
       eprint = {2211.07333},
 primaryClass = {astro-ph.GA},
       adsurl = {https://ui.adsabs.harvard.edu/abs/2022PASA...39...59D}
}

@ARTICLE{Kreckel2025,
       author = {{Kreckel}, K. and {Rickards Vaught}, R.~J. and {Egorov}, O.~V. and {M{\'e}ndez-Delgado}, J.~E. and {Belfiore}, F. and {Brazzini}, M. and {Egorova}, E. and {Congiu}, E. and {Dale}, D.~A. and {Dlamini}, S. and {Glover}, S.~C.~O. and {Grasha}, K. and {Klessen}, R.~S. and {Liang}, F.-H. and {Pan}, H.-A. and {S{\'a}nchez-Bl{\'a}zquez}, P. and {Williams}, T.~G.},
        title = "{Temperature-based radial metallicity gradients in nearby galaxies}",
      journal = {\aap},
         year = 2025,
        month = nov,
       volume = {703},
          eid = {A42},
        pages = {A42},
          doi = {10.1051/0004-6361/202556017},
archivePrefix = {arXiv},
       eprint = {2507.20744},
 primaryClass = {astro-ph.GA},
       adsurl = {https://ui.adsabs.harvard.edu/abs/2025A&A...703A..42K}
}

@ARTICLE{Brazzini2024,
       author = {{Brazzini}, Matilde and {Belfiore}, Francesco and {Ginolfi}, Michele and {Groves}, Brent and {Kreckel}, Kathryn and {Rickards Vaught}, Ryan J. and {Baron}, Dalya and {Bigiel}, Frank and {Blanc}, Guillermo A. and {Dale}, Daniel A. and {Grasha}, Kathryn and {Habjan}, Eric and {Klessen}, Ralf S. and {M{\'e}ndez-Delgado}, Jose Eduardo and {Sandstrom}, Karin and {Williams}, Thomas G.},
        title = "{Metallicity calibrations based on auroral lines from PHANGS{\textendash}MUSE data}",
      journal = {\aap},
         year = 2024,
        month = nov,
       volume = {691},
          eid = {A173},
        pages = {A173},
          doi = {10.1051/0004-6361/202451007},
archivePrefix = {arXiv},
       eprint = {2410.00106},
 primaryClass = {astro-ph.GA},
       adsurl = {https://ui.adsabs.harvard.edu/abs/2024A&A...691A.173B}
}

@ARTICLE{Emsellem2022,
       author = {{Emsellem}, Eric and {Schinnerer}, Eva and {Santoro}, Francesco and {Belfiore}, Francesco and {Pessa}, Ismael and {McElroy}, Rebecca and {Blanc}, Guillermo A. and {Congiu}, Enrico and {Groves}, Brent and {Ho}, I.-Ting and {Kreckel}, Kathryn and {Razza}, Alessandro and {Sanchez-Blazquez}, Patricia and {Egorov}, Oleg and {Faesi}, Chris and {Klessen}, Ralf S. and {Leroy}, Adam K. and {Meidt}, Sharon and {Querejeta}, Miguel and {Rosolowsky}, Erik and {Scheuermann}, Fabian and {Anand}, Gagandeep S. and {Barnes}, Ashley T. and {Be{\v{s}}li{\'c}}, Ivana and {Bigiel}, Frank and {Boquien}, M{\'e}d{\'e}ric and {Cao}, Yixian and {Chevance}, M{\'e}lanie and {Dale}, Daniel A. and {Eibensteiner}, Cosima and {Glover}, Simon C.~O. and {Grasha}, Kathryn and {Henshaw}, Jonathan D. and {Hughes}, Annie and {Koch}, Eric W. and {Kruijssen}, J.~M. Diederik and {Lee}, Janice and {Liu}, Daizhong and {Pan}, Hsi-An and {Pety}, J{\'e}r{\^o}me and {Saito}, Toshiki and {Sandstrom}, Karin M. and {Schruba}, Andreas and {Sun}, Jiayi and {Thilker}, David A. and {Usero}, Antonio and {Watkins}, Elizabeth J. and {Williams}, Thomas G.},
        title = "{The PHANGS-MUSE survey. Probing the chemo-dynamical evolution of disc galaxies}",
      journal = {\aap},
         year = 2022,
        month = mar,
       volume = {659},
          eid = {A191},
        pages = {A191},
          doi = {10.1051/0004-6361/202141727},
archivePrefix = {arXiv},
       eprint = {2110.03708},
 primaryClass = {astro-ph.GA},
       adsurl = {https://ui.adsabs.harvard.edu/abs/2022A&A...659A.191E}
}

@ARTICLE{Kang2025,
       author = {{Kang}, Xiaoyu and {Kudritzki}, Rolf-Peter and {Gong}, Xiaobo and {Zhang}, Fenghui},
        title = "{Evolution and star formation history of NGC 300 from a chemical evolution model with radial gas inflows}",
      journal = {\aap},
         year = 2025,
        month = sep,
       volume = {701},
          eid = {A2},
        pages = {A2},
          doi = {10.1051/0004-6361/202554108},
archivePrefix = {arXiv},
       eprint = {2507.10245},
 primaryClass = {astro-ph.GA},
       adsurl = {https://ui.adsabs.harvard.edu/abs/2025A&A...701A...2K}
}

@ARTICLE{Portinari2000,
       author = {{Portinari}, L. and {Chiosi}, C.},
        title = "{On radial gas flows, the Galactic Bar and chemical evolution in the Galactic Disc}",
      journal = {\aap},
         year = 2000,
        month = mar,
       volume = {355},
        pages = {929-948},
          doi = {10.48550/arXiv.astro-ph/0002145},
archivePrefix = {arXiv},
       eprint = {astro-ph/0002145},
 primaryClass = {astro-ph},
       adsurl = {https://ui.adsabs.harvard.edu/abs/2000A&A...355..929P}
}

@ARTICLE{Spitoni2011,
       author = {{Spitoni}, E. and {Matteucci}, F.},
        title = "{Effects of the radial flows on the chemical evolution of the Milky Way disk}",
      journal = {\aap},
         year = 2011,
        month = jul,
       volume = {531},
          eid = {A72},
        pages = {A72},
          doi = {10.1051/0004-6361/201015749},
archivePrefix = {arXiv},
       eprint = {1104.4881},
 primaryClass = {astro-ph.GA},
       adsurl = {https://ui.adsabs.harvard.edu/abs/2011A&A...531A..72S}
}

@ARTICLE{BP2000,
       author = {{Boissier}, S. and {Prantzos}, N.},
        title = "{Chemo-spectrophotometric evolution of spiral galaxies - II. Main properties of present-day disc galaxies}",
      journal = {\mnras},
         year = 2000,
        month = feb,
       volume = {312},
       number = {2},
        pages = {398-416},
          doi = {10.1046/j.1365-8711.2000.03133.x},
       adsurl = {https://ui.adsabs.harvard.edu/abs/2000MNRAS.312..398B}
}

@ARTICLE{Bresolin2019,
       author = {{Bresolin}, Fabio},
        title = "{Metallicity gradients in small and nearby spiral galaxies}",
      journal = {\mnras},
         year = 2019,
        month = sep,
       volume = {488},
       number = {3},
        pages = {3826-3843},
          doi = {10.1093/mnras/stz1947},
archivePrefix = {arXiv},
       eprint = {1907.05071},
 primaryClass = {astro-ph.GA},
       adsurl = {https://ui.adsabs.harvard.edu/abs/2019MNRAS.488.3826B}
}

@ARTICLE{Bresolin2025,
       author = {{Bresolin}, Fabio and {Kudritzki}, Rolf-Peter and {Urbaneja}, Miguel A. and {Sextl}, Eva and {Riess}, Adam G.},
        title = "{Blue Supergiants in the Pinwheel Galaxy M101: Comparison with H II Region Chemical Abundances, Spectroscopic Distance, and an Independent Determination of the Hubble Constant}",
      journal = {\apj},
         year = 2025,
        month = oct,
       volume = {991},
       number = {2},
          eid = {151},
        pages = {151},
          doi = {10.3847/1538-4357/adfc4c},
archivePrefix = {arXiv},
       eprint = {2508.11837},
 primaryClass = {astro-ph.GA},
       adsurl = {https://ui.adsabs.harvard.edu/abs/2025ApJ...991..151B}
}

@ARTICLE{Ho2015,
       author = {{Ho}, I.-Ting and {Kudritzki}, Rolf-Peter and {Kewley}, Lisa J. and {Zahid}, H. Jabran and {Dopita}, Michael A. and {Bresolin}, Fabio and {Rupke}, David S.~N.},
        title = "{Metallicity gradients in local field star-forming galaxies: insights on inflows, outflows, and the coevolution of gas, stars and metals}",
      journal = {\mnras},
         year = 2015,
        month = apr,
       volume = {448},
       number = {3},
        pages = {2030-2054},
          doi = {10.1093/mnras/stv067},
archivePrefix = {arXiv},
       eprint = {1501.02668},
 primaryClass = {astro-ph.GA},
       adsurl = {https://ui.adsabs.harvard.edu/abs/2015MNRAS.448.2030H}
}

@ARTICLE{Kubryk2015,
       author = {{Kubryk}, M. and {Prantzos}, N. and {Athanassoula}, E.},
        title = "{Evolution of the Milky Way with radial motions of stars and gas. I. The solar neighbourhood and the thin and thick disks}",
      journal = {\aap},
         year = 2015,
        month = aug,
       volume = {580},
          eid = {A126},
        pages = {A126},
          doi = {10.1051/0004-6361/201424171},
archivePrefix = {arXiv},
       eprint = {1412.0585},
 primaryClass = {astro-ph.GA},
       adsurl = {https://ui.adsabs.harvard.edu/abs/2015A&A...580A.126K}
}

@ARTICLE{Kudritzki2015,
       author = {{Kudritzki}, Rolf-Peter and {Ho}, I.-Ting and {Schruba}, Andreas and {Burkert}, Andreas and {Zahid}, H. Jabran and {Bresolin}, Fabio and {Dima}, Gabriel I.},
        title = "{The chemical evolution of local star-forming galaxies: radial profiles of ISM metallicity, gas mass, and stellar mass and constraints on galactic accretion and winds}",
      journal = {\mnras},
         year = 2015,
        month = jun,
       volume = {450},
       number = {1},
        pages = {342-359},
          doi = {10.1093/mnras/stv522},
archivePrefix = {arXiv},
       eprint = {1503.01503},
 primaryClass = {astro-ph.GA},
       adsurl = {https://ui.adsabs.harvard.edu/abs/2015MNRAS.450..342K}
}

@ARTICLE{ZhangK2017,
       author = {{Zhang}, Kai and {Yan}, Renbin and {Bundy}, Kevin and {Bershady}, Matthew and {Haffner}, L. Matthew and {Walterbos}, Ren{\'e} and {Maiolino}, Roberto and {Tremonti}, Christy and {Thomas}, Daniel and {Drory}, Niv and {Jones}, Amy and {Belfiore}, Francesco and {S{\'a}nchez}, Sebastian F. and {Diamond-Stanic}, Aleksandar M. and {Bizyaev}, Dmitry and {Nitschelm}, Christian and {Andrews}, Brett and {Brinkmann}, Jon and {Brownstein}, Joel R. and {Cheung}, Edmond and {Li}, Cheng and {Law}, David R. and {Roman Lopes}, Alexandre and {Oravetz}, Daniel and {Pan}, Kaike and {Storchi Bergmann}, Thaisa and {Simmons}, Audrey},
        title = "{SDSS-IV MaNGA: the impact of diffuse ionized gas on emission-line ratios, interpretation of diagnostic diagrams and gas metallicity measurements}",
      journal = {\mnras},
         year = 2017,
        month = apr,
       volume = {466},
       number = {3},
        pages = {3217-3243},
          doi = {10.1093/mnras/stw3308},
archivePrefix = {arXiv},
       eprint = {1612.02000},
 primaryClass = {astro-ph.GA},
       adsurl = {https://ui.adsabs.harvard.edu/abs/2017MNRAS.466.3217Z}
}

@ARTICLE{Pichon2011,
       author = {{Pichon}, C. and {Pogosyan}, D. and {Kimm}, T. and {Slyz}, A. and {Devriendt}, J. and {Dubois}, Y.},
        title = "{Rigging dark haloes: why is hierarchical galaxy formation consistent with the inside-out build-up of thin discs?}",
      journal = {\mnras},
         year = 2011,
        month = dec,
       volume = {418},
       number = {4},
        pages = {2493-2507},
          doi = {10.1111/j.1365-2966.2011.19640.x},
archivePrefix = {arXiv},
       eprint = {1105.0210},
 primaryClass = {astro-ph.CO},
       adsurl = {https://ui.adsabs.harvard.edu/abs/2011MNRAS.418.2493P}
}

@ARTICLE{2016Ap&SS.361...61D,
       author = {{Dopita}, Michael A. and {Kewley}, Lisa J. and {Sutherland}, Ralph S. and {Nicholls}, David C.},
        title = "{Chemical abundances in high-redshift galaxies: a powerful new emission line diagnostic}",
      journal = {\apss},
         year = 2016,
        month = feb,
       volume = {361},
          eid = {61},
        pages = {61},
          doi = {10.1007/s10509-016-2657-8},
archivePrefix = {arXiv},
       eprint = {1601.01337},
 primaryClass = {astro-ph.GA},
       adsurl = {https://ui.adsabs.harvard.edu/abs/2016Ap&SS.361...61D}
}

@BOOK{Vaucouleurs1991,
       author = {{de Vaucouleurs}, Gerard and {de Vaucouleurs}, Antoinette and {Corwin}, Jr., Herold G. and {Buta}, Ronald J. and {Paturel}, Georges and {Fouque}, Pascal},
        title = "{Third Reference Catalogue of Bright Galaxies}",
         year = 1991,
       volume = {1},
        publisher= {Springer-Verlag: New York},
       adsurl = {https://ui.adsabs.harvard.edu/abs/1991rc3..book.....D}
}

@ARTICLE{Iles2025,
       author = {{Iles}, Elizabeth J. and {Bland-Hawthorn}, Joss and {Crawford}, Courtney and {Croom}, Scott M. and {Davis}, Hillary and {Pedersen}, May Gade and {Green}, Anne and {Gunawardhana}, Madusha and {Icaza-Lizaola}, Miguel and {Johnston}, Helen and {Kerrison}, Emily F. and {Mai}, Yifan and {Montet}, Benjamin T. and {Rose}, Kovi and {Rutherford}, Tomas and {Saraf}, Manasvee and {Sirks}, Ellen L. and {Spalding}, Eckhart and {Tuntipong}, Sujeeporn and {van de Sande}, Jesse and {Yamsiri}, Pavadol},
        title = "{findAbar: How astronomers may perceive the bar in galaxies differently}",
      journal = {\pasa},
         year = 2025,
        month = nov,
       volume = {42},
          eid = {e166},
        pages = {e166},
          doi = {10.1017/pasa.2025.10126},
archivePrefix = {arXiv},
       eprint = {2511.09908},
 primaryClass = {astro-ph.GA},
       adsurl = {https://ui.adsabs.harvard.edu/abs/2025PASA...42..166I}
}

@ARTICLE{Buta2015,
       author = {{Buta}, Ronald J. and {Sheth}, Kartik and {Athanassoula}, E. and {Bosma}, A. and {Knapen}, Johan H. and {Laurikainen}, Eija and {Salo}, Heikki and {Elmegreen}, Debra and {Ho}, Luis C. and {Zaritsky}, Dennis and {Courtois}, Helene and {Hinz}, Joannah L. and {Mu{\~n}oz-Mateos}, Juan-Carlos and {Kim}, Taehyun and {Regan}, Michael W. and {Gadotti}, Dimitri A. and {Gil de Paz}, Armando and {Laine}, Jarkko and {Men{\'e}ndez-Delmestre}, Kar{\'\i}n and {Comer{\'o}n}, S{\'e}bastien and {Erroz Ferrer}, Santiago and {Seibert}, Mark and {Mizusawa}, Trisha and {Holwerda}, Benne and {Madore}, Barry F.},
        title = "{A Classical Morphological Analysis of Galaxies in the Spitzer Survey of Stellar Structure in Galaxies (S4G)}",
      journal = {\apjs},
         year = 2015,
        month = apr,
       volume = {217},
       number = {2},
          eid = {32},
        pages = {32},
          doi = {10.1088/0067-0049/217/2/32},
archivePrefix = {arXiv},
       eprint = {1501.00454},
 primaryClass = {astro-ph.GA},
       adsurl = {https://ui.adsabs.harvard.edu/abs/2015ApJS..217...32B}
}

@ARTICLE{Scheuermann2022,
       author = {{Scheuermann}, Fabian and {Kreckel}, Kathryn and {Anand}, Gagandeep S. and {Blanc}, Guillermo A. and {Congiu}, Enrico and {Santoro}, Francesco and {Van Dyk}, Schuyler D. and {Barnes}, Ashley T. and {Bigiel}, Frank and {Glover}, Simon C.~O. and {Groves}, Brent and {Klessen}, Ralf S. and {Kruijssen}, J.~M. Diederik and {Rosolowsky}, Erik and {Schinnerer}, Eva and {Schruba}, Andreas and {Watkins}, Elizabeth J. and {Williams}, Thomas G.},
        title = "{Planetary nebula luminosity function distances for 19 galaxies observed by PHANGS-MUSE}",
      journal = {\mnras},
         year = 2022,
        month = apr,
       volume = {511},
       number = {4},
        pages = {6087-6109},
          doi = {10.1093/mnras/stac110},
archivePrefix = {arXiv},
       eprint = {2201.04641},
 primaryClass = {astro-ph.GA},
       adsurl = {https://ui.adsabs.harvard.edu/abs/2022MNRAS.511.6087S}
}

@ARTICLE{Jacoby2024,
       author = {{Jacoby}, George H. and {Ciardullo}, Robin and {Roth}, Martin M. and {Arnaboldi}, Magda and {Weilbacher}, Peter M.},
        title = "{Toward Precision Cosmology with Improved Planetary Nebula Luminosity Function Distances Using VLT-MUSE. II. A Test Sample from Archival Data}",
      journal = {\apjs},
         year = 2024,
        month = apr,
       volume = {271},
       number = {2},
          eid = {40},
        pages = {40},
          doi = {10.3847/1538-4365/ad2166},
archivePrefix = {arXiv},
       eprint = {2309.11603},
 primaryClass = {astro-ph.CO},
       adsurl = {https://ui.adsabs.harvard.edu/abs/2024ApJS..271...40J}
}

@ARTICLE{Shaya2017,
       author = {{Shaya}, Edward J. and {Tully}, R. Brent and {Hoffman}, Yehuda and {Pomar{\`e}de}, Daniel},
        title = "{Action Dynamics of the Local Supercluster}",
      journal = {\apj},
         year = 2017,
        month = dec,
       volume = {850},
       number = {2},
          eid = {207},
        pages = {207},
          doi = {10.3847/1538-4357/aa9525},
archivePrefix = {arXiv},
       eprint = {1710.08935},
 primaryClass = {astro-ph.CO},
       adsurl = {https://ui.adsabs.harvard.edu/abs/2017ApJ...850..207S}
}

@ARTICLE{PP04,
       author = {{Pettini}, Max and {Pagel}, Bernard E.~J.},
        title = "{[OIII]/[NII] as an abundance indicator at high redshift}",
      journal = {\mnras},
         year = 2004,
        month = mar,
       volume = {348},
       number = {3},
        pages = {L59-L63},
          doi = {10.1111/j.1365-2966.2004.07591.x},
archivePrefix = {arXiv},
       eprint = {astro-ph/0401128},
 primaryClass = {astro-ph},
       adsurl = {https://ui.adsabs.harvard.edu/abs/2004MNRAS.348L..59P}
}

@ARTICLE{PowerBin2025,
       author = {{Cappellari}, Michele},
        title = "{PowerBin: fast adaptive data binning with Centroidal Power Diagrams}",
      journal = {\mnras},
         year = 2025,
        month = dec,
       volume = {544},
       number = {2},
        pages = {1432-1446},
          doi = {10.1093/mnras/staf1726},
archivePrefix = {arXiv},
       eprint = {2509.06903},
 primaryClass = {astro-ph.IM},
       adsurl = {https://ui.adsabs.harvard.edu/abs/2025MNRAS.544.1432C}
}

@ARTICLE{Cappellari2023,
       author = {{Cappellari}, Michele},
        title = "{Full spectrum fitting with photometry in PPXF: stellar population versus dynamical masses, non-parametric star formation history and metallicity for 3200 LEGA-C galaxies at redshift z {\ensuremath{\approx}} 0.8}",
      journal = {\mnras},
         year = 2023,
        month = dec,
       volume = {526},
       number = {3},
        pages = {3273-3300},
          doi = {10.1093/mnras/stad2597},
archivePrefix = {arXiv},
       eprint = {2208.14974},
 primaryClass = {astro-ph.GA},
       adsurl = {https://ui.adsabs.harvard.edu/abs/2023MNRAS.526.3273C}
}

@article{Davidson2008,
title = {The wild bootstrap, tamed at last},
journal = {Journal of Econometrics},
volume = {146},
number = {1},
pages = {162-169},
year = {2008},
issn = {0304-4076},
doi = {https://doi.org/10.1016/j.jeconom.2008.08.003},
url = {https://www.sciencedirect.com/science/article/pii/S0304407608000833},
author = {Russell Davidson and Emmanuel Flachaire}
}

@ARTICLE{Sanchez2006,
       author = {{S{\'a}nchez-Bl{\'a}zquez}, P. and {Peletier}, R.~F. and {Jim{\'e}nez-Vicente}, J. and {Cardiel}, N. and {Cenarro}, A.~J. and {Falc{\'o}n-Barroso}, J. and {Gorgas}, J. and {Selam}, S. and {Vazdekis}, A.},
        title = "{Medium-resolution Isaac Newton Telescope library of empirical spectra}",
      journal = {\mnras},
         year = 2006,
        month = sep,
       volume = {371},
       number = {2},
        pages = {703-718},
          doi = {10.1111/j.1365-2966.2006.10699.x},
archivePrefix = {arXiv},
       eprint = {astro-ph/0607009},
 primaryClass = {astro-ph},
       adsurl = {https://ui.adsabs.harvard.edu/abs/2006MNRAS.371..703S}
}

@ARTICLE{Barroso2011,
       author = {{Falc{\'o}n-Barroso}, J. and {S{\'a}nchez-Bl{\'a}zquez}, P. and {Vazdekis}, A. and {Ricciardelli}, E. and {Cardiel}, N. and {Cenarro}, A.~J. and {Gorgas}, J. and {Peletier}, R.~F.},
        title = "{An updated MILES stellar library and stellar population models}",
      journal = {\aap},
         year = 2011,
        month = aug,
       volume = {532},
          eid = {A95},
        pages = {A95},
          doi = {10.1051/0004-6361/201116842},
archivePrefix = {arXiv},
       eprint = {1107.2303},
 primaryClass = {astro-ph.CO},
       adsurl = {https://ui.adsabs.harvard.edu/abs/2011A&A...532A..95F}
}

@ARTICLE{Veronese2025,
       author = {{Veronese}, S. and {de Blok}, W.~J.~G. and {Fraternali}, F. and {Maccagni}, F.~M. and {Healy}, J. and {Kleiner}, D. and {Oosterloo}, T.~A. and {Morganti}, R.},
        title = "{The inside-out quenching of the MHONGOOSE galaxy NGC 1371}",
      journal = {\aap},
         year = 2025,
        month = nov,
       volume = {703},
          eid = {A249},
        pages = {A249},
          doi = {10.1051/0004-6361/202555735},
archivePrefix = {arXiv},
       eprint = {2509.18728},
 primaryClass = {astro-ph.GA},
       adsurl = {https://ui.adsabs.harvard.edu/abs/2025A&A...703A.249V}
}

@ARTICLE{Calzetti2000,
       author = {{Calzetti}, Daniela and {Armus}, Lee and {Bohlin}, Ralph C. and {Kinney}, Anne L. and {Koornneef}, Jan and {Storchi-Bergmann}, Thaisa},
        title = "{The Dust Content and Opacity of Actively Star-forming Galaxies}",
      journal = {\apj},
         year = 2000,
        month = apr,
       volume = {533},
       number = {2},
        pages = {682-695},
          doi = {10.1086/308692},
archivePrefix = {arXiv},
       eprint = {astro-ph/9911459},
 primaryClass = {astro-ph},
       adsurl = {https://ui.adsabs.harvard.edu/abs/2000ApJ...533..682C}
}

@ARTICLE{CidFernandes2013,
       author = {{Cid Fernandes}, R. and {P{\'e}rez}, E. and {Garc{\'\i}a Benito}, R. and {Gonz{\'a}lez Delgado}, R.~M. and {de Amorim}, A.~L. and {S{\'a}nchez}, S.~F. and {Husemann}, B. and {Falc{\'o}n Barroso}, J. and {S{\'a}nchez-Bl{\'a}zquez}, P. and {Walcher}, C.~J. and {Mast}, D.},
        title = "{Resolving galaxies in time and space. I. Applying STARLIGHT to CALIFA datacubes}",
      journal = {\aap},
         year = 2013,
        month = sep,
       volume = {557},
          eid = {A86},
        pages = {A86},
          doi = {10.1051/0004-6361/201220616},
archivePrefix = {arXiv},
       eprint = {1304.5788},
 primaryClass = {astro-ph.CO},
       adsurl = {https://ui.adsabs.harvard.edu/abs/2013A&A...557A..86C}
}

@ARTICLE{Easeman2024,
       author = {{Easeman}, Bethan and {Schady}, Patricia and {Wuyts}, Stijn and {Yates}, Robert M.},
        title = "{Optimal metallicity diagnostics for MUSE observations of low-z galaxies}",
      journal = {\mnras},
         year = 2024,
        month = jan,
       volume = {527},
       number = {3},
        pages = {5484-5502},
          doi = {10.1093/mnras/stad3464},
archivePrefix = {arXiv},
       eprint = {2311.03514},
 primaryClass = {astro-ph.GA},
       adsurl = {https://ui.adsabs.harvard.edu/abs/2024MNRAS.527.5484E}
}

@ARTICLE{Teimoorinia2021,
       author = {{Teimoorinia}, Hossen and {Jalilkhany}, Mansoureh and {Scudder}, Jillian M. and {Jensen}, Jaclyn and {Ellison}, Sara L.},
        title = "{A reassessment of strong line metallicity conversions in the machine learning era}",
      journal = {\mnras},
         year = 2021,
        month = may,
       volume = {503},
       number = {1},
        pages = {1082-1095},
          doi = {10.1093/mnras/stab466},
archivePrefix = {arXiv},
       eprint = {2102.07058},
 primaryClass = {astro-ph.GA},
       adsurl = {https://ui.adsabs.harvard.edu/abs/2021MNRAS.503.1082T}
}

@ARTICLE{Eldridge2017,
       author = {{Eldridge}, J.~J. and {Stanway}, E.~R. and {Xiao}, L. and {McClelland}, L.~A.~S. and {Taylor}, G. and {Ng}, M. and {Greis}, S.~M.~L. and {Bray}, J.~C.},
        title = "{Binary Population and Spectral Synthesis Version 2.1: Construction, Observational Verification, and New Results}",
      journal = {\pasa},
         year = 2017,
        month = nov,
       volume = {34},
          eid = {e058},
        pages = {e058},
          doi = {10.1017/pasa.2017.51},
archivePrefix = {arXiv},
       eprint = {1710.02154},
 primaryClass = {astro-ph.SR},
       adsurl = {https://ui.adsabs.harvard.edu/abs/2017PASA...34...58E}
}

@ARTICLE{Smith2002,
       author = {{Smith}, Linda J. and {Norris}, Richard P.~F. and {Crowther}, Paul A.},
        title = "{Realistic ionizing fluxes for young stellar populations from 0.05 to 2 Z$_{solar}$}",
      journal = {\mnras},
         year = 2002,
        month = dec,
       volume = {337},
       number = {4},
        pages = {1309-1328},
          doi = {10.1046/j.1365-8711.2002.06042.x},
archivePrefix = {arXiv},
       eprint = {astro-ph/0207554},
 primaryClass = {astro-ph},
       adsurl = {https://ui.adsabs.harvard.edu/abs/2002MNRAS.337.1309S}
}

@ARTICLE{Lancon2000,
       author = {{Lan{\c{c}}on}, A. and {Wood}, P.~R.},
        title = "{A library of 0.5 to 2.5 mu m spectra of luminous cool stars}",
      journal = {\aaps},
         year = 2000,
        month = oct,
       volume = {146},
        pages = {217-249},
          doi = {10.1051/aas:2000269},
       adsurl = {https://ui.adsabs.harvard.edu/abs/2000A&AS..146..217L}
}

@ARTICLE{Rauch2003,
       author = {{Rauch}, T.},
        title = "{A grid of synthetic ionizing spectra for very hot compact stars from NLTE model atmospheres}",
      journal = {\aap},
         year = 2003,
        month = may,
       volume = {403},
        pages = {709-714},
          doi = {10.1051/0004-6361:20030412},
archivePrefix = {arXiv},
       eprint = {astro-ph/0303464},
 primaryClass = {astro-ph},
       adsurl = {https://ui.adsabs.harvard.edu/abs/2003A&A...403..709R}
}

@ARTICLE{Aringer2009,
       author = {{Aringer}, B. and {Girardi}, L. and {Nowotny}, W. and {Marigo}, P. and {Lederer}, M.~T.},
        title = "{Synthetic photometry for carbon rich giants. I. Hydrostatic dust-free models}",
      journal = {\aap},
         year = 2009,
        month = sep,
       volume = {503},
       number = {3},
        pages = {913-928},
          doi = {10.1051/0004-6361/200911703},
archivePrefix = {arXiv},
       eprint = {0905.4415},
 primaryClass = {astro-ph.SR},
       adsurl = {https://ui.adsabs.harvard.edu/abs/2009A&A...503..913A}
}

@ARTICLE{Barnes2026,
       author = {{Barnes}, A.~T. and {Chandar}, R. and {Kreckel}, K. and {Belfiore}, F. and {Pathak}, D. and {Thilker}, D. and {Leroy}, A.~K. and {Groves}, B. and {Glover}, S.~C.~O. and {McClain}, R. and {Amiri}, A. and {Bazzi}, Z. and {Boquien}, M. and {Congiu}, E. and {Dale}, D.~A. and {Egorov}, O.~V. and {Emsellem}, E. and {Grasha}, K. and {Gonzalez Lobos}, J. and {Henny}, K. and {He}, H. and {Indebetouw}, R. and {Lee}, J.~C. and {Li}, J. and {Liang}, F.-H. and {Larson}, K. and {Maschmann}, D. and {Meidt}, S.~E. and {Eduardo M{\'e}ndez-Delgado}, J. and {Neumann}, J. and {Pan}, H.-A. and {Querejeta}, M. and {Rosolowsky}, E. and {Sarbadhicary}, S.~K. and {Scheuermann}, F. and {{\'U}beda}, L. and {Williams}, T.~G. and {Weinbeck}, T.~D. and {Whitmore}, B. and {Wofford}, A. and {PHANGS Collaborationn}},
        title = "{The PHANGS-MUSE/HST-H{\ensuremath{\alpha}} nebulae catalogue: Parsec-scale resolved structure, physical conditions, and stellar associations across nearby galaxies}",
      journal = {\aap},
         year = 2026,
        month = feb,
       volume = {706},
          eid = {A95},
        pages = {A95},
          doi = {10.1051/0004-6361/202555751},
archivePrefix = {arXiv},
       eprint = {2510.11778},
 primaryClass = {astro-ph.GA},
       adsurl = {https://ui.adsabs.harvard.edu/abs/2026A&A...706A..95B}
}

@ARTICLE{Dotter2016,
       author = {{Dotter}, Aaron},
        title = "{MESA Isochrones and Stellar Tracks (MIST) 0: Methods for the Construction of Stellar Isochrones}",
      journal = {\apjs},
         year = 2016,
        month = jan,
       volume = {222},
       number = {1},
          eid = {8},
        pages = {8},
          doi = {10.3847/0067-0049/222/1/8},
archivePrefix = {arXiv},
       eprint = {1601.05144},
 primaryClass = {astro-ph.SR},
       adsurl = {https://ui.adsabs.harvard.edu/abs/2016ApJS..222....8D}
}

@software{Conroy2010,
       author = {{Conroy}, Charlie and {Gunn}, James E.},
        title = "{FSPS: Flexible Stellar Population Synthesis}",
 howpublished = {Astrophysics Source Code Library, record ascl:1010.043},
         year = 2010,
        month = oct,
          eid = {ascl:1010.043},
archivePrefix = {ascl},
       eprint = {1010.043},
       adsurl = {https://ui.adsabs.harvard.edu/abs/2010ascl.soft10043C}
}

\end{document}